\documentclass[prd,amssymb,amsmath,amsfonts,nofootinbib,reprint,longbibliography,superscriptaddress, floatfix]{revtex4-2}

\usepackage{graphicx}
\usepackage{lmodern}
\usepackage{amsmath,amssymb}
\usepackage{mathrsfs}
\usepackage{amsfonts}
\usepackage[utf8]{inputenc}
\usepackage{url}
\usepackage[colorlinks]{hyperref}
\usepackage[dvipsnames,x11names,svgnames,rgb,table]{xcolor}
\usepackage{multirow}
\usepackage[normalem]{ulem}
\usepackage{float}
\usepackage{marvosym}
\usepackage{enumerate}
\usepackage{color,soul}
\usepackage{placeins}
\usepackage{bm}
\usepackage{color}
\usepackage{commath}
\allowdisplaybreaks
\usepackage{multirow}
\usepackage{cleveref}
\usepackage{rotating} 
\usepackage{orcidlink}
\usepackage{makecell,tabularx,colortbl}
\usepackage{booktabs,cellspace}
\usepackage[caption=false]{subfig}

\usepackage{float}
\def\TEOBResumS{\texttt{TEOBResumS}}

\DeclareMathOperator{\sign}{sign}

\definecolor{cyan}{rgb}{0,0.9,0.9}
\definecolor{orange}{rgb}{0.9,0.5,0}
\definecolor{magenta}{rgb}{1,0,1}
\definecolor{purple}{rgb}{0.8,0.4,0.8}
\definecolor{gray}{rgb}{0.8242,0.8242,0.8242}
\definecolor{amethyst}{HTML}{a45ee5}
\definecolor{aquamarine}{rgb}{0,0.7,0.6}
\definecolor{dodgerblue}{HTML}{1E90FF}
\definecolor{viennared}{HTML}{DA0A14}
\definecolor{ctorange}{HTML}{FF6C0C}
\definecolor{wales}{HTML}{ff0038}
\definecolor{benettongreen}{HTML}{009421}
\definecolor{valenciacfred}{HTML}{ee3524}
\definecolor{barcelonafcgold}{HTML}{edbb00}
\definecolor{jam}{HTML}{A50B5E}
\definecolor{austriawien}{HTML}{441678}
\definecolor{italia90green}{HTML}{009966}
\definecolor{ferrarired}{HTML}{ff2800}
\definecolor{gray}{HTML}{F0F0F0}
\definecolor{LightCyan}{rgb}{0.88,1,1}

\AtBeginDocument{
  \hypersetup{
    citecolor=dodgerblue,
    linkcolor=dodgerblue,   
    urlcolor=dodgerblue
    }
}

\newcommand{\nnrsim}{28}

\newcolumntype{C}{>{\centering\arraybackslash}X}
\newcolumntype{L}{>{\arraybackslash}X}

\newcolumntype{a}{>{\columncolor{gray}}c}
\newcolumntype{b}{>{\columncolor{white}}c}

\def\TwoPunctures{\texttt{TwoPunctures}~}
\def\McLachlan{\texttt{McLachlan}~}
\def\WeylScal{\texttt{WeylScal4}~}
\def\AHFinderDirect{\texttt{AHFinderDirect}~}

\def\Carpet{\texttt{Carpet}~}

\def\QuasiLocalMeasures{\texttt{QuasiLocalMeasures}~}

\def\TEOBDALI{\texttt{TEOBResumS-Dal\'i}}

\newcommand{\bham}{\affiliation{School of Physics and Astronomy and Institute for Gravitational Wave Astronomy, University of Birmingham, Edgbaston, Birmingham, B15 2TT, United Kingdom}}

\hypersetup{
    colorlinks=true,
    linkcolor=dodgerblue, 
    urlcolor =dodgerblue}

\begin{document}

%~~~~~~~~~~~~~~~ Title ~~~~~~~~~~~~~~~
\title{Mapping eccentricity evolutions between numerical relativity and effective-one-body gravitational waveforms}

\author{Alice Bonino \orcidlink{0000-0001-6502-284X}}
\email{axb1612@student.bham.ac.uk}
\bham
\author{Patricia Schmidt \orcidlink{0000-0003-1542-1791}}
\email{P.Schmidt@bham.ac.uk}
\bham
\author{Geraint Pratten \orcidlink{0000-0003-4984-0775}}
\email{G.Pratten@bham.ac.uk}
\bham

%~~~~~~~~~~~~~~~ Abstract ~~~~~~~~~~~~~~~
\begin{abstract}
Orbital eccentricity in compact binaries is considered to be a key tracer of their astrophysical origin, and can be inferred from gravitational-wave observations due to its imprint on the emitted signal. 
For a robust measurement, accurate waveform models are needed. 
However, ambiguities in the definition of eccentricity can obfuscate the physical meaning and result in seemingly discrepant measurements. 
In this work we present a suite of \nnrsim{} new numerical relativity simulations of eccentric, aligned-spin binary black holes with mass ratios between $1$ and $6$ and initial post-Newtonian eccentricities between $0.05$ and $0.3$. 
We then develop a robust pipeline for measuring the eccentricity evolution as a function of frequency from gravitational-wave observables that is applicable even to signals that span at least $\gtrsim 7$ orbits. 
We assess the reliability of our procedure and quantify its robustness under different assumptions on the data.
Using the eccentricity measured at the first apastron, we initialise effective-one-body waveforms and quantify how the precision in the eccentricity measurement, and therefore the choice of the initial conditions, impacts the agreement with the numerical data. We find that even small deviations in the initial eccentricity can lead to non-negligible differences in the phase and amplitude of the waveforms. However, we demonstrate that we can reliably map the eccentricities between the simulation data and analytic models, which is crucial for robustly building eccentric hybrid waveforms, and to improve the accuracy of eccentric waveform models in the strong-field regime. 
\end{abstract}

\date{\today}
\maketitle

%~~~~~~~~~~~~~~~ Introduction ~~~~~~~~~~~~~~~
\section{Introduction}
Gravitational-wave (GW) observations are poised to become an instrumental tool in constraining the formation and evolutionary pathways of stellar mass black holes. To date, a wide range of evolutionary mechanisms have been proposed, see~\cite{Mandel:2018hfr,Mapelli:2020vfa,Mandel:2021smh,Mapelli:2021taw} for recent reviews, and it is crucial that we develop the tools to robustly distinguish between the mechanism. Identifying the dominant mechanisms, and understanding their respective branching ratios, is expected to yield significant insight into stellar physics, binary interactions, and relativistic astrophysics \cite{Rodriguez:2016vmx,Farr:2017uvj,Zevin:2017evb,Zevin:2020gbd}. Unfortunately, the predicted mass distribution of stellar mass black hole can be highly degenerate between the many different formation channels, so we must appeal to other observables to help distinguish between the channels. The spin of the black holes and the eccentricity of their orbits will leave characteristic imprints in the emitted gravitational-wave signal, making them powerful tracers of the underlying formation channel. 

Binaries that form through isolated evolution ~\cite{Belczynski:2001uc,Dominik:2013tma,Mandel:2015qlu,Belczynski:2016obo,Mapelli:2017hqk,Giacobbo:2017qhh} are typically expected to circularize through the emission of gravitational radiation ~\cite{Peters:1963ux,Peters:1964zz}. This process is particularly efficient, such that these binaries will appear quasi-circular by the time they enter the sensitivity band of the current generation of ground-based detectors. However, binaries that are formed through recent dynamical interactions are expected to have non-negligible eccentricities \cite{Naoz:2016klo,Samsing:2017xmd,Zevin:2018kzq,Tagawa:2020jnc,Romero-Shaw:2022xko}.
This is particularly expected for dense stellar environments, such as globular clusters and active galactic nuclei, where the rate of dynamical interactions and captures can be enhanced. A robust measurement of eccentricity is therefore though to be a particularly clean tracer for such dynamical interactions \cite{Zevin:2021rtf}. Spin misalignment \cite{Apostolatos:1994mx,Kidder:1995zr,Schmidt:2012rh} is also thought to be characteristic signature for such dynamically formed binaries, though this may again be degenerate with other formation channels~\cite{Gerosa:2013laa,Gerosa:2018wbw,Steinle:2020xej}. 

A current limitation to extracting information on eccentricity from the current generation of ground-based GW detectors is the lack of sensitivity below $20$ Hz. 
Recent studies suggest that eccentricities $\geq 0.05$ at a gravitational-wave frequency of $10$ Hz will be necessary to extract eccentric information based on the current generation of detectors at design sensitivity \cite{Lower:2018seu}. 
Improving the low-frequency sensitivity of the next-generation of ground-based detectors, including Cosmic Explorer~\cite{Reitze:2019iox} and the Einstein Telescope~\cite{Maggiore:2019uih}, could lead to significant improvements and allow us to resolve eccentricities down to the $\sim 10^{-3}$ level, e.g. \cite{Lenon:2021zac,Saini:2023wdk}.

Similarly, multiband observations between space-based and ground-based detectors may be a potential route to extracting the natal eccentricities of stellar mass black holes \cite{Breivik:2016ddj,Nishizawa:2016eza,Klein:2022rbf}, though the event rates are highly uncertain \cite{Gerosa:2019dbe,Ewing:2020brd} and they require a robust framework for tracking and modelling the evolution of eccentricity over a period of year(s). 

In recent years, significant effort has been invested in incorporating eccentricity into the current generation of waveform models. Within the post-Newtonian (PN) approximation \cite{Blanchet:2013haa}, which assumes weak-fields and small velocities $[\epsilon \sim G M / (c^2 r) \sim (v / c)^2 \ll 1]$, the conservative dynamics for generic orbits are known up to 5PN \cite{Damour:1985der,Lincoln:1990cfw,Damour:2004bz,Memmesheimer:2004cv,Konigsdorffer:2006zt,Damour:2016abl,Bini:2019nra,Bini:2020wpo,Blumlein:2021txe,Dlapa:2021vgp,Henry:2023sdy} and radiative effects up to 3PN \cite{Arun:2007rg,Arun:2007sg,Arun:2009mc,Mishra:2015bqa,Loutrel:2016cdw,Boetzel:2019nfw,Ebersold:2019kdc,Khalil:2021txt,Paul:2022xfy}. 
This has led to the construction of numerous eccentric waveform approximants within the PN framework, e.g. \cite{Yunes:2009yz,Tessmer:2010sh,Tanay:2016zog,Klein:2018ybm,Moore:2019xkm,Tiwari:2020hsu}. Likewise, there have been numerous developments within the gravitational self-force (GSF) and scattering paradigms, e.g. \cite{Hopper:2015icj,Forseth:2015oua,Bini:2017wfr,Kavanagh:2017wot,Munna:2020iju,Lynch:2021ogr,Albertini:2023aol}.
However, whilst these provide an accurate description of the gravitational-wave signals throughout the inspiral, they are only valid up to moderate eccentricities $(e \sim 0.1)$ and are not reliable as the binary approaches merger.

To mitigate against such limitations, one can appeal to numerical relativity (NR) simulations to help model the complete inspiral-merger-ringdown (IMR) signal from eccentric binaries \cite{Ramos-Buades:2019uvh,Sperhake:2019wwo,Ramos-Buades:2022lgf,Healy:2022wdn,Joshi:2022ocr,Ferguson:2023vta}. Such catalogs have enabled the construction of eccentric NR surrogates ~\cite{Islam:2021mha,Huerta:2017kez} as well as PN-NR hybrids that blend analytical information from PN evolutions with the NR simulations ~\cite{Hinder:2017sxy,Ramos-Buades:2019uvh,Tiwari:2020hsu,Cho:2021oai,Chattaraj:2022tay}. Even though NR simulations cover a comparatively restricted region of the parameter space, they do provide the full non-linear solutions of the Einstein field equations and play a crucial role in accurately modelling the remnant black hole and providing an accurate representation of the GW signal through merger. As such, whilst eccentric NR surrogates and hybrid models are highly accurate, they are necessarily constrained by the poor parameter space coverage, accuracy, and finite length of available NR simulations, which often do not capture the early inspiral regime.

One alternative approach is the so called effective-one-body (EOB) framework \cite{Buonanno:1998gg,Buonanno:2000ef,Damour:2000we,Damour:2001tu}, which is a successful paradigm for modelling the complete GW signal emitted by compact binaries on arbitrarily eccentric orbits. Some of the first studies on incorporating eccentricity into the EOB framework were outlined in \cite{Bini:2012ji,Hinderer:2017jcs,Cao:2017ndf}. 
There have been numerous subsequent technical developments that have dramatically improved the accuracy of eccentric EOB models \cite{Liu:2019jpg,Chiaramello:2020ehz,Nagar:2021gss,Placidi:2021rkh,Albertini:2021tbt,Albanesi:2022ywx,Albanesi:2022xge,Ramos-Buades:2021adz,Liu:2021pkr,Yun:2021jnh,Henry:2023tka,Liu:2023dgl}. Of the available models, \texttt{SEOBNRv4EHM}, \cite{Ramos-Buades:2021adz} \TEOBDALI{} \cite{Chiaramello:2020ehz, Nagar:2021xnh, Nagar:2021gss, Albanesi:2021rby, Nagar:2020xsk, Placidi:2021rkh}, and \texttt{SEOBNRE} \cite{Cao:2017ndf} are the most mature.  
An extension of \texttt{SEOBNRE} to eccentric binaries with precessing spins has recently been presented \cite{Liu:2023ldr}.

However, a key limitation in comparing eccentric waveform models is that eccentricity itself is not gauge invariant in general relativity. This means that we need to exercise due caution if we are to ensure that we are comparing equivalent binary configurations. A number of different estimators for eccentricity have been introduced with \cite{Ramos-Buades:2022lgf} having recently identified an estimator that can be applied directly to the GW signal and has the correct Newtonian limit. If we are able to construct a map between models that allows us to set up self-consistent initial conditions (ICs), we can start to meaningfully compare the physics content of each model and the subsequent evolution of the eccentricity. 
This also has implications for understanding astrophysical formation scenarios, in which recent work has highlighted the need for self-consistent definitions of eccentricity between gravitational-wave data analysis and astrophysical population simulations \cite{Vijaykumar:2024piy}. 

The aim of this work is to (i) extend the parameter space coverage in available catalogs of NR simulations for eccentric binary black holes; (ii) to build a robust pipeline for measuring the eccentricity evolution from numerical relativity waveforms that is applicable even for short-duration signals; and (iii) to identify a map between the measured eccentricity evolution and initial conditions. We revisit the framework introduced in \cite{Bonino:2022hkj} and demonstrate its robustness against our suite of new aligned-spin eccentric NR simulations and \TEOBDALI{}.

The paper is organised as follows. Section~\ref{sec:numericalsimulations} provides an extensive overview of the new suite of NR simulations we have performed. Section~\ref{sec:eccestimation} furnishes the methodology we use to measure the eccentricity from the numerical data and assesses the impact of numerical errors on the procedure. Section~\ref{sec:comp} provides a detailed description and assessment of the mapping between initial conditions from numerical simulations and one particular EOB waveform model, \TEOBDALI{}. In Sec.~\ref{sec:discussion} we discuss and summarize the main results of this work. Throughout the paper, we use $G = c = 1$ unless otherwise stated.

%~~~~~~~~~~~~~~~ Numerical Simulations ~~~~~~~~~~~~~~~
\section{Numerical Simulations}
\label{sec:numericalsimulations}
\subsection{Overview}
In this paper, we present a catalog of \nnrsim{} eccentric NR simulations performed with the open-source Einstein Toolkit \cite{Loffler:2011ay, roland_haas_2022_7245853} within the moving punctures framework \cite{Campanelli:2005dd,Baker:2005vv}. We use conformally flat Bowen-York initial data \cite{Bowen:1980yu,Brandt:1997tf} as computed using the \TwoPunctures thorn \cite{Ansorg:2004ds}. Evolution is performed using the $W$-variant \cite{Marronetti:2007wz} of the Baumgarte-Shapiro-Shibata-Nakamura (BSSN) formulation of the Einstein field equations \cite{Shibata:1995we,Baumgarte:1998te} as implemented by the \McLachlan thorn. The black holes are evolved using the moving punctures gauge conditions \cite{Campanelli:2005dd,Baker:2005vv}. The lapse is evolved using a $1+\log$ slicing condition \cite{Bona:1994b} and the shift is evolved using the hyperbolic $\tilde{\Gamma}$-driver equation \cite{Alcubierre:2002kk}. Spatial derivatives are computed using an 8th order accurate finite differencing scheme together with Kreiss-Oliger dissipation \cite{Kreiss:1973}. The apparent horizons (AH) are computed using the \AHFinderDirect \cite{Thornburg:2004dv}. The black hole spins are estimated through a dynamical horizon formalism as implemented in \QuasiLocalMeasures where  \cite{Ashtekar:2003hk,Dreyer:2003bv,Schnetter:2006yt}
\begin{align}
S &= \frac{1}{8 \pi} \oint_{\rm AH} n^a \varphi^b K_{ab} \, d^2 A,
\end{align}
where $\varphi^a$ denotes an (approximate) axial Killing vector field, $n^a$ a spacelike unit normal to the horizon, and $K_{ab}$ the extrinsic curvature. 
Adaptive mesh refinement is provided by \Carpet \cite{Schnetter:2003rb}, with the near-zone being covered by high-resolution Cartesian grids. 
The wave extraction zone is computed using the Llama infrastructure~\cite{Pollney:2009yz,Reisswig:2012nc}, whose grids are adapted to the spherical topology of the zone and allow for high-resolution extraction out to comparatively large radii relative to standard Cartesian grids.
Gravitational waves are extracted from the simulations using the Newman-Penrose Weyl scalar $\Psi_4$, as provided by the \WeylScal thorn \cite{Newman:1961qr}. As is conventional, $\Psi_4$ is decomposed into spin-weighted spherical harmonics of spin weight $s=-2$, such that
\begin{equation}
    r\Psi_{4,\ell m} = \int_0^{2\pi} \int_0^\pi {}^{-2}Y^*_{\ell m}(\theta, \phi) \Psi_4 (\theta, \phi) \sin \theta \, d\theta \, d\phi, 
\label{eq:lm}
\end{equation}
where ${}^{-2}Y^*_{\ell m}$ denotes the complex conjugate of the spin-weighted spherical harmonics and $(\theta, \phi)$ the spherical coordinates on the unit sphere. The gravitational-wave strain $h$ can then be derived from $\Psi_4 \equiv -\ddot{h}$ through fixed frequency integration (FFI) \cite{Reisswig:2010di}.

To help mitigate against near-field gauge effects, we calculate the asymptotic waveform at future null infinity $\mathscr{I}^{+}$ using extrapolation, e.g. \cite{Boyle:2009vi,Hinder:2013oqa}. In this framework, we extract a quantity $\mathcal{Q}$ on a series of spheres of increasing radii and perform a least-squares fit to find the asymptotic behaviour of the non-oscillatory function $\mathcal{Q}$ 
\begin{align}
\mathcal{Q}(u_i,r) &= \displaystyle\sum_{n=0}^{N} \frac{\mathcal{Q}_n}{r^n},
\end{align}
such that the leading order coefficient $\mathcal{Q}_0$ represents the asymptotic value at $\mathscr{I}^+$ at an order $N$. For a subset of simulations, we also use the perturbative framework introduced in \cite{Nakano:2015pta} to extrapolate $r \Psi_4$, finding good agreement between the two methods. 

%~~~~~~~~~~~~~~~ Eccentric Initial Conditions ~~~~~~~~~~~~~~~
\subsection{Eccentric Initial Conditions}
In order to produce eccentric initial data, we use a PN approximation to estimate the momenta of the black holes given a specific eccentricity at some reference separation. We follow the approach outlined in \cite{Ramos-Buades:2018azo} in which the tangential momentum is perturbed away from the quasi-circular value by a correction factor that is derived within a quasi-Keplerian parametrization \cite{Mishra:2015bqa}. This procedure was used in \cite{Ramos-Buades:2018azo} to iteratively construct low-eccentricity initial data, and in \cite{Ramos-Buades:2019uvh} to construct initial data with a prescribed eccentricity. As detailed in \cite{Ramos-Buades:2018azo, Ramos-Buades:2019uvh}, the correction factor for the tangential momenta $\lambda_t$ in the low eccentricity limit is given by 
\begin{equation}
\label{eq:lambdat}
    \lambda_{t} (D, e_0, \nu, \sign) = 1+\frac{e_0}{2} \times \sign \times  \left[ 1-\frac{1}{D}(\nu+2) \right], 
\end{equation}
where $\nu = m_1 m _2 / M^2$ is the symmetric mass ratio, $M = m_1 + m_2$ the total mass, $D$ the initial orbital separation, $e_0$ is the initial eccentricity, and $\sign = \pm 1$ depends on the initial phase. However, as discussed in \cite{Ramos-Buades:2019uvh}, the correction factor that we apply to the tangential momentum is taken to be the average between the inverse of $\lambda_t$ with a plus sign plus and $\lambda_t$ with a minus sign, i.e. 
\begin{equation}
\label{eq:barlabdat}
    \bar\lambda_{t}^{0} (D, e_0, \nu) = \frac{1}{2} \left[ \lambda_{t}(D, e_0, \nu, +1)^{-1} + \lambda_{t}(D, e_0, \nu, -1) \right].
\end{equation}

\subsection{Catalog of Simulations}
The suite of eccentric aligned-spin NR simulations presented here spans a range of parameters. The mass ratio $q=m_1/m_2 \geq 1$ of the binaries' components ranges from $q=1$ to $q=6$ with initial eccentricities $e_0$ up to moderate values of $0.3$. The largest dimensionless black hole spin that we consider for the $i$-th black hole is $|\chi_i| \simeq 0.5$. The simulations cover between 3 and 15 complete orbits before merger, depending on the binary parameters, typically starting at an initial separation of $D \sim 12.5M$. 

The Cartesian grids are implemented using the Carpet thorn, which provides a multiblock scheme that implements Berger-Oliger adaptive mesh refinement. 
The grids are nested, such that there are $L^k$ levels of increasing resolution, i.e., $\Delta x^k = \Delta x^{k-1} / 2$. 
The innermost Cartesian grids are initialized such that their radii are $\sim 1.2 \, r_{\rm AH}$, where $r_{\rm AH}$ is the radius of the coordinate apparent horizon. 
Likewise, the outermost Cartesian grid, $L^0$, spans a domain that covers both black holes with a resolution set by the transition to the spherical grids managed by the Llama thorn. 
The resolution of the radial grids when extracting the gravitational waves is guided by the expected wavelength of higher multipole moments in the ringdown, typically resulting in a resolution $\Delta x < 1.0M$. 

\begin{figure}[t]
\includegraphics[width=\columnwidth]{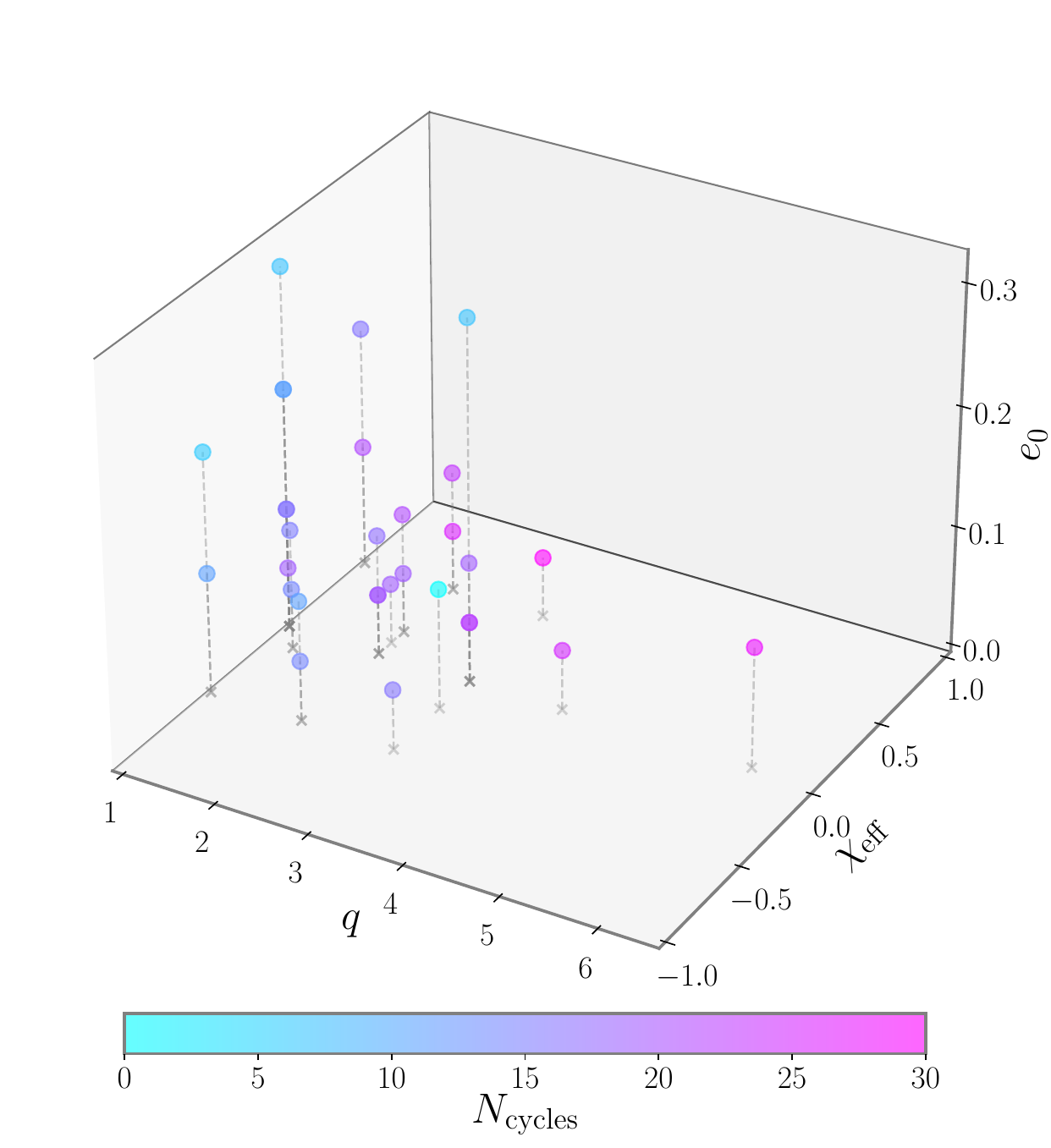}
\caption{
\label{fig:catalog}Parameter space coverage of our NR simulations listed in Tab~\ref{tab:simulations}. We show the mass ratio $q$, the effective aligned-spin $\chi_{\rm eff}$, and the initial PN eccentricity $e_0$. The colour represents the number of complete GW cycles.} 
\end{figure}

In Tab.~\ref{tab:simulations}, we provide a summary of the simulation properties for the catalog presented here, including the number of complete GW cycles up to merger $N_{\rm cycles}$ computed from $\Psi_4$, where we define the merger time as the time of the maximum rms amplitude of the $(\ell,m)=(2, \pm 2)$ modes and remove all data before junk radiation plus $50M$ when computing the number of GW cycles. 
We also give the Christodolou mass of the remnant black hole, $M_f$ as well as its dimensionless spin $\chi_f = S_f / M^2_f$, with the Christodolou mass given by \cite{Christodoulou:1970wf}
\begin{align}
M_{f} = \sqrt{M^2_{\rm irr} + \frac{S^2_f}{4 M^2_{\rm irr}}}, \quad M_{\rm irr} = \sqrt{ \frac{A}{16 \pi}},
\end{align}
where $M_{\rm irr}$ is the irreducible mass and is related to the area $A$ of the BH horizon. The last column in Tab.~\ref{tab:simulations} is the angular frequency measured at the first periastron $M\omega_0^p$, see Sec.~\ref{sec:eccestimation} for further details. 
The $\lbrace q, \chi_{{\rm eff}}, e_0 \rbrace$ parameter space covered by our simulations is shown in Fig.~\ref{fig:catalog}. 

In addition to the final mass and spin of the remnant, we also compute the Christodoulou mass and spin associated with each black hole horizon after equilibrium is reached at a simulation time of $t=100M$. This allows us to evaluate the remnant spin fit of Ref.~\cite{Jimenez-Forteza:2016oae} for quasi-circular non-precessing BBHs and compare it to our measurements. We find that for all our simulations the absolute difference between the quasi-circular prediction and our eccentric mergers is on the order of $\mathcal{O}(10^{-3})$, consistent with previous findings for equal-mass, moderately eccentric systems~\cite{Islam:2021mha}.

We also compute the peak luminosity taking into account all multipoles of $\Psi_4$ up to $\ell = 6$ \cite{Ruiz:2007yx}
\begin{equation}
  \mathcal{L}_{\rm peak} = \max_t \frac{1}{16\pi} \sum_{\ell = 2}^6 \sum_{m = -\ell}^{\ell} |\dot{h}^{\infty}_{\ell m}(t)|^2,  
\end{equation}
where ${h}^{\infty}_{\ell m}$ denotes the strain modes obtained via FFI and extrapolated to infinity. We note that we exclude the $(m=0)-$modes from our calculation. The obtained peak luminosity for each simulation is given in the final column of Tab.~\ref{tab:simulations}. Comparing to the fit for quasi-circular simulations from Ref.~\cite{Keitel:2016krm}, we find small differences of the order of $10^{-3}$, suggesting that the binaries have circularised sufficiently before the peak emission is reached. 

\begin{figure}[t]
\includegraphics[width=\columnwidth]{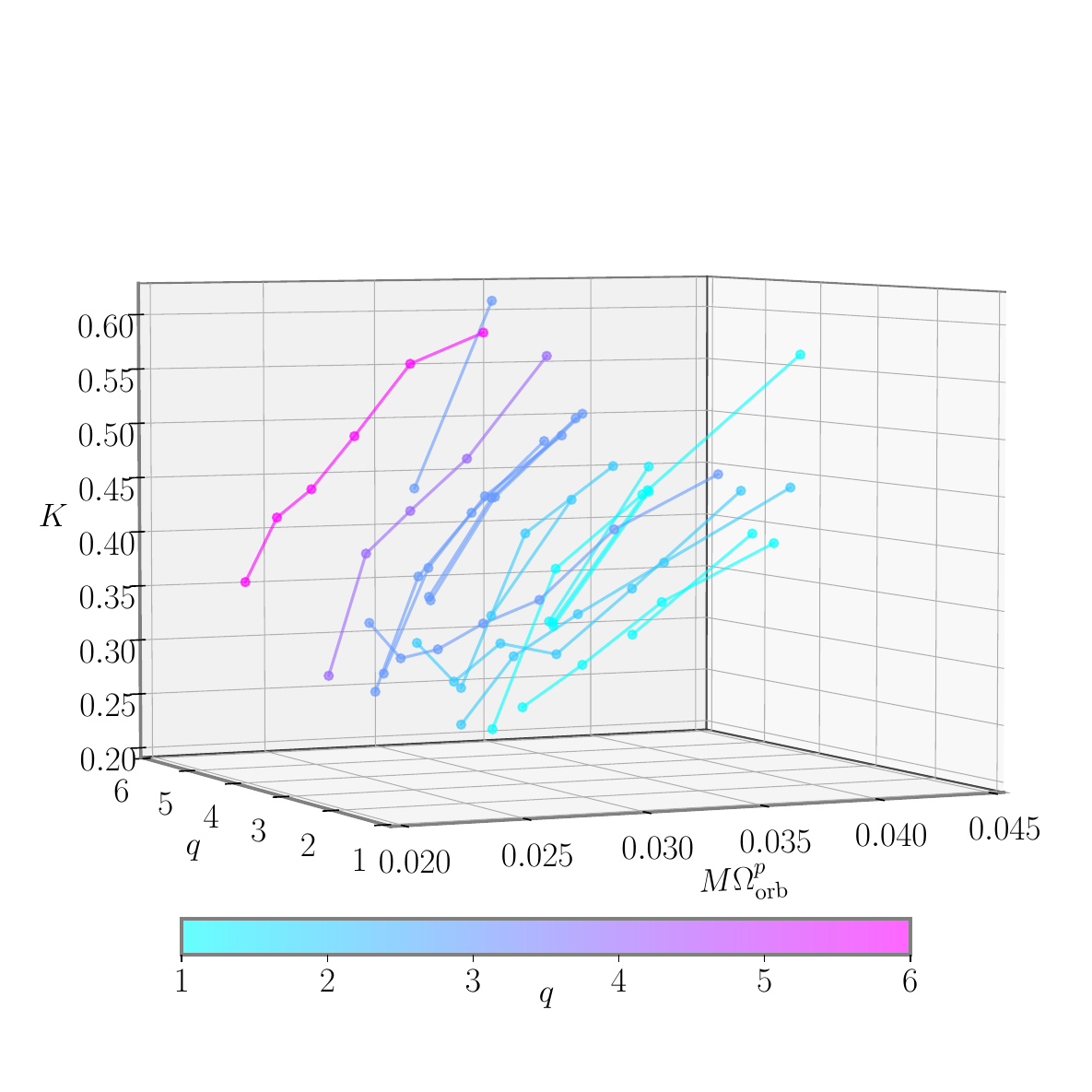}
\caption{
\label{fig:periastronprecession} Periastron advance per orbit $K$ as a function of orbital frequency at periastra passages (marked by the dots) for our NR simulations that span at least $16$ complete GW cycles, coloured by mass ratio $q$.} 
\end{figure}

Finally, we compute the fractional periastron advance per orbit following Eq.~(29) of \cite{Mroue:2010re} \begin{equation}
 \frac{\Omega^p_{\rm orb}}{\Omega^p_r} = \frac{\phi_{\rm orb} (t^p_{i+1})-\phi_{\rm orb} (t^p_{i})}{2 \pi},
\end{equation}
where $t^p_i$ and $t^p_{i+1}$ indicate two consecutive periastra passages, $\Omega^p_{\rm orb}$ and $\Omega^p_r$ are respectively the orbital and the radial frequencies at periastron passages, and $\phi_{\rm orb}$ is the orbital phase of the binary. 
We define the periastron advance per orbit as 
\begin{equation}
K \equiv \frac{\Omega^p_{\rm orb}}{\Omega^p_r} -1. 
\label{eq:k}
\end{equation}
Figure~\ref{fig:periastronprecession} shows the periastron advance for all simulations that have at least $16$ complete GW cycles. The periastron precession $K$ across all simulations spans $\sim 0.2 - 0.63$, in agreement with Fig.~(7) of \cite{Mroue:2010re}.

% TABLE I
\begin{table*}[th!]
\centering
\begin{tabularx}{0.975\linewidth}{ SC@{} | SC@{} | SC@{} | SC@{} | SC@{} | SC@{} | SC@{} | SC@{} | SC@{} | SC@{} | SC@{} | SC@{} } 
\hline
\hline
ID  & $q$ & $\chi_{1,z}$ & $\chi_{2,z}$ & $\chi_{\rm eff}$ & $e_0$
&$D\, [M]$ & $N_{\rm cycles}$ & $M_f$ & $\chi_{f}$ & $M\omega_{22}^p$ & $\mathcal{L}_{\rm peak}$ \\
\hline
% equal mass
\rowcolor{austriawien!05}
0001& 1 & 0 & 0 & 0 & 0.05 & $12.5$ & 19 & $0.9515$ & $0.6861$ & $0.052$ & 0.00105 \\
0002& 1 & $0$ & $0$ & 0 & $0.1$ & $12.5$ & 16 & $0.9513$& $0.6864$ & $0.062$ & 0.00106 \\
\rowcolor{austriawien!05}
0003& 1 & 0.5 & $-0.5$ & 0 & $0.1$ & $12.5$ & 16 & $0.9511$ & $0.6855$ & $0.060$ & 0.00107 \\
0004& 1 & $-0.5$ & $-0.5$ & $-0.5$ & $0.1$ & $12.5$ & 11 & $0.9620$ & $0.5261$ & $0.068$ & 0.00086 \\ 
\rowcolor{austriawien!05}
0005& 1 & 0.5 & 0.5 & 0.5 & $0.1$ & $12.5$ & 21 & $0.9327$ & $0.8312$ & $0.058$ & 0.00137 \\
0006 & 1 & 0 & 0 & 0 & $0.2$ & $12.5$ & 11 & $0.9516$ & $0.6839$ & $0.080$ & 0.00103 \\
\rowcolor{austriawien!05}
0007 & 1 & 0.5 & $-0.5$ & 0 & $0.2$ & $12.5$ & 11 & $0.9514$ & $0.6830$ & $0.070$ & 0.00104 \\
0008 & 1 & $-0.5$ & $-0.5$ & $-0.5$ & $0.2$ & $12.5$ & 6 & $0.9622$ & $0.5323$ & 0.060 & 0.00090\\
\rowcolor{austriawien!05}
0009 & 1 & 0.5 & 0.5 & 0.5 & $0.2$ & $12.5$ & 16 & $0.9330$ & $0.8299$ & $0.060$ & 0.00135 \\
0010 & 1 & 0 & 0 & 0 & 0.3 & $12.5$ & 7 & -- & -- & 0.060 & 0.00109 \\
\rowcolor{austriawien!05}
% GW150914-like
0011 & 1.24 & 0.31 & $-0.46$ & $-0.1163$ & 0.05 & 11.05 & 14 & 0.9527 & 0.6869 & 0.060 & 0.00102 \\ 
0012 & 1.24 & 0.31 & $-0.46$ & $-0.1163$ & 0.1 & 11.05 & 15 & 0.9524 & 0.6879 & 0.062 & 0.00104 \\ 
\rowcolor{austriawien!05}
% mass ratio q=2
0013 & 2 & 0 & 0 & 0 & 0.05 & $12.5$ & 20 & 0.9613 & 0.6236 & 0.050 & 0.00080\\
0014 & 2 & 0 & 0 & 0 & 0.05 & 12.1579 & 18 & 0.9611 & 0.6235 & $0.054$ & 0.00080 \\
\rowcolor{austriawien!05}
0015 & 2 & $-0.5$ & $-0.5$ & $-0.5$ & 0.05 & $12.5$ & 14 & 0.9700 & 0.4273 &  $0.054$ & 0.00064 \\
%16 & 2 & 0.25 & 0 & 0.0833 & 0.05 & 12.1579 & 19 & 0.9609 & 0.6278 & \\
%\rowcolor{austriawien!05}
%17 & 2 & 0.5 & 0 & 0.1667 & 0.05 & 12.1579 & 19 & 0.9596 & 0.6376 & \\
0016 & 2 & 0.5 & 0.5 & 0.5 & 0.05 & $12.5$ & 26 & 0.9448 & 0.8045 & $0.050$ & 0.00107 \\
\rowcolor{austriawien!05}
0017 & 2 & 0 & 0 & 0 & 0.1 & $12.5$ & 17 & $0.9614$ & $0.6226$ & $0.060$ & 0.00079 \\
0018 & 2 & $-0.5$ & $-0.5$ & $-0.5$ & 0.1 & $12.5$ & 11 & $0.9699$ & $0.4264$ & $0.068$ & 0.00064 \\
\rowcolor{austriawien!05}
%21 & 2 & 0.5 & 0.0 & 0.1667 & 0.1 & 13.4444 & 21 & 0.9597 & 0.6376 & \\
0019 & 2 & 0.5 & 0.5 & 0.5 & 0.1 & $12.5$ & 23 & $0.9445$ & $0.8048$ & $0.060$ & 0.00108 \\
%\rowcolor{austriawien!05}
% mass ratio q=3 and higher
0020 & 3 & $0$ & $0$ & 0 & $0.05$ & $12.5$ & 22 & 0.9713 & 0.5402 & 0.052 & 0.00054 \\
\rowcolor{austriawien!05}
0021 & 3 & 0 & 0 & 0 & 0.05 & 12.379 & 22 & 0.9712 & 0.5399 & $0.052$ & 0.00054 \\
0022 & 3 & $-0.5$ & $-0.5$ & $-0.5$ & 0.05 & 12.379 & 16 & 0.9780 & 0.2991 & $0.054$ & 0.00043 \\
\rowcolor{austriawien!05}
0023 & 3 & 0.5 & 0.5 & 0.5 & 0.05 & 12.379 & 30 & 0.9582 & 0.7685 & $0.050$ & %$0.07104^*$
--
\\
0024 & 3 & $0$ & $0$ & 0 & $0.1$ & $12.5$ & 19 & -- & -- & $0.062$ & 0.00054\\
\rowcolor{austriawien!05}
0025 & 3 & $0.1$ & $-0.3$ & $-0.2$ & $0.1$ & $12.5$ & 19 & 0.9703 & 0.5747 & 0.064 & 0.00056 \\
0026 & 3 & $0$ & $0$ & 0 & $0.3$ & $12.5$ & 7 & -- & -- & 0.100 & 0.00056 \\
\rowcolor{austriawien!05}
0027 & 4 & $0$ & $0$ & 0 & $0.05$ & 12.379 & 25 & 0.9778 & 0.4715 & $0.052$ & 0.00038\\
%\rowcolor{austriawien!05}
0028 & 6 & $0$ & $0$ & 0 & 0.1 & 12.83 & 28 & 0.9854 & 0.3727 & 0.060 & 0.00022 \\
%\textcolor{blue}{32} & 1 & 0 & 0 & 0 & 0.1 & 16 & R &  &  & \\
\hline
\hline
\end{tabularx}
\caption{\label{tab:simulations} Summary of the eccentric NR simulations presented in this work. The first column gives the identifier of the simulation. The following columns indicate the mass ratio $q$ , the $z$-components of the dimensionless spins $\chi_{1,z}$ and $\chi_{2,z}$, the effective spin $\chi_{\rm eff}$, the initial PN eccentricity $e_0$, the initial separation $D/M$, the number of complete GW cycles up to the rms peak of the $(2,\pm 2)$-modes $N_{\rm cycles}$ as well as the mass $M_f$ and spin $\chi_f$ of the final black hole, the angular frequency at the first periastron $M\omega^p_{22}$ measured from the $(2,2)$-mode of $\Psi_4$ and the peak luminosity.
}
\end{table*} 

%~~~~~~~~~~~~~~~ Eccentricity Estimation ~~~~~~~~~~~~~~~
\section{Eccentricity Estimation}
\label{sec:eccestimation}
\subsection{Introduction}
As has been extensively discussed in the literature, e.g. see \cite{Mora:2002gf,Mora:2003wt,Loutrel:2018ydu, Shaikh:2023ypz}, eccentricity is not a uniquely defined quantity in general relativity and gauge ambiguities make comparisons between measurements difficult. A number of estimators have been proposed in the literature, including a more recent definition of eccentricity that resolves some of these gauge ambiguities \cite{Ramos-Buades:2019uvh}. We are particularly interested in inferring the evolution of the eccentricity directly from the gravitational-wave signal, be it on real data, (semi-)analytical waveform models, or numerical relativity simulations. This requires a number of key ingredients: i) a robust estimator for the eccentricity, ii) a parameter that describes the position of the compact objects on the orbit, iii) a parameter that characterizes the size of the orbit, and iv) a mapping between the time and some appropriately averaged frequency. In tackling this challenge, two highly complementary approaches have recently been presented in \cite{Bonino:2022hkj} and \cite{Shaikh:2023ypz}. The framework outlined in \cite{Shaikh:2023ypz} provides a very thorough and detailed discussion on the challenges of inferring eccentricity and the robustness of the implementation. To fully characterize an eccentric orbit, \cite{Shaikh:2023ypz} adopts the eccentricity estimator introduced in \cite{Ramos-Buades:2022lgf} $e_{\rm gw} (t)$ together with a generalization of the Newtonian mean anomaly $l_{\rm gw}(t)$ over an interval between two consecutive periastron passages $t^p_i$ and $t^p_{i+1}$ defined as
\begin{align}
l_{\rm gw}(t) &= 2 \pi \frac{t - t^p_i}{t^p_{i+1} - t^p_i}.
\end{align} 
In order to estimate the reference frequency, \cite{Shaikh:2023ypz} uses an orbit-averaged frequency $\hat{\omega}_{22}$ that is again defined between two consecutive periastron passages~\cite{Ramos-Buades:2022lgf,Ramos-Buades:2023yhy}. The framework presented in \cite{Bonino:2022hkj} also adopts the eccentricity estimator defined in \cite{Ramos-Buades:2022lgf} but uses the value of the eccentricity at an initial frequency such that the system always starts at the apastron. In this instance, the joint specification of the initial eccentricity and initial frequency replaces the eccentricity and mean anomaly. The final ingredient introduced in \cite{Bonino:2022hkj} is to use the mean of the frequencies between the periastra $\omega^p_{22} (t)$ and apastra $\omega^a_{22}(t)$ to estimate a monotonically increasing mean frequency $\bar{\omega}(t)$. 
Here we build on the methodology introduced in \cite{Bonino:2022hkj}. We note that the choice of using the mean frequency $\bar{\omega}(t)$ rather than $\hat{\omega}_{22}$ is motivated by the comparison of our numerical simulations against the EOB model \TEOBDALI{} in Sec.~\ref{sec:comp}. However, the framework that will be presented below can also be applied to $\hat{\omega}_{22}$. 

\subsection{Methodology}
\label{sec:methodology}
The starting point for our discussion is the choice of the eccentricity estimator. Whilst this issue has previously been outlined in the literature, e.g. \cite{Damour:1988mr,Mora:2002gf,Loutrel:2018ydu,Ramos-Buades:2019uvh, Ramos-Buades:2022lgf,Shaikh:2023ypz,Bonino:2022hkj,Nagar:2021gss, Carullo:2023kvj, Gamba:2021gap}, we include some discussion here for completeness. In the Newtonian limit, the eccentricity is uniquely defined by the separation of the binary at its periastron $r^p$ and apastron $r^a$
\begin{align}
e_{\rm Newt.} (t) &= \frac{r^a -r^p}{r^a + r^p}. 
\label{eq:e_newt}
\end{align}
However, the Newtonian expression rapidly breaks down as PN corrections are included with additional parameters being required to fully characterize the eccentricity.
Moreover, an eccentricity defined through the trajectories of the binary is gauge dependent, see \cite{Mroue:2010re,Purrer:2012wy,Ramos-Buades:2022lgf,Shaikh:2023ypz}. The second key drawback is that the estimator defined in Eq.~\eqref{eq:e_newt} depends on quantities that are not directly observable through GWs, which impedes the applicability of the estimator to GW observations. 

To mitigate against some of these issues, an estimator defined from the \textit{orbital} frequencies at the periastron $\Omega^p_{\rm orb}$ and apastron passages $\Omega^a_{\rm orb}$ was introduced in \cite{Mora:2002gf} and is given by
\begin{align}
e_{\Omega_{\rm orb}} (t) &= \frac{\sqrt{\Omega^p_{\rm orb}(t)} - \sqrt{\Omega^a_{\rm orb}(t)}}{\sqrt{\Omega^p_{\rm orb}(t)} + \sqrt{\Omega^a_{\rm orb}(t)}}.
\end{align}
This estimator has the correct Newtonian limit Eq.~\eqref{eq:e_newt} but is also estimated from orbital trajectories, leading to the same caveats as before. An estimator based on the GW frequency of the $(2,2)$-mode, $\omega_{22} = d \phi_{22} / dt$, was introduced in Ref.~\cite{Ramos-Buades:2020noq} and is given by 
\begin{align}
\label{eq:e22}
e_{\omega_{22}} (t) &= \frac{ \sqrt{\omega^p_{22}(t)} - \sqrt{\omega^a_{22}(t)} }{ \sqrt{\omega^p_{22}(t)} + \sqrt{\omega^a_{22}(t)} } .
\end{align}
%
%FIG 3
\begin{figure}[t]
\includegraphics[width=\columnwidth]{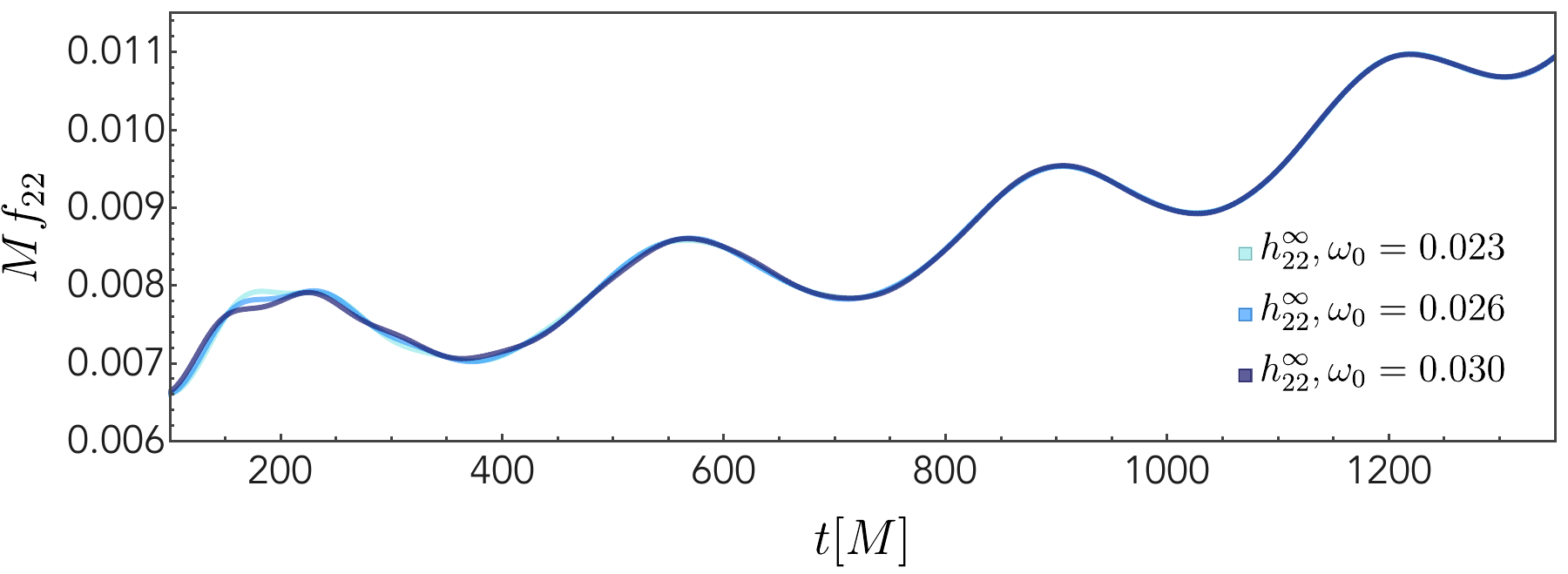}
\caption{
\label{fig:FFIfreqs} Gravitational-wave frequency $f_{22} =\omega_{22}/(2\pi)$ obtained from $h^\infty_{22}$ using different values of $\omega_0$ for the simulation with \texttt{ID:0001}. Depending on the choice of $\omega_0$ that is used to carry out the FFI, the location of the first peak varies.
}
\end{figure}  
This estimator has the joint benefit of being gauge-independent and directly observable from the GW signal alone. However, it was noted in \cite{Ramos-Buades:2022lgf} that Eq.~\eqref{eq:e22} does not obey the Newtonian limit, see also the discussion in \cite{Shaikh:2023ypz}. To correct for this, \cite{Ramos-Buades:2022lgf} proposed a modified estimator defined by
\begin{align}
\label{eq:egw}
e_{\rm gw} (t) &= \cos \left(\frac{\Psi(t)}{ 3} \right) - \sqrt{3} \sin \left(\frac{\Psi(t)}{ 3} \right),
\end{align}
where 
\begin{align}
\label{eq:egw_psi}
\Psi(t) = \arctan \left( \frac{1 - e^2_{\omega_{22}}(t)}{2 e_{\omega_{22}}(t)} \right).
\end{align}
Unless otherwise stated, we will adopt the eccentricity estimator $e_{\rm gw}(t)$ given in Eq.~\eqref{eq:egw} for the remained of this work. 

In order to measure the eccentricity from the NR data, we first need to determine $\omega_{22}(t)$. For each simulation in Tab.~\ref{tab:simulations}, we follow the steps described in Sec.~\ref{sec:numericalsimulations} to obtain the extrapolated multipole moments $\Psi_{4,\ell m}$, from which we calculate the extrapolated strain modes $h^\infty_{\ell m}$.
This requires us to choose a physically motivated low-frequency cutoff $\omega_0$ for the FFI that suppresses spurious non-linear drifts arising from the numerical integration and the amplification of unphysical frequencies resulting from spectral leakage \cite{Reisswig:2010di}. The choice of $\omega_0$ can impact the location of the first peak in the GW frequency, which in turn will impact the measured eccentricity (see App.~\ref{sec:appA}). An example of this is shown in Fig.~\ref{fig:FFIfreqs} for three different values of $\omega_0$. 
As a consistency check on the above procedure, we compare against the strain $h$ that is directly calculated from the metric perturbations via the Regge-Wheeler-Zerilli equations \cite{Regge:1957td,Zerilli:1970se,Sarbach:2001qq}, as provided in the SXS waveform catalog \cite{Boyle:2019kee}. Using \texttt{SXS:BBH:1360}, we find good agreement between the two methods for estimating $h$, with small variations in $\omega_0$ having a subdominant impact on the $h$ inferred using the FFI \textcolor{blue}{(see App.~\ref{sec:appA})}. 
For our analyses, we choose $\omega_0= \omega^{p}_{22}/2$, where $\omega^{p}_{22}$ is the frequency of $\Psi_{4,22}$ at the time of the first periastron.

After performing the time integration, we remove the junk radiation in all the simulations being mindful of preserving the first peak of $\omega_{22}(t)$. We have checked that this procedure is robust across all of our simulations and adopt $t_{\rm min} \equiv 100M$. 

Next we need an algorithm to identify the peaks in the GW frequency, which correspond to the periastron and apastron passages. 
Rather than fitting the first and second derivatives to identify the peaks, as was done in \cite{Bonino:2022hkj}, we instead opt to use a rational function to calculate the residual between the frequency and its secular trend. We find this to be more robust than using derivatives due to the presence of numerical noise. After isolating the peaks, we fit the maxima and the minima of the frequency using a linear fit if only two peaks are available and a nonlinear rational fit of the form $(a+bx)/(1+cx)$ when more peaks are available. We typically find rational functions to be more robust than an equivalent high-order polynomial. In order for the procedure to work, we require at least two peaks to be present, though we caution against the accuracy of the estimated eccentricity in this limit, see the discussion below in Sec.~\ref{sec:sim_length}. 

% FIGURE 4
\begin{figure*}[t]
\center
\includegraphics[width=0.92\textwidth]{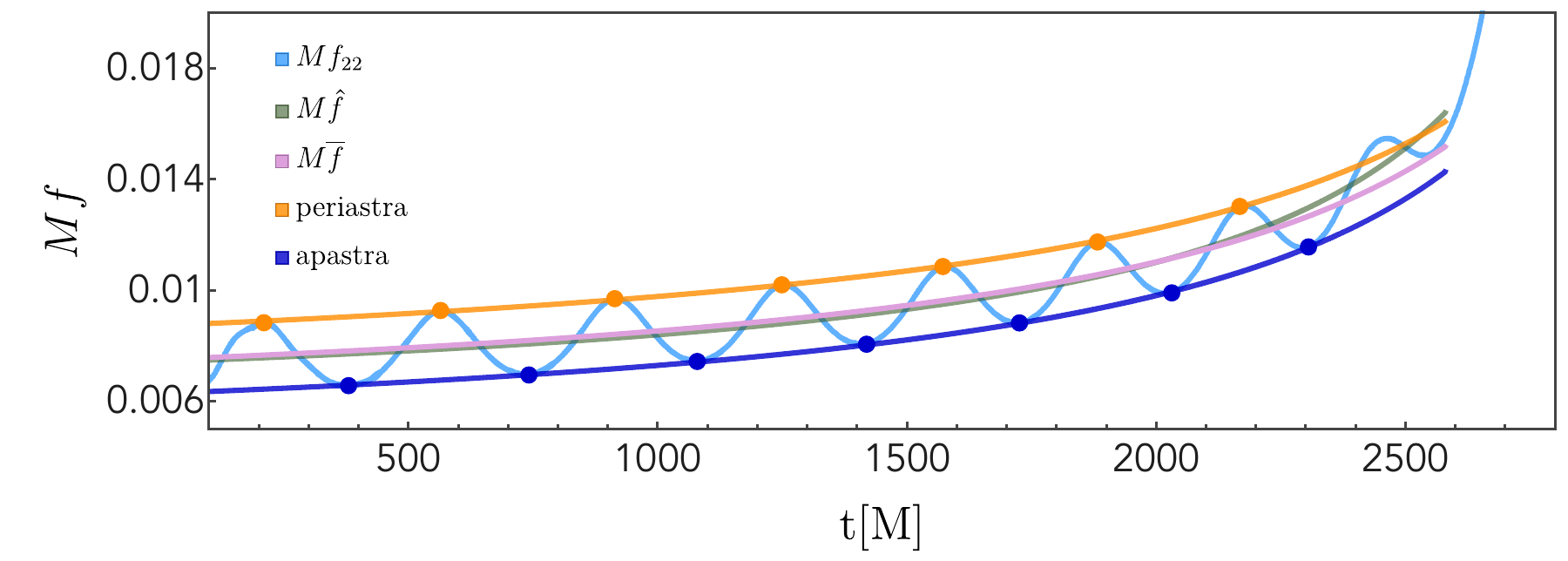}
\includegraphics[width=0.92\textwidth]{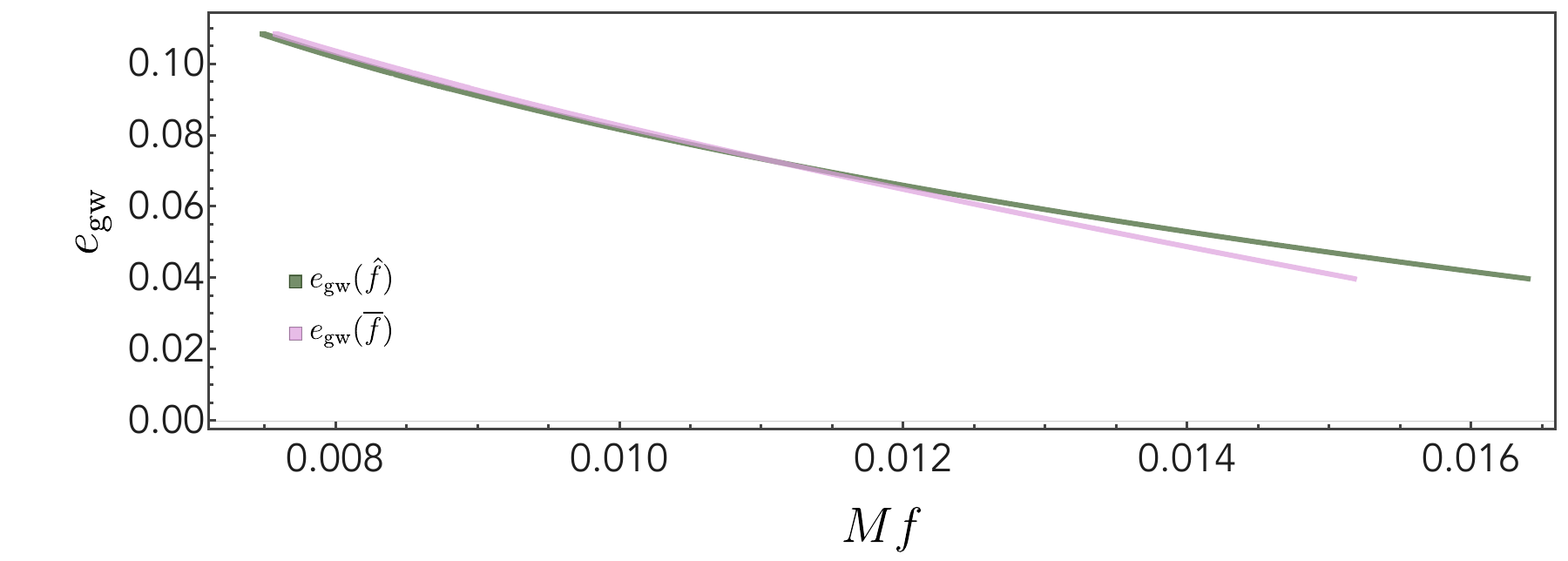}
\caption{\label{fig:fit}
Illustration of the end-to-end procedure to measure the eccentricity from a GW signal for our $q=6$ simulation (\texttt{ID:0028}). The top panel shows the GW frequency (light blue), the fits of the apastra (orange) and periastra (blue) following the procedure described in the main text. We also show the two monotonic frequencies $\bar{f}$ (pink) and $\hat{f}$ (green).  
The bottom panel shows the measured evolution of the eccentricity as a function of either $\hat{f}$ (green) or $\bar{f}$ (pink). 
}
\end{figure*} 

% \begin{figure*}[!h]
% \center
% \includegraphics[width=\columnwidth]{./figures/diffsfreqs.pdf}
% \\
% \includegraphics[width=\columnwidth]{./figures/diffs.pdf}
% \caption{\label{fig:egwdiffs
% }}
% \end{figure*}

As discussed in \cite{Bonino:2022hkj}, it is convenient to define the eccentricity as a function of a monotonically increasing frequency, providing a bijective correspondence between the eccentricity and the frequency. There are two possible choices for defining such a frequency that have been adopted in the literature: The first is to define a \emph{mean} GW frequency from the $(2,2)$-mode~\cite{Bonino:2022hkj}, 
as
\begin{align}
\label{eq:meanf}
\bar{f} \equiv \frac{1}{2} \left( f^{p}_{22} + f^{a}_{22} \right),
\end{align}
where $f^X_{22} = \omega^X_{22}/(2\pi)$ and $X$ denotes the periastron ($p$) or apastron ($a$) respectively.

The second choice is to introduce an \emph{average} GW frequency $\hat{f}$ that provides an orbit average between any two consecutive periastra or apastra~\cite{Ramos-Buades:2022lgf,Shaikh:2023ypz} enumerated by the index $i$
\begin{align}
\hat{f}^{X}_{22,i} (t) &= \frac{1}{t^X_{i+1} - t^X_i} \int^{t^X_{i+1}}_{t^X_i} \, f_{22} (t) dt , \nonumber \\ 
&= \frac{\phi_{22} (t^X_{i+1}) - \phi_{22}(t^X_i)}{t^X_{i+1} - t^X_i}.
\end{align}
The average frequency can then be defined as 
\begin{align}
\label{eq:avgf}
\hat{f}_i (t) \equiv \hat{f}^{p}_{22,i} (t) + \hat{f}^{a}_{22,i} (t),
\end{align}
over the interval $t_i \leq t < t_{i+1}$.
We refer the reader to~\cite{Ramos-Buades:2021adz,Ramos-Buades:2022lgf,Shaikh:2023ypz,Ramos-Buades:2023yhy} for a detailed discussion on the orbit-averaging procedure.
We interpolate $\hat{f}(t)$ with either a rational function or a linear fit. 
As also discussed in \cite{Ramos-Buades:2022lgf,Shaikh:2023ypz}, we observe that the average frequency $\hat{f}$ and the mean frequency $\bar{f}$ start to diverge on their approach to merger but agree in the early stages of the inspiral. 
This suggests that the impact of the explicit choice of the frequency parameter is subdominant in the determination of the eccentricity for simulations that have at least $3-4$ complete orbits prior to the plunge. 

% FIGURE 5
\begin{figure*}[t]
\center
\includegraphics[width=\textwidth]{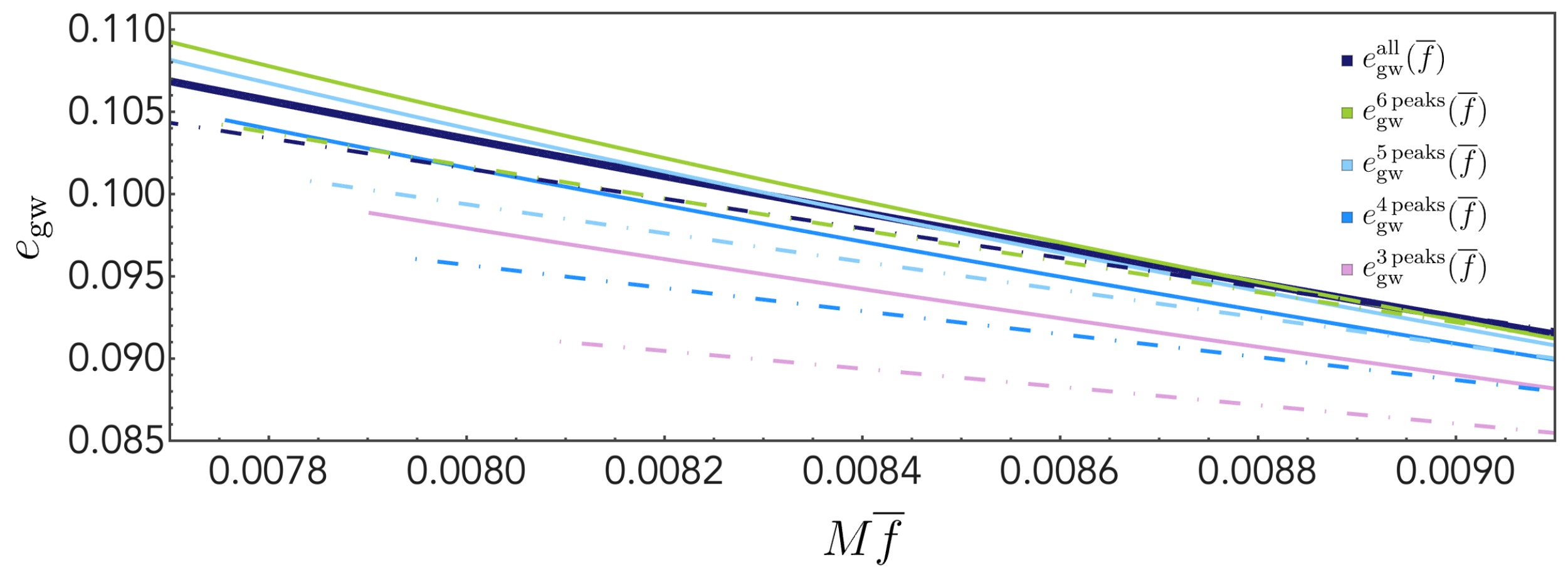}
\caption{\label{fig:egw_q6_peaks_number} {Eccentricity evolution for \texttt{ID:0028} as a function of mean frequency for different numbers of peaks. The dot-dashed lines show the evolution obtained retaining the last peak, the solid lines the evolution obtained when removing the last peak. The solid dark blue line is the reference baseline which includes all peaks except the last one. We observe that retaining the last peak (dot-dashed) usually leads to lower values of the inferred eccentricity reflecting the uncertainty in determining the eccentricity.}
}
\end{figure*}

The end-to-end procedure to calculate the eccentricity evolution $e_{{\rm{gw}}}$ is illustrated for our $q=6$ simulation (\texttt{ID:0028}) in Fig.~\ref{fig:fit}: 
In the upper panel, we show the highly oscillatory GW frequency (light blue), the periastra (orange dots) and apastra (blue dots), as well as the nonlinear rational fits, and the mean (pink) and average frequency (green). The bottom panel shows the resulting eccentricity measurement as a function of the mean frequency (pink) and the average frequency (green). 

In the next section, we will discuss the robustness of this method to measure the eccentricity from NR simulations and possible sources of error.

%~~~~~~~~~~~~~~~ Assessment of Robustness ~~~~~~~~~~~~~~~
\subsection{Assessment of Robustness}
\label{sec:robustness}
%~~~~~~~~~~~~~~~ Simulation Length ~~~~~~~~~~~~~~~
\subsubsection{Simulation Length}
\label{sec:sim_length}
A crucial part of the eccentricity estimation are the fits to the apastra and periastra. 
However, the outcome depends on the choice of the domain that the fits are performed over. 
To obtain a robust measurement, and to safeguard against using the eccentricity estimator in a regime where it is no longer valid, we need to truncate the data before merger. 
We opt to remove the last $\sim 4$  GW cycles before merger from the data in agreement with \cite{Shaikh:2023ypz}, which allows us to remove all spurious peaks close to merger whilst retaining as much information as possible.

We assess the impact of the number of peaks included to infer the eccentricity evolution, which we find to be non-trivial. For the vast majority of our simulations, we found it beneficial to remove the last periastron and apastron. This is demonstrated in Fig.~\ref{fig:egw_q6_peaks_number}, where we show the eccentricity evolution, $e_{{\rm gw}}(\bar{f})$, inferred with a varying number of peaks. 
Using our longest simulation with 28 GW cycles (\texttt{ID:0028}), we initially include all frequency peaks and sequentially omit peaks starting from the beginning. 
The solid (dot-dashed) lines represent the inferred eccentricity evolution when the last peak has been removed (retained). 
By comparing to the eccentricity curve that uses the most information (black line), we note that removing the last peak typically leads to a smaller spread in the estimated eccentricity relative to the evolution inferred when retaining the last peak . This is particularly prominent for short simulations that span fewer than $\sim 15$ GW cycles as the peaks closest to merger correspond to the regime where the GW frequency changes rapidly and can therefore shift the inferred eccentricity to higher or lower values, introducing inaccuracies in the inferred quantities. As we do not trust the estimator close to merger, and given the observed lack of robustness, we always remove the last peak when estimating the eccentricity.

%%%%%%%%%%%%%%%%%%%%%%%%%%%%%%%%%%%%%%%%%%%%%%%%%%
\subsubsection{Extrapolation vs. finite radius}
\label{sec:extrap}
Numerical relativity simulations often extract the GW signal at a series of finite radii, which are then extrapolated to future null infinity. This procedure can introduce errors, which can impact the eccentricity measurement. 

\begin{figure}[!h]
\center
\includegraphics[width=\columnwidth]{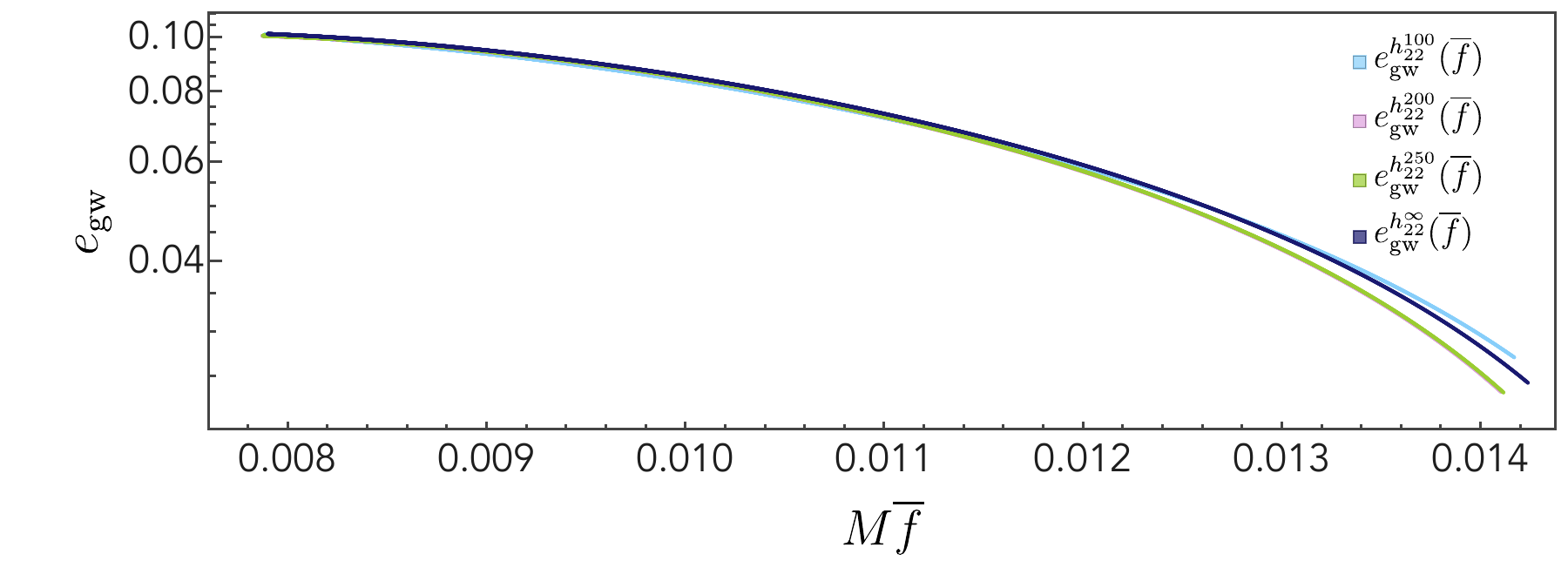}
\includegraphics[width=\columnwidth]{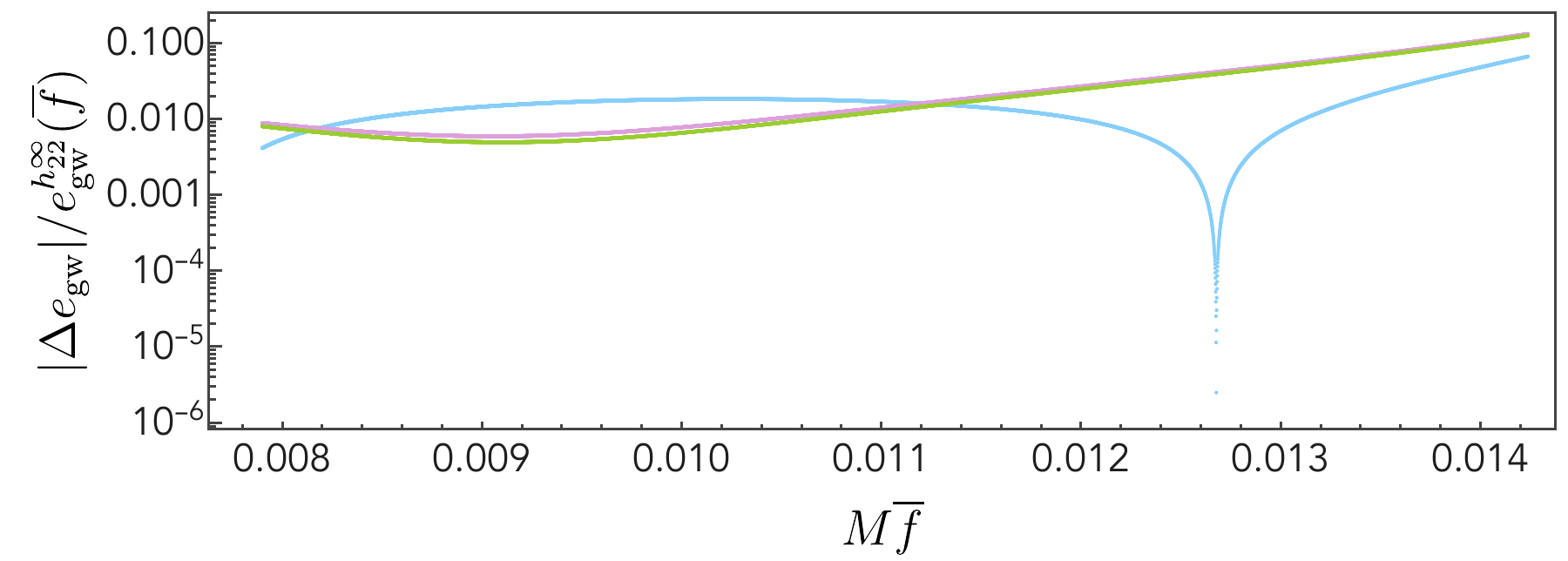}
\caption{\label{fig:egwextrradii}
{Top panel: Eccentricity evolution computed for simulation \texttt{ID:0002} for a series of finite extraction radii compared to the extrapolated waveform (dark blue). Bottom panel: The relative error, where $\Delta e_{\rm gw} =  e^{X}_{\rm gw} - e^{h^{\infty}_{22}}_{\rm gw}$, between the eccentricity inferred from the extrapolated waveform and the finite-radii waveforms.
The curves agree for most of the inspiral phase up to around $4$ orbits before merger, where we cut the data.
}}
\end{figure}

In Fig.~\ref{fig:egwextrradii} we compare the eccentricity curves $e_{gw}(\bar{f})$ obtained from different extraction radii against those inferred using the extrapolated waveforms. We find very good agreement throughout the inspiral phase but see more differences during the later stages close to merger, as shown in the bottom panel. Our results suggest that one can use either finite radius waveforms or extrapolated waveforms to reliably compute the eccentricity evolution up to $\sim 4$ orbits before merger.

%FIGURE 6

%%%%%%%%%%%%%%%%%%%%%%%%%%%%%%%%%
\subsubsection{$h$ vs. $\psi_4$}
\label{sec:hvspsi4}
Until now, we have investigated the eccentricity estimated only from the $(2,2)$-mode of the gravitational-wave strain $h$. In this section, we use $\Psi_4$ to estimate the eccentricity, as this is often a directly measured quantity in NR data. 

% FIGURE 7
\begin{figure}[t]
\center
\includegraphics[width=\columnwidth]{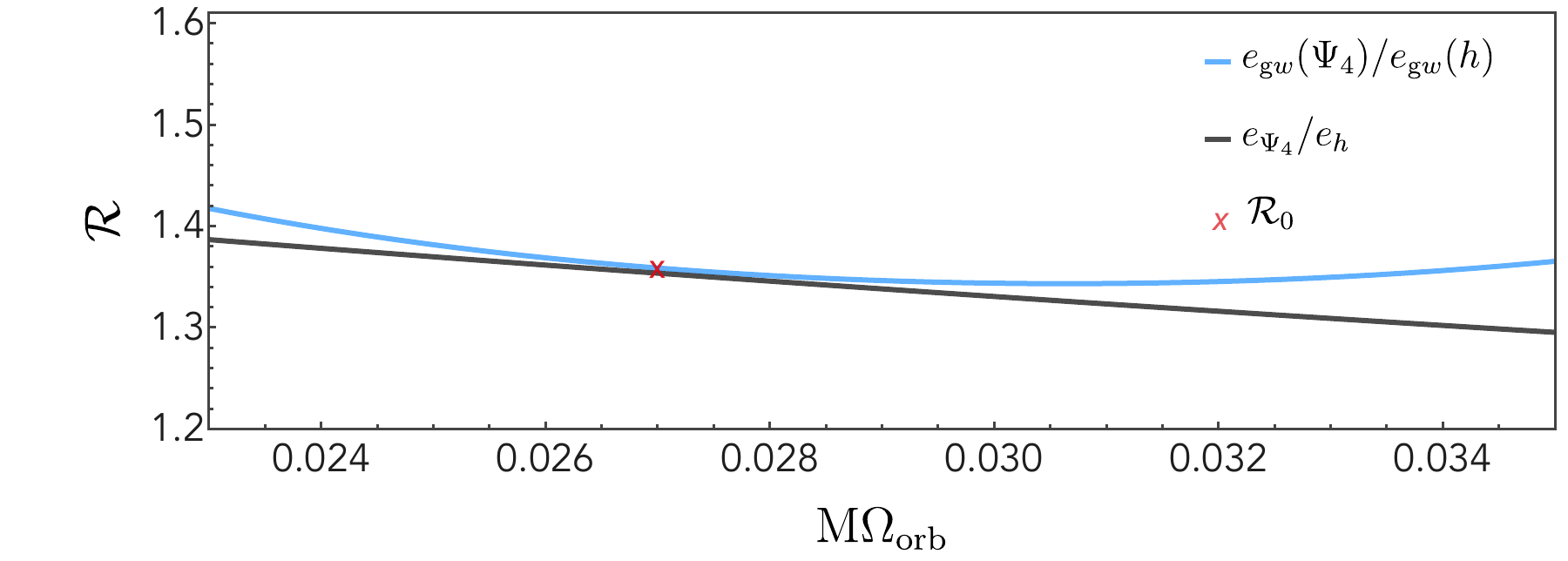}
\caption{\label{fig:ratio}
Ratio $\mathcal{R}$ between the eccentricity evolution computed employing $\Psi_4$ and eccentricity evolution computed employing $h$. The blue curve shows the ratio $e_{\rm gw}(\Psi_4)/e_{\rm gw}(h)$ as estimated from the numerical data. 
The black curve shows the ratio calculated from Eq.~\eqref{eq:egwpsi41pn} and Eq.~\eqref{eq:egwh1pn}.
} 
\end{figure}

To estimate the eccentricity evolution from $\Psi_4$, we use the same procedure as for $h$, described in Sec.~\ref{sec:methodology}. 
However, when using $\Psi_4$, we apply a \textit{trimmed mean} filter to the data to mitigate against numerical noise when calculating the derivative of the phase, which can induce errors in determining the positions of the apastra and periastra. 
Care needs to be taken when filtering the data to ensure that we do not introduce unphysical features, e.g. artificially shifting the time of the merger (see Appendix \ref{sec:appA}). 

An important subtlety is that the eccentricity measured from $\Psi_{4}$ differs from the eccentricity measured from $h$, e.g., see \cite{Purrer:2012wy}. 
% %
The eccentricity derived from the GW frequency
of $\Psi_4$ and $h$ can be analytically estimated to leading order in terms of an eccentric parameter $e_t$ defined in the quasi-Keplerian parametrization \cite{Memmesheimer:2004cv}. 
At 1PN, we have  
 \begin{align}
 \label{eq:egwpsi41pn}
 e_{\Psi_4} &= \frac{7}{4} e_t - \epsilon \, x \left(\frac{125 + 52 \nu}{168} \right) e_t,
\\
 e_{h} &= e_t + \epsilon \, x \left( \frac{115-16 \nu}{42} \right) e_t,
\label{eq:egwh1pn}
 \end{align}
where $e_{\Psi_4}$ is derived in the quasi-Keplerian parametrization in Appendix \ref{sec:appB} and $e_{h}$ is derived using Eq. (19) and (23) of \cite{Ramos-Buades:2022lgf}, expanded for small eccentricities. 
In Eq.~\eqref{eq:egwpsi41pn} and Eq.~\eqref{eq:egwh1pn} $x = ( M \Omega_{\rm orb})^{2/3}$, and $\epsilon^2 = c^{-2}$ tracks the PN order.
We can calculate the ratio between the two as a consistency check on the agreement between the eccentricity inferred from $h$ and $\Psi_4$,
\begin{align}
\label{eq:ratio}
\mathcal{R} &\equiv \frac{e_{\Psi_4}}{e_h}
\end{align}
and we see how this agrees at Newtonian order, with Eq. (C25) of \cite{Purrer:2012wy}.
In Fig.~\ref{fig:ratio}, we calculate the eccentricity evolution of \texttt{ID:0016} using $\Psi_{4}$ and $h$ and numerically calculate the ratio (blue curve). The black curve shows the analytical ratio in Eq.~\eqref{eq:ratio} derived from Eq.~\eqref{eq:egwpsi41pn} and Eq.~\eqref{eq:egwh1pn}.
The red cross indicates the value $\{M\Omega_{\rm orb} = 0.027$, $\mathcal{R}_0 = 1.36\}$ calculated in \cite{Purrer:2012wy}, see Sec. II and Fig.(20) therein. 
We see how the value agrees with our computed ratios. 
The numerically and analytically estimated ratios agree reasonably well within a range of frequencies spanning $0.025 \lesssim M \Omega_{\rm orb} \lesssim 0.029$. The disagreement at low- and high-frequencies is likely driven by a number of factors including numerical noise and the comparatively low PN order used for the estimators.

%%%%%%%%%%%%%%%%%%%%%%
%%%%%%% Table 2 %%%%%%
\begin{table*}[t!]
\centering
\renewcommand{\arraystretch}{1.3}
\begin{tabular}{c|>{\centering}p{2cm}|>{\centering}p{2cm}|>{\centering}p{2cm}|>{\centering}p{2cm}|c|c}
    \hline
    \hline
 \multicolumn{3}{c|}{Numerical Relativity}
 
  &\multicolumn{2}{c|}{\TEOBDALI{}}
  &\multicolumn{1}{c|}{$\rm \Delta{\phi}$ [rad]}
  &\multicolumn{1}{c}{$\Delta t\, [M]$}\\
\hline
ID & $M\bar{f}_{\rm NR}$ & $e_{\rm NR}$  & $M\bar{f}_{\rm DALI}$& $e_{\rm DALI}$& & \\
\hline
%BBH$\_$q1$\_$e0.05$\_$N80
\texttt{0001} & $0.0076203$ & $0.0510486$ &$0.00762334$ & $0.0510313$ & $0.439$ & $1.64$\\
%BBH$\_$q1$\_$e1$\_$N80$\_$v2 
\texttt{0002} & $0.00853042$ & $0.0980399$ & $0.00852527$ & $0.0980725$ & $0.346$ & $-1.78$\\ 
%BBH$\_$q1$\_$e0.1$\_$a+0.5$\_$a-0.5
\texttt{0003} & $0.00853016$ & $0.0993008$ & $0.00853787$ & $0.0992382$ & $0.383$ & $2.59$\\
%BBH$\_$q1$\_$e0.1$\_$a-0.5$\_$a-0.5 
\texttt{0004} & $0.00949267$ & $0.116386$ & $0.00937463$ & $0.119877$ & $0.319$ & $-16.92$ \\
%BBH$\_$q1$\_$e0.1$\_$a0.5$\_$a0.5
\texttt{0005} & $0.00807595$ & $0.0876158$ & $0.00811259$ & $0.0873528$ & $-0.012$ & $18.17$ \\
%BBH$\_$q1$\_$e0.2$\_$a0$\_$a0 
\texttt{0006} & $0.0113294$ & $0.186742$ & $0.0111475$ & $0.192453$ & $0.053$ & $-13.96$ \\
%BBH$\_$q1$\_$e0.2$\_$a+0.5$\_$a-0.5 
\texttt{0007} & $0.0113196$ & $0.1871$ & $0.0111526$ & $0.192687$ & $0.098$ & $-12.70$ \\
%BBH$\_$q1$\_$e0.2$\_$a-0.5$\_$a-0.5 
\texttt{0008} & -- & -- & -- & -- & -- & --\\
%BBH$\_$q1$\_$e0.2$\_$a0.5$\_$a0.5
\texttt{0009} & $0.00987994$ & $0.170253$ &$0.00998654$ & $0.16836$ & $-0.426$ & $24.35$ \\
%BBH$\_$q1$\_$e3$\_$N80 
\texttt{0010} &-- &-- &-- &-- &-- &--\\
%GW150914$\_$N32$\_$e0p05 
\texttt{0011} & $0.00914328$ & $0.0398909$ & $0.00910584$ & $0.0405297$ & $0.195$ & $-8.14$ \\
%GW150914$\_$N32$\_$e0p10
\texttt{0012} & $0.00862177$ & $0.0869213$ & $0.00863143$ & $0.0868591$ & $0.490$ & $3.17$ \\
\texttt{0013} & -- & -- & -- & -- & -- & --\\
%q2$\_$e0p05$\_$a1$\_$0p0 
\texttt{0014} & $0.00791276$  & $0.053303$& $0.00793832$ & $0.0531359$ & $0.463$ & $12.90$\\
%BBH$\_$q2$\_$e0.05$\_$a-0.5$\_$a-0.5 
\texttt{0015} & $0.00799565$ & $0.0613589$ & $0.00794926$ & $0.0622875$ & $0.211$ & $-16.89$\\
%BBH$\_$q2$\_$e0.05$\_$a0.5$\_$a0.5 
\texttt{0016} & $0.00736959$ & $0.0457956$ & $0.0073548$& $0.0458236$ & $-0.345$ & $-16.40$\\
%BBH$\_$q2$\_$e0.1$\_$a0$\_$a0 
\texttt{0017} & $0.00845326$ & $0.101092$& $0.00844698$ & $0.101141$ & $0.306$ & $2.29$\\
%BBH$\_$q2$\_$e0.1$\_$a-0.5$\_$a-0.5 
\texttt{0018} & $0.00936701$& $0.11873$& $0.00925574$ & $0.121864$ & $0.018$ & $-18.90$ \\
%BBH$\_$q2$\_$e0.1$\_$a0.5$\_$a0.5 
\texttt{0019} & $0.00801989$ & $0.0871244$ & $0.00803822$ & $0.087002$ & $-0.40$ & $11.03$ \\
\texttt{0020} & $0.0074702$ & $0.0533001$ & $0.00747325$ & $0.0532801$ & $0.654$ & $2.31$ \\
%q3$\_$a1$\_$0p0$\_$a2$\_$0p0$\_$e0p05$\_$D12p379$\_$vf (first apastron)
\texttt{0021} & $0.00757658$ & $0.0547304$ & $0.00756598$ & $0.0548007$ & $0.725$ & $-7.71$\\
%q3$\_$a1$\_$m0p5$\_$a2$\_$m0p5$\_$e0p05$\_$D12p379$\_$vf 
\texttt{0022} & $0.00800259$ & $0.0617299$ & $0.00797527$ & $0.0619813$ & $0.452$ & $-14.36$\\
%q3$\_$a1$\_$0p5$\_$a2$\_$0p5$\_$e0p05$\_$D12p379$\_$vf 
\texttt{0023} & $0.00741632$ & $0.0477771$ & $0.00738662$ & $0.0479505$ & $-0.459$ & $-30.35$ \\
%BBH$\_$q3$\_$e1$\_$N80 
\texttt{0024} & $0.00834383$ & $0.102726$ & $0.00835499$ & $0.102646$ & $0.586$ & $5.60$\\
\texttt{0025} & $0.00834383$ & $0.102726$ & $0.00835499$ & $0.102646$ & $0.586$ & $5.60$\\
%BBH$\_$q3$\_$e3$\_$N80
\texttt{0026} & -- & -- & -- & -- & -- & --\\
%q4$\_$e0p05$\_$D12p39 (first apastron) 
\texttt{0027} & $0.00751769$& $0.0542868$& $0.00751769$ & $0.0542868$ & $0.499$ & $-15.40$ \\
%q6$\_$a1$\_$0p0$\_$a2$\_$0p0$\_$e$\_$0p1$\_$D12p83 
\texttt{0028} & $0.00781405$ & $0.105506$ & $0.00777383$ & $0.105979$ & $1.326$ & $-0.60$\\
\hline
\texttt{SXS:BBH:1358} & $0.00592098$ & $0.152821$ &  $0.00595224$ & $0.152033$ & $0.509$ & $52.10$ \\ 
\texttt{SXS:BBH:1360} & $0.0065153$ & $0.207213$  & $0.00664508$ & $0.202663$ & $0.605$ & $136.59$\\ 
\hline
\hline
\end{tabular}
\caption{\label{tab:comparison} Summary of the eccentric initial conditions calculated for our set of NR simulations (were applicable) and for two SXS simulations. The first column indicates the simulation ID, the second and third columns give the eccentric initial conditions at first apastron in the NR data. The fourth and fifth columns give the values of the eccentric initial conditions calculated following our procedure. In last two columns we give the phase differences at merger as well as the time difference $\Delta t$.
Simulations without values are too short to reliably determine the eccentricity evolution.}
\end{table*}

%~~~~~~~~~~~~~~~ Comparison to Analytical Models ~~~~~~~~~~~~~~~
\section{Comparison to Analytical Models}
\label{sec:comp}

In this section, we compare our suite of NR simulations to an eccentric extension of the \TEOBResumS{} model~\cite{Damour:2014sva,Nagar:2019wds,Nagar:2020pcj,Riemenschneider:2021ppj} known as \TEOBDALI{} \cite{Chiaramello:2020ehz, Nagar:2021xnh, Nagar:2021gss, Albanesi:2021rby, Nagar:2020xsk, Placidi:2021rkh}. 
In this model, eccentricity is incorporated by dressing the circular azimuthal radiation reaction term with the Newtonian (leading-order) non-circular correction \cite{Chiaramello:2020ehz}. 
This modelling of eccentricity has been subsequently improved by incorporating higher order post-Newtonian information in a factorized and resummed form \cite{Placidi:2021rkh,Albanesi:2022ywx,Albanesi:2022xge}, building on the 2PN results presented in \cite{Khalil:2021txt}. 

In the previous section we introduced a robust method for identifying pairs of mean frequency and eccentricity, $\{M\bar{f}, e_{\rm gw}\}_{\rm NR}$ at the first apastron from NR simulations. 
In this section we will now use the eccentricity evolution as a function of mean frequency to develop a robust procedure for directly comparing eccentric NR simulations with other waveform models that may use a different definition of eccentricity, which can obfuscate 
comparisons between different waveform models. 
The advantage of this is that we can perform these direct comparisons without the need for numerical optimisations over the eccentricity as was done in previous works, e.g~\cite{Nagar:2021gss,Ramos-Buades:2021adz,Placidi:2021rkh,Albanesi:2022xge,Knee:2022hth}. 
We will demonstrate the efficacy of our procedure for the new simulations listed in Tab.~\ref{tab:simulations} as well as for a selected number of simulations from the \texttt{SXS} catalog~\cite{Boyle:2019kee} to compare against \TEOBDALI{}.

\subsection{Determining Initial Conditions}
\label{sec:ICs}
We start by calculating the mean frequency and eccentricity evolution of the NR simulations that have at least two clear peaks in the waveform frequency using the method described in Sec.~\ref{sec:methodology}. 
We use the first apastron to set a reference time, $t^a_{\rm NR}$, and denote the values of the mean frequency and the eccentricity at the first apastron as $\{M\bar{f}, e\}_{\rm NR}$.
We then initialise \TEOBDALI{} with these values to generate the waveform to which we apply a time shift such that the merger occurs at the same time as in the corresponding NR waveform. 
Note that we choose $M\bar{f}$ for consistency with the definition of the monotonic frequency in \TEOBDALI{}, which is defined as $ \bar{f}_{\rm{TEOB}} =(f^{a} +f^{p})/2$. 
However, the same procedure can be applied utilizing $M\hat{f}$.

After generating the \TEOBDALI{} waveform, we find that the time of its first apastron does not exactly correspond to $t^a_{\rm NR}$. 
This offset is due to the following: 1) \TEOBDALI{} uses a different definition of the eccentricity, which is based on orbital quantities, while the gauge invariant estimator in Eq.~\eqref{eq:egw} only depends on the waveform; 
2) the eccentric initial conditions in \TEOBDALI{} are estimated using a post-adiabatic expansion \cite{Nagar:2018gnk} (see also Appendix A of \cite{Bonino:2022hkj}), which makes the model sensitive to changes in the frequency and eccentricities at the $\sim \mathcal{O}(10^{-6})$ level \cite{Nagar:2021gss}. 
We determine the difference in the time of the apastra as
\begin{equation}
\Delta t = t^{a}_{\rm NR} - t^{a}_{\rm TEOB}, 
\label{eq:timediff}   
\end{equation} 
and calculate a \textit{new} pair of initial conditions defined at the time $ t^{a}_{\rm NR}+\Delta t$ such that $ t^{a}_{\rm TEOB}\simeq t^{ a}_{\rm NR}$. The procedure can be schematically written as:
\begin{enumerate}
    \item Calculate $\{M\bar{f}, e_{\rm gw}\}_{\rm NR}$ at the time of the first apastron $t^a_{\rm NR}$ using the methodology described in Sec.~\ref{sec:methodology}. 
    \item Use $\{M\bar{f}, e_{\rm gw}\}_{\rm NR}$ to initialise \TEOBDALI{} and perform a time shift such that the \TEOBDALI{} merger time corresponds to the NR merger time.
    \item Identify the time difference $\Delta t$ between the first apastron in NR vs. \TEOBDALI{}.
    \item Read off a new pair values at $t^{a}_{\rm NR} + \Delta t$ and denote these $\{M\bar{f}, e_{\rm gw} \}_{\rm DALI}$.
    \item Use $\{M\bar{f}, e_{\rm gw}\}_{\rm DALI}$ to initialise \TEOBDALI{}.
    \item Perform an overall phase shift to align the phases at $ t_{\rm TEOB}=0$. 
\end{enumerate}
We note that by specifying the initial conditions at the time of the first apastron for a given frequency, we are effectively fixing the mean anomaly. This provides a geometrically motivated picture for generating a set of self-consistent initial conditions. Other approaches, such as \cite{Nagar:2021gss}, numerically optimize over the eccentricity and frequency, making it difficult to meaningfully compare the dynamics in a transparent manner.  

\begin{figure}[t]
\center
\includegraphics[width=\columnwidth]{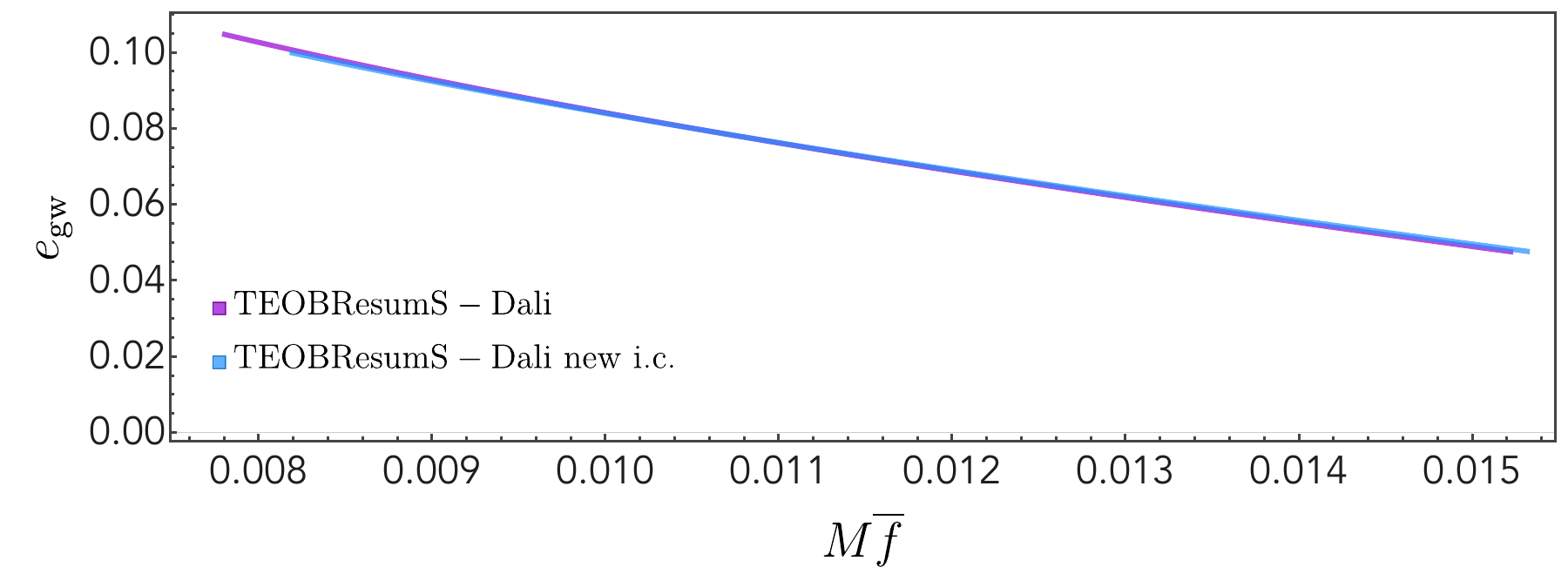}\\
\includegraphics[width=\columnwidth]{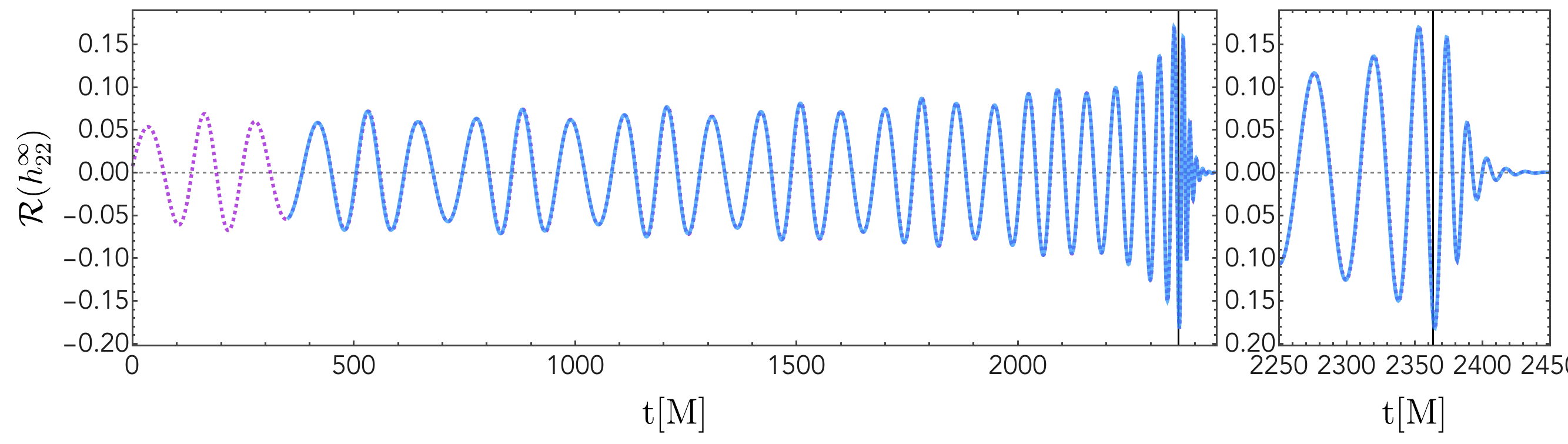}\\
\includegraphics[width=\columnwidth]{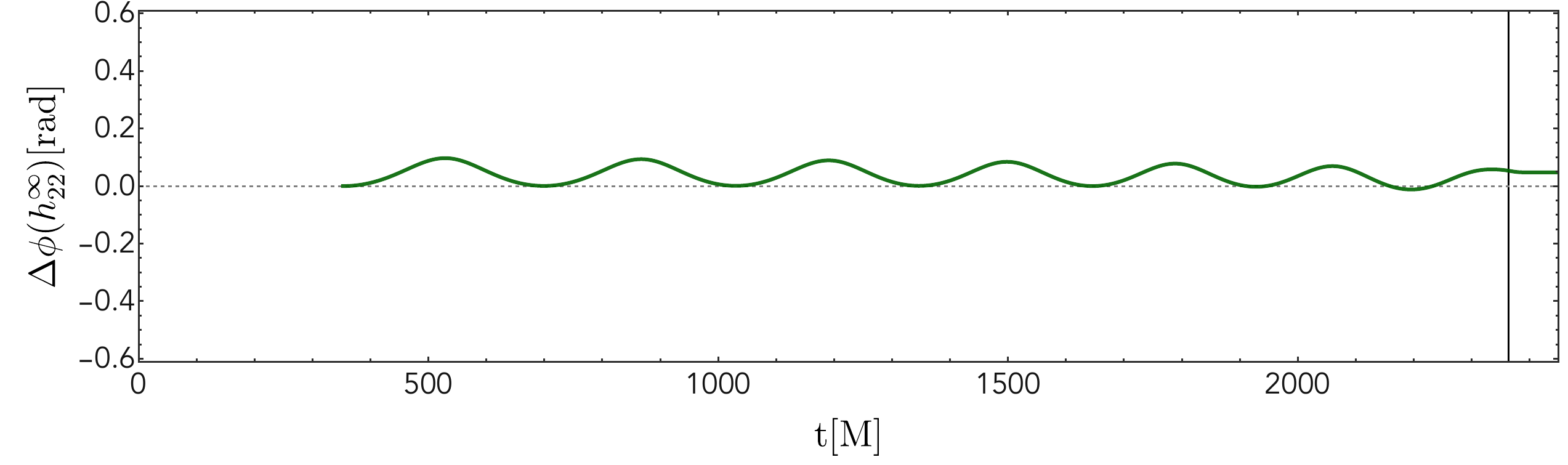}
\caption{Comparison between two \TEOBDALI{} waveforms to verify the proposed procedure to determine the initial conditions. The upper panel shows the eccentricity curves $e_{\rm gw}$, the middle panel the waveforms and the bottom panel the phase differences. We find almost identical waveforms and oscillatory phase difference smaller than $0.2$ rad.}
\label{fig:teob_teob_comparison}
\end{figure}

To validate our procedure for determining self-consistent initial conditions we apply it to a \TEOBDALI{} waveform with $\{q=6, \chi_{1,z}=\chi_{2,z}=0\}$ as shown in Fig.~\ref{fig:teob_teob_comparison}. 
In the upper panel, we see that the eccentricity evolution generated using the initial conditions calculated through our procedure (blue) closely matches the original evolution (purple) with phase differences smaller than 0.2 rad (bottom panel).

The result of applying this procedure to the suite of NR simulations is summarised in Tab.~\ref{tab:comparison}. 
The initial conditions used to start the EOB dynamics closely resemble the initial conditions originally computed, with differences ranging from $10^{-6}$ to $10^{-4}$ for the frequency and $10^{-4}$ to $10^{-3}$ for the eccentricity. 
In Tab.~\ref{tab:comparison} we also indicate the dephasing $\Delta \phi = \phi_{\rm NR} - \phi_{\rm TEOB}$ at the time of merger, which acts as a measure of the accumulated error.  

% FIG. 8

% FIG 9.
%---- Waveform Panel Figure ----
\begin{figure*}[p!]
\center
\includegraphics[width=\columnwidth]{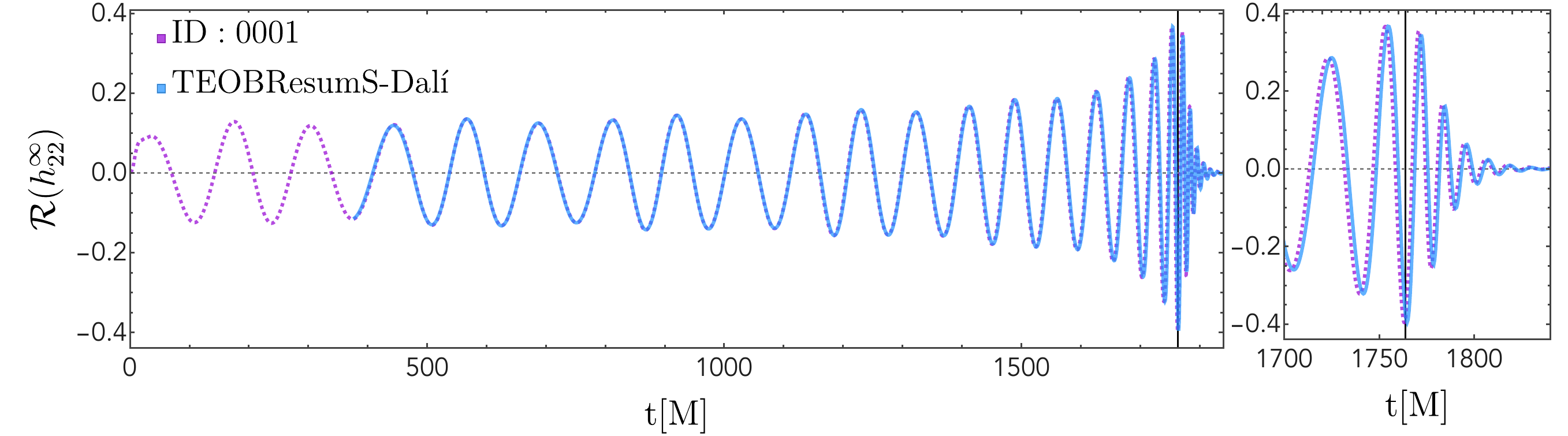}
\includegraphics[width=\columnwidth]{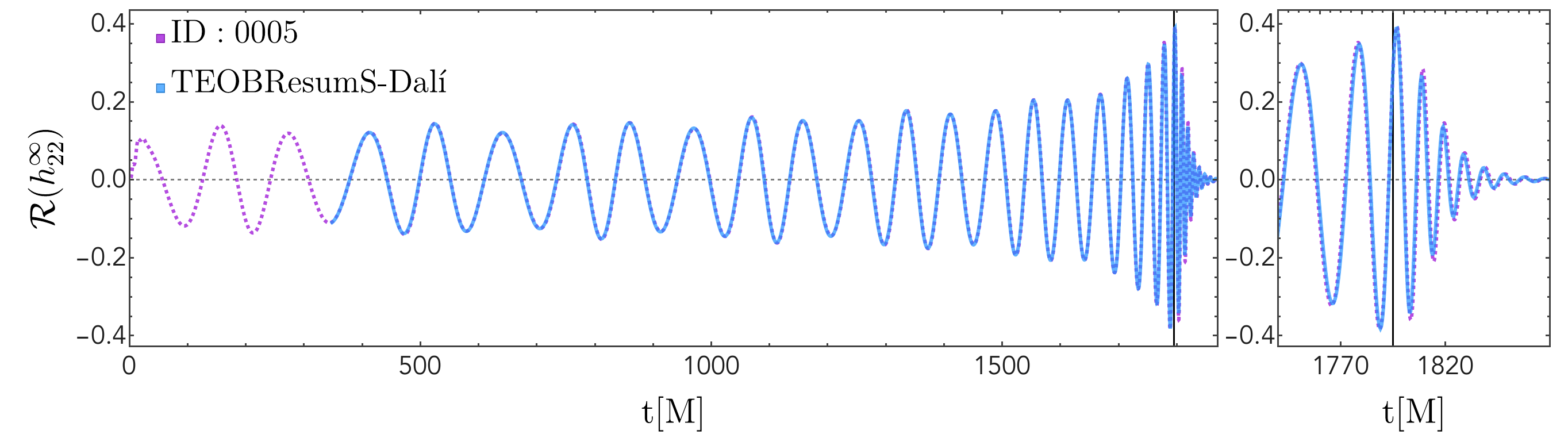}\\
%\vspace{-0.3cm}
\includegraphics[width=\columnwidth]{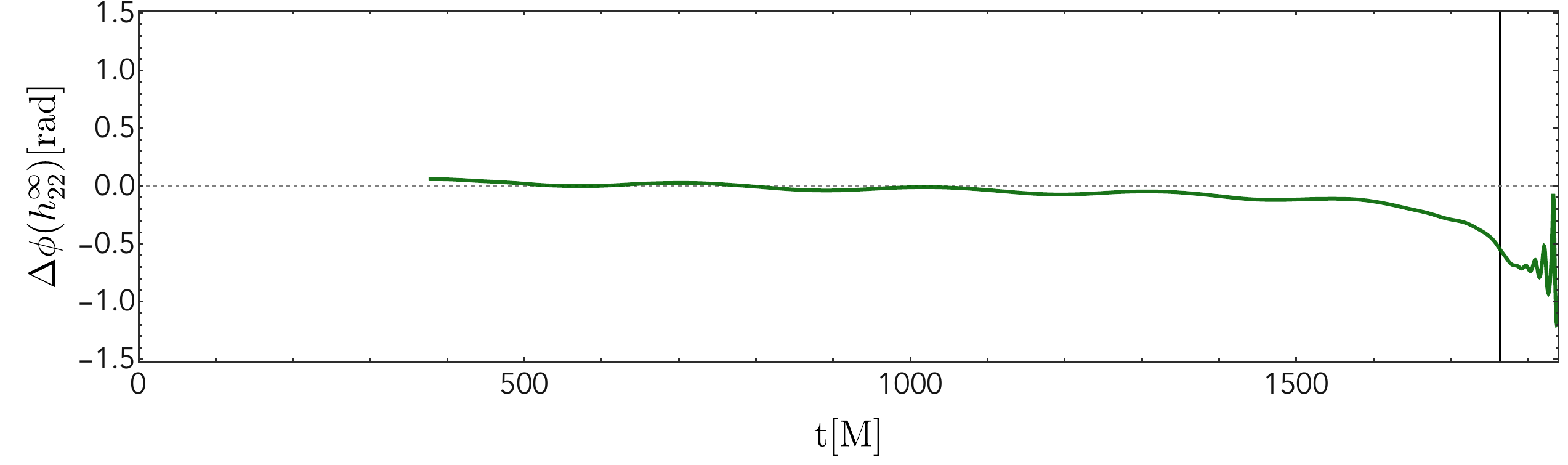}
\includegraphics[width=\columnwidth]{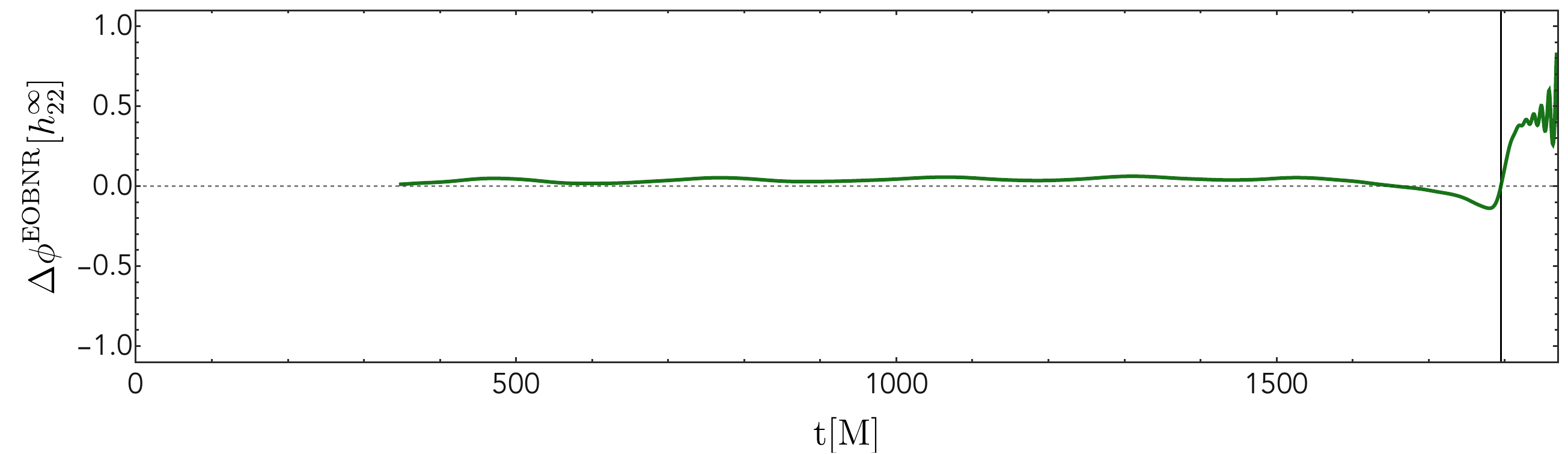}
\\
\vspace{0.5cm}
\includegraphics[width=\columnwidth]{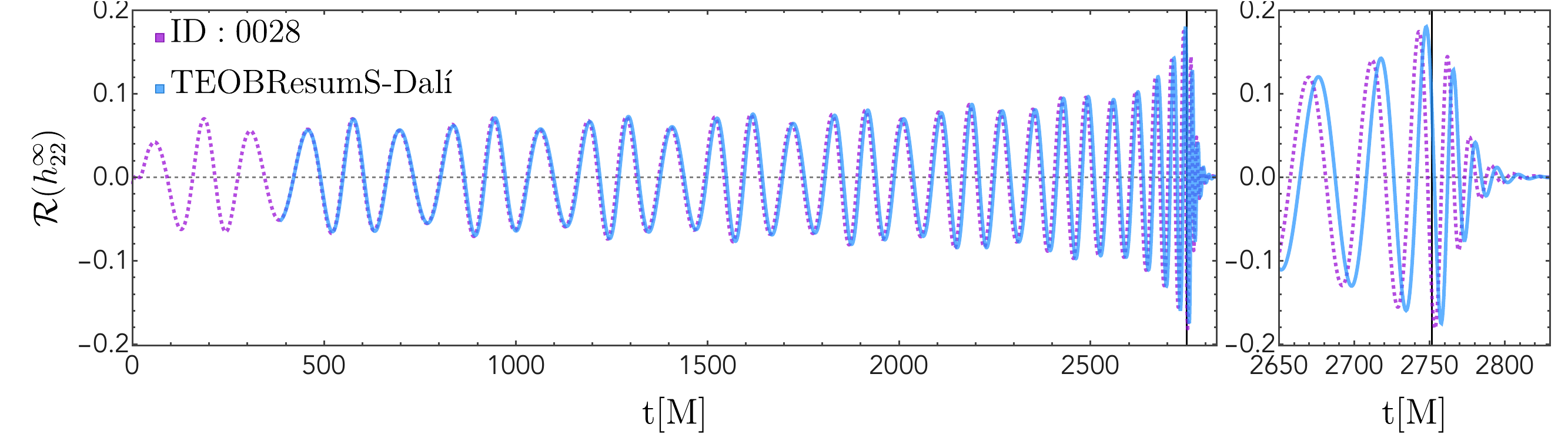}
\includegraphics[width=\columnwidth]{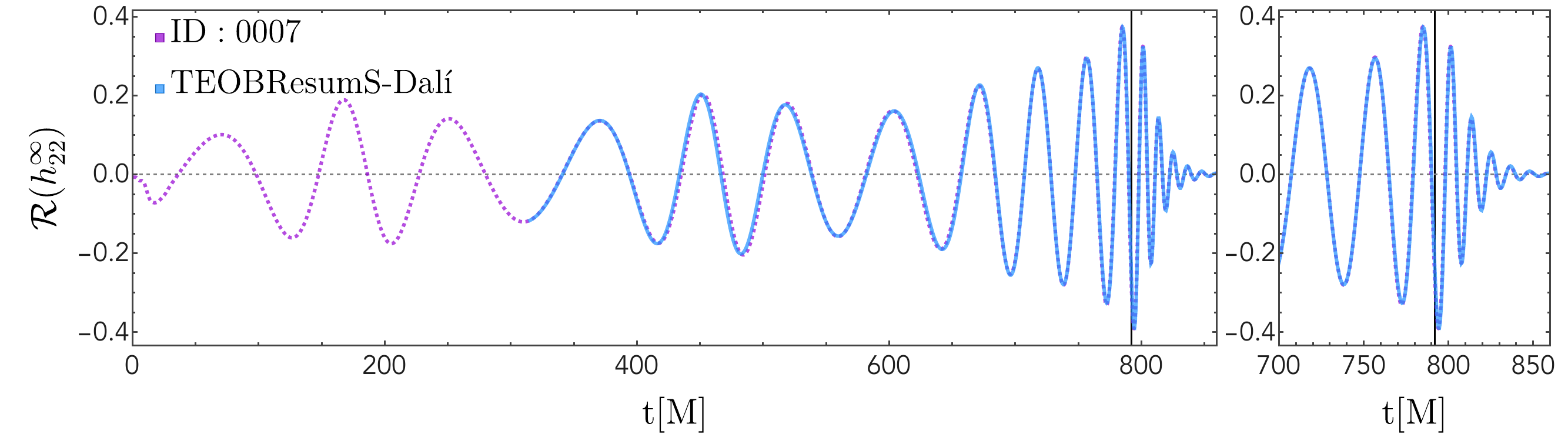}\\
%\vspace{-0.3cm}
\includegraphics[width=\columnwidth]{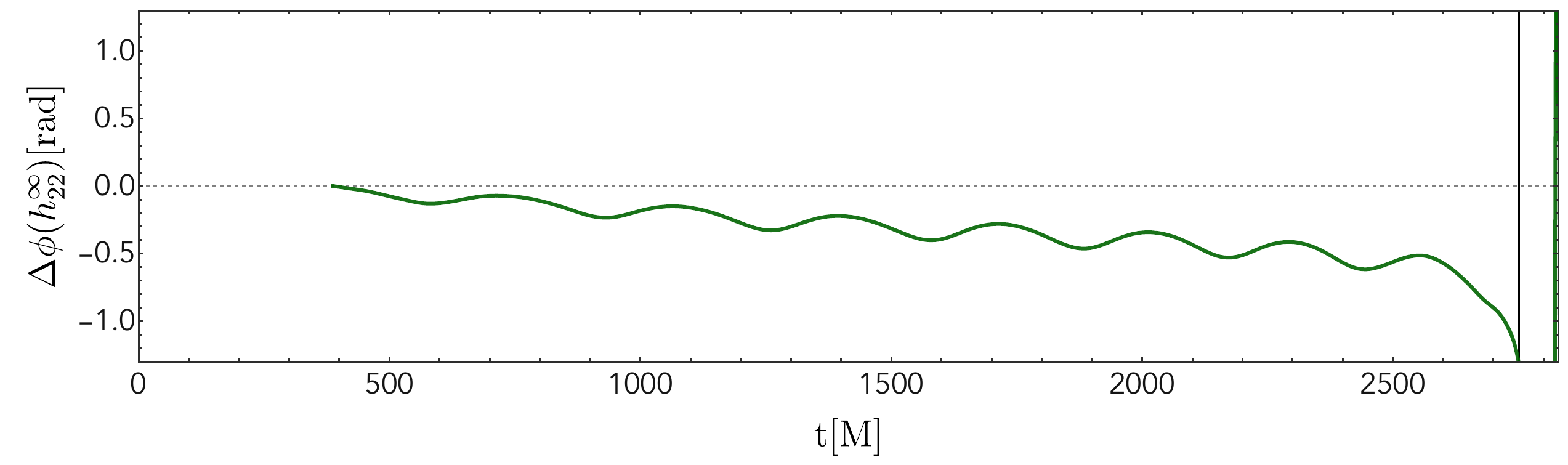}
\includegraphics[width=\columnwidth]{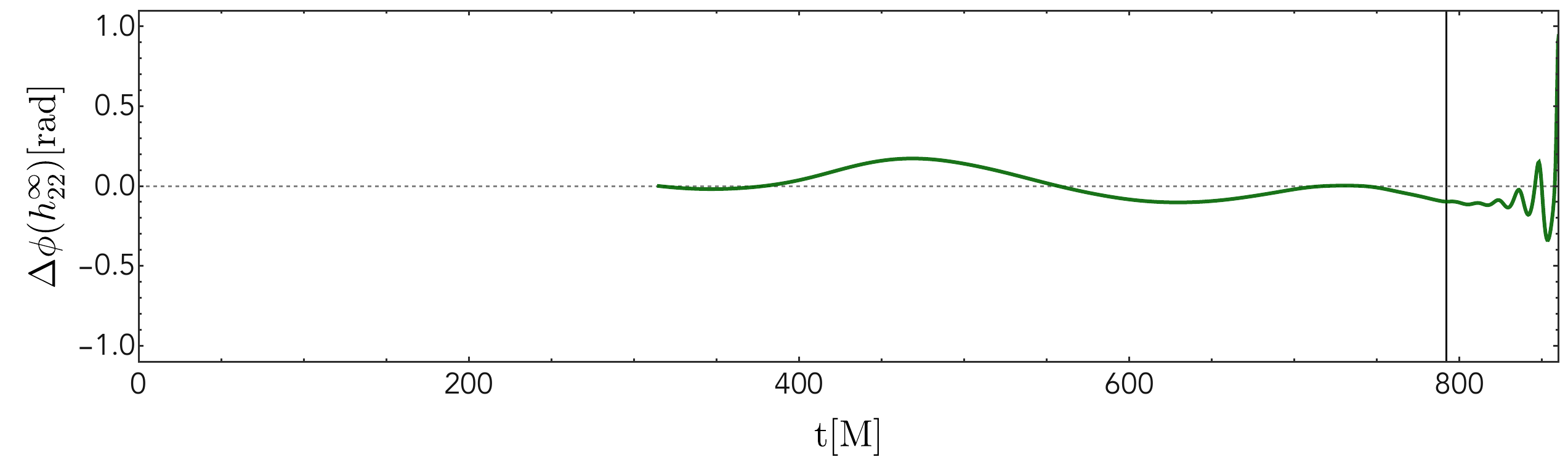} \\
\vspace{0.5cm}
\includegraphics[width=\columnwidth]{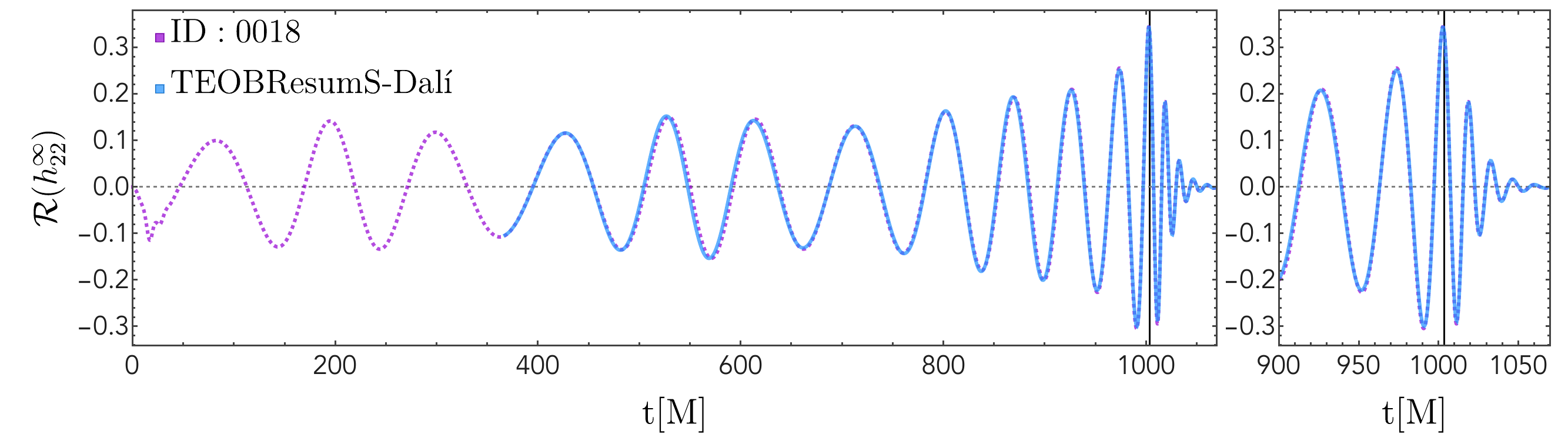}
\includegraphics[width=\columnwidth]{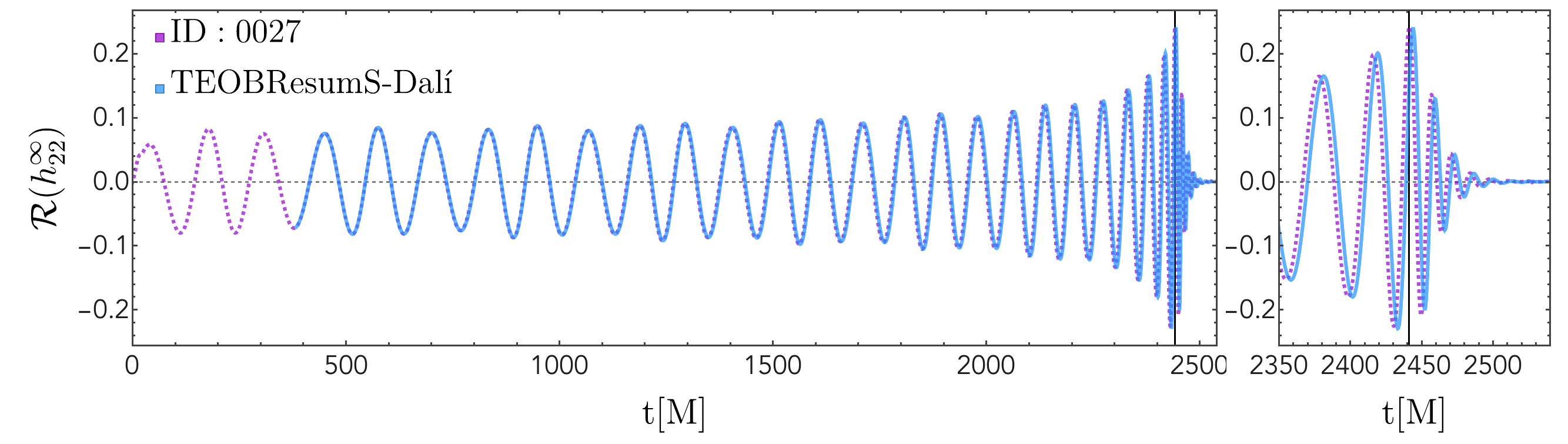}\\
%\vspace{0.0cm}
\includegraphics[width=\columnwidth]{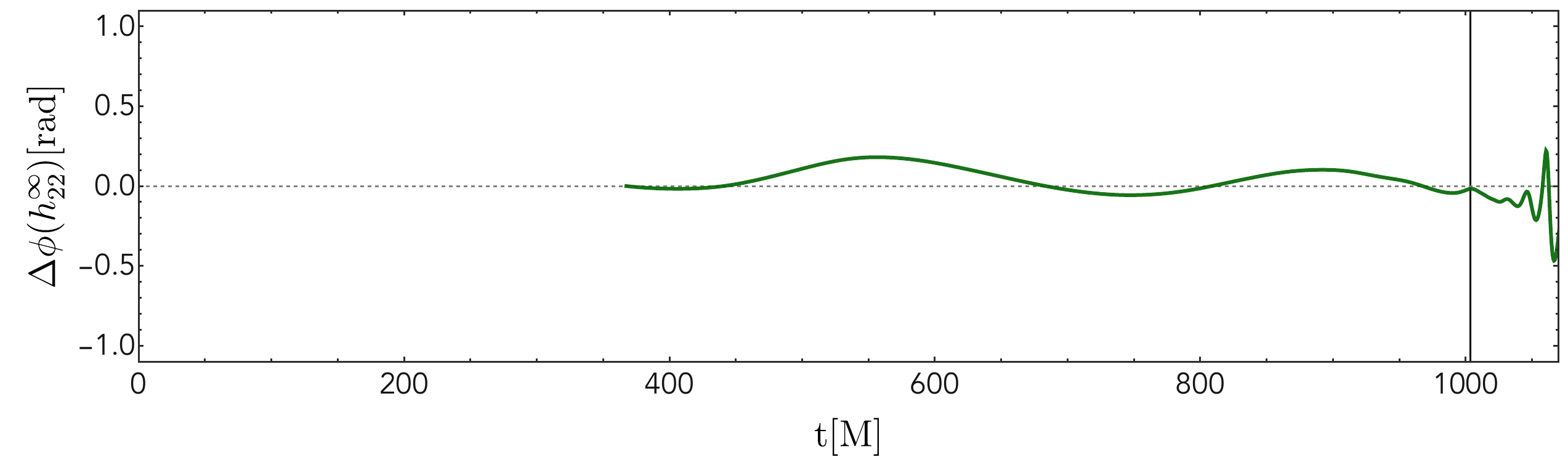}
\includegraphics[width=\columnwidth]{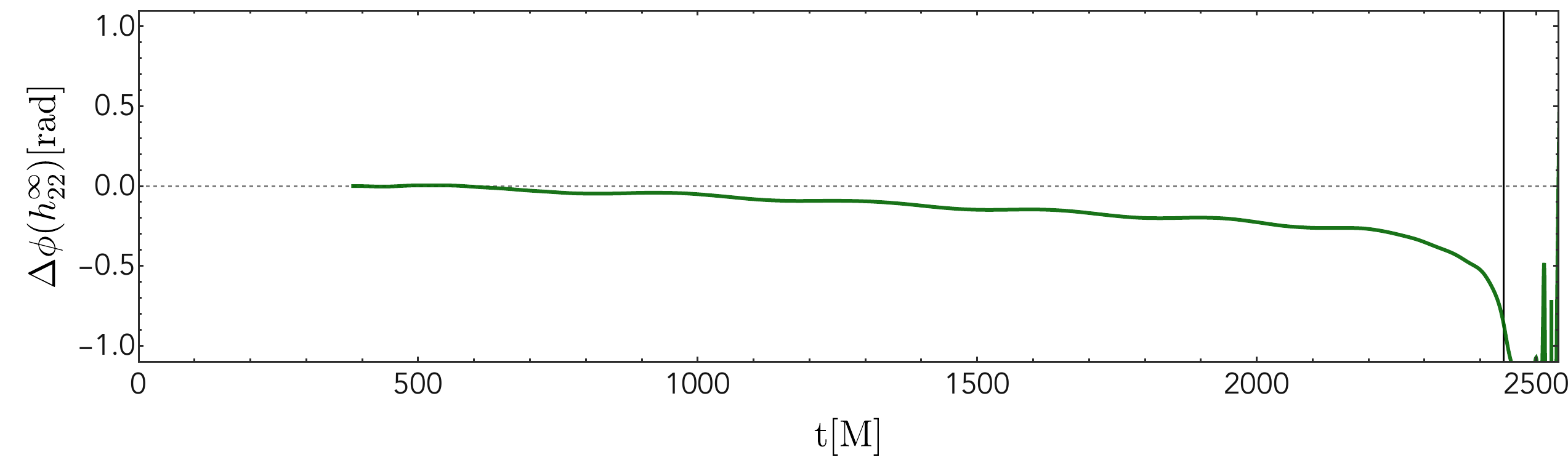}\\
\vspace{0.5cm}
\includegraphics[width=\columnwidth]{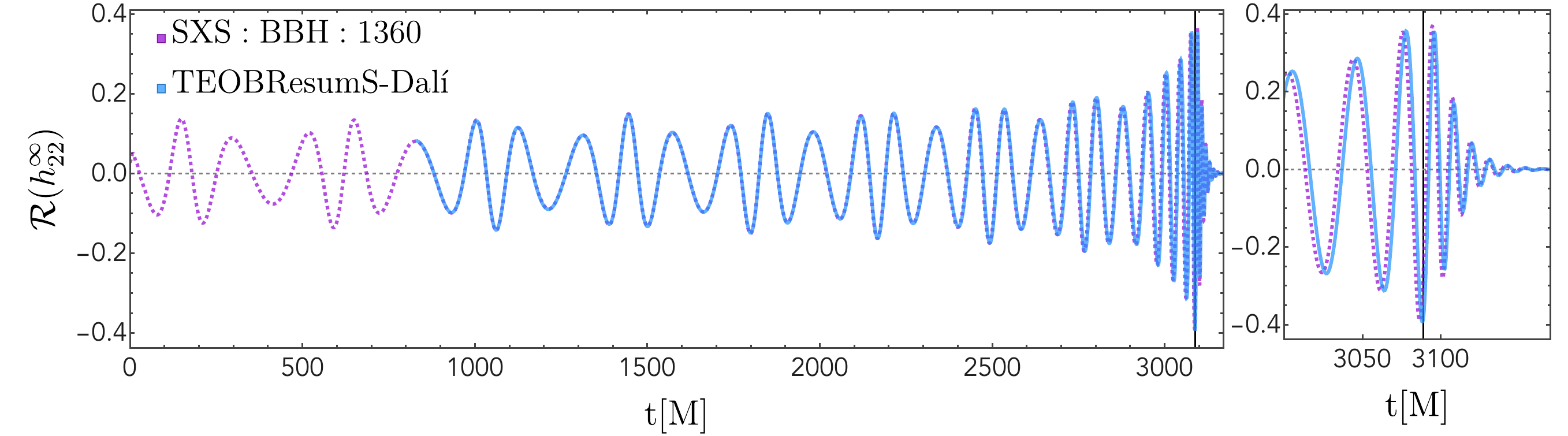}
\includegraphics[width=\columnwidth]{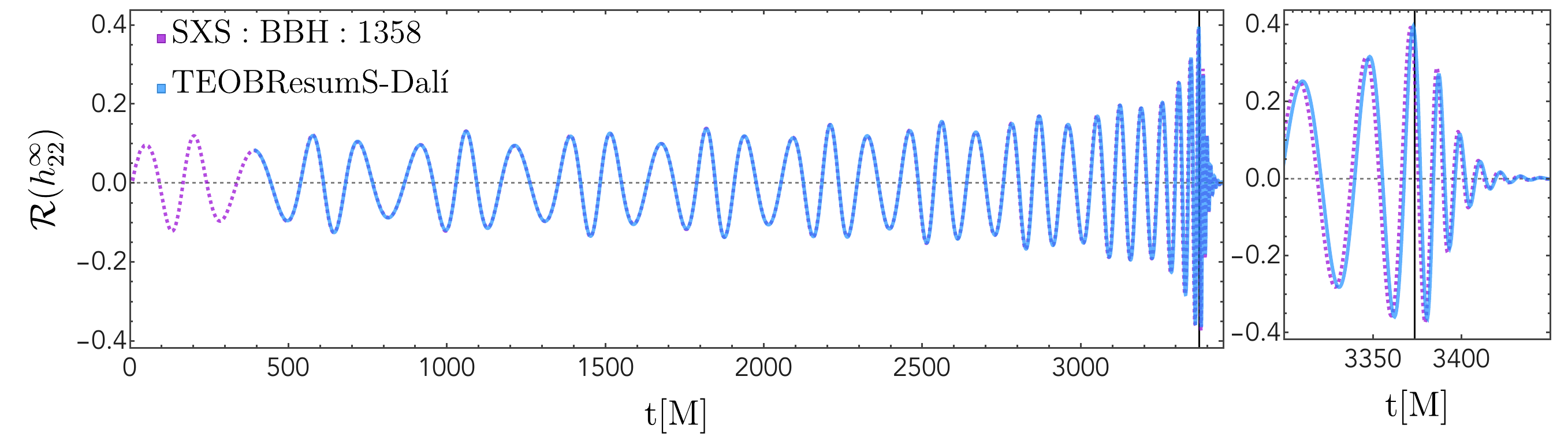}\\
%\vspace{-0.3cm}
\includegraphics[width=\columnwidth]{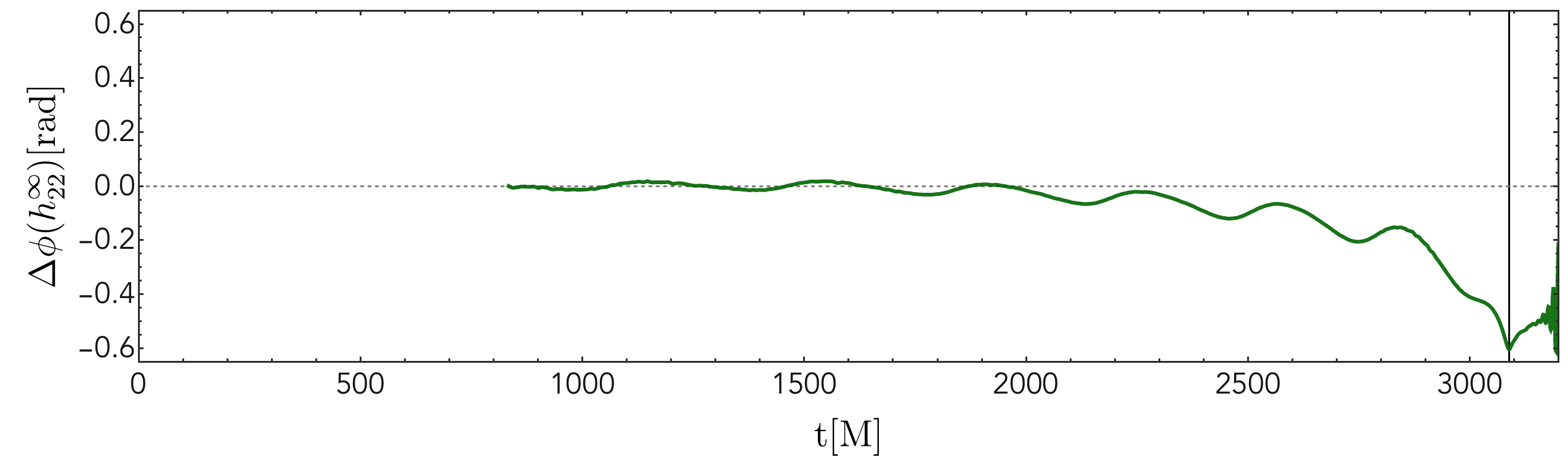}
\includegraphics[width=\columnwidth]{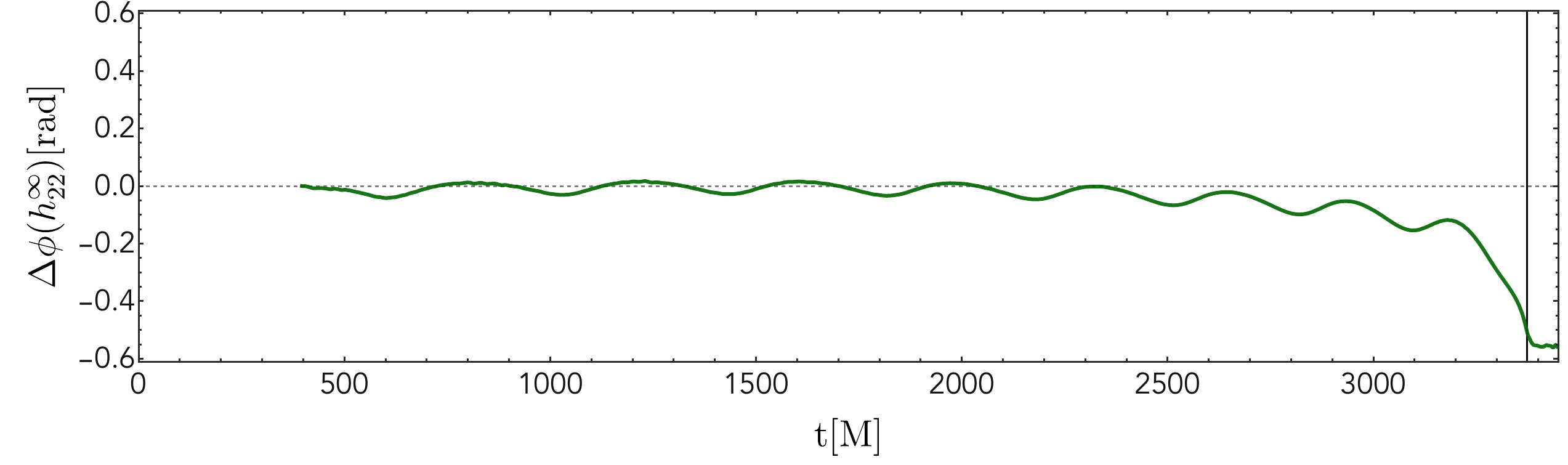}
\caption{
\label{fig:waveformpanel} 
Illustration of the ICs-finding procedure for a selection of NR simulations. For each simulation we show the NR (purple, dashed) and \TEOBDALI{} (blue, solid) waveforms in the top panel with a zoom-in of the merger part on the right. The bottom panel shows the phase difference. We see that the EOB waveforms generated with the initial conditions as determined by the procedure described in Sec.~\ref{sec:ICs} closely match the NR waveforms; the phase differences are less than $\sim 1$ rad at merger. 
}
\end{figure*}

A comparison between a subset of the NR simulations across the parameter space and \TEOBDALI{} waveforms is shown in Fig.~\ref{fig:waveformpanel}. 
For each configuration the top panel shows the real part of $h_{22}^\infty$ of the NR waveform (purple, dashed) and the corresponding \TEOBDALI{} waveform (blue, solid); the bottom panel shows the dephasing $\Delta \phi$ with the maximum dephasing on the order of $\sim 1.326$ rad. 

\subsection{Robustness}
\begin{figure*}[!t]
\subfloat[\label{subfig:a}]{%
   \begin{minipage}{\columnwidth}
   \includegraphics[width=\columnwidth]{./q6_a1_0p0_a2_0p0_e_0p1_D12p83_waveform.pdf} \\
   \includegraphics[width=\columnwidth]{./q6_a1_0p0_a2_0p0_e_0p1_D12p83_phase.pdf}
   \end{minipage}
}
\subfloat[\label{subfig:b}]{%
\begin{minipage}{\columnwidth}
    \includegraphics[width=\columnwidth]{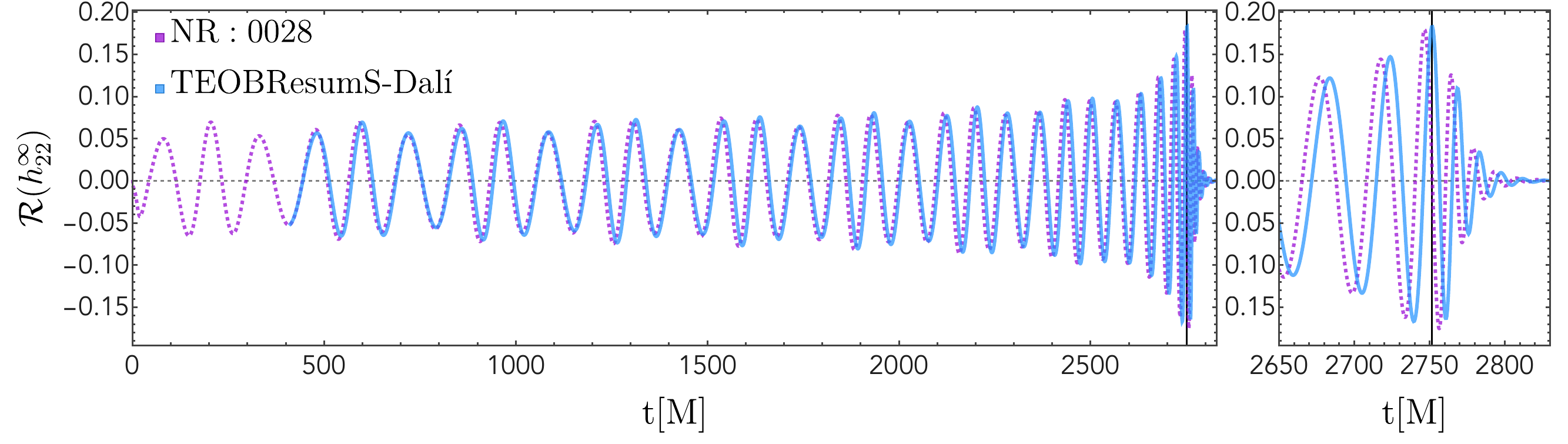} \\
    \includegraphics[width=\columnwidth]{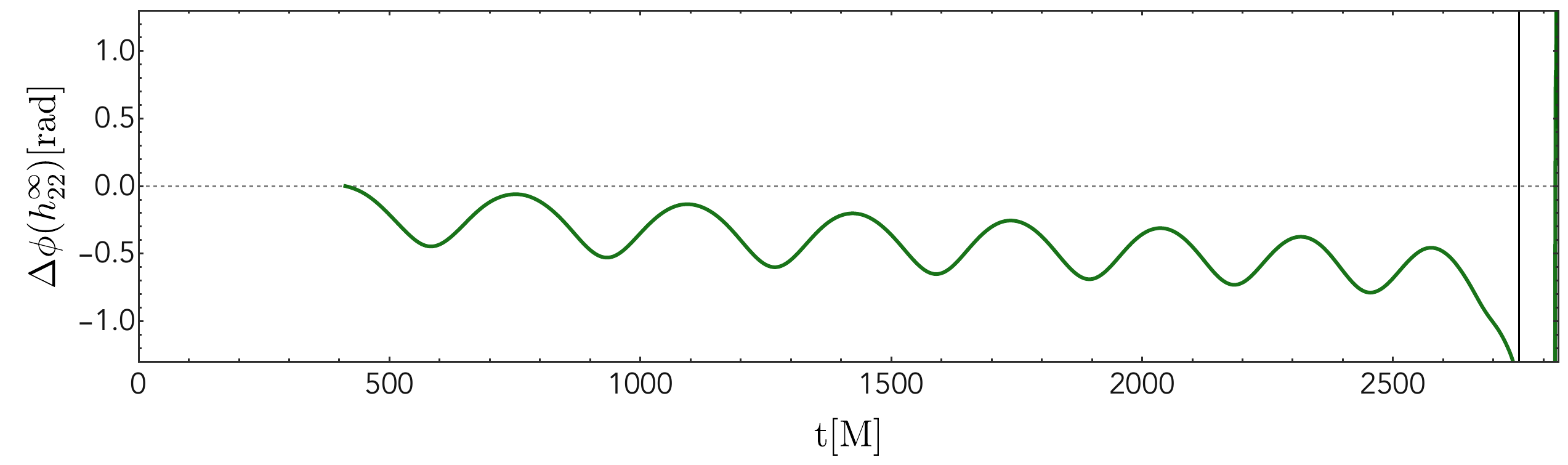}
\end{minipage}
}

\subfloat[\label{subfig:c}]{%
\begin{minipage}{\columnwidth}
    \includegraphics[width=\columnwidth]{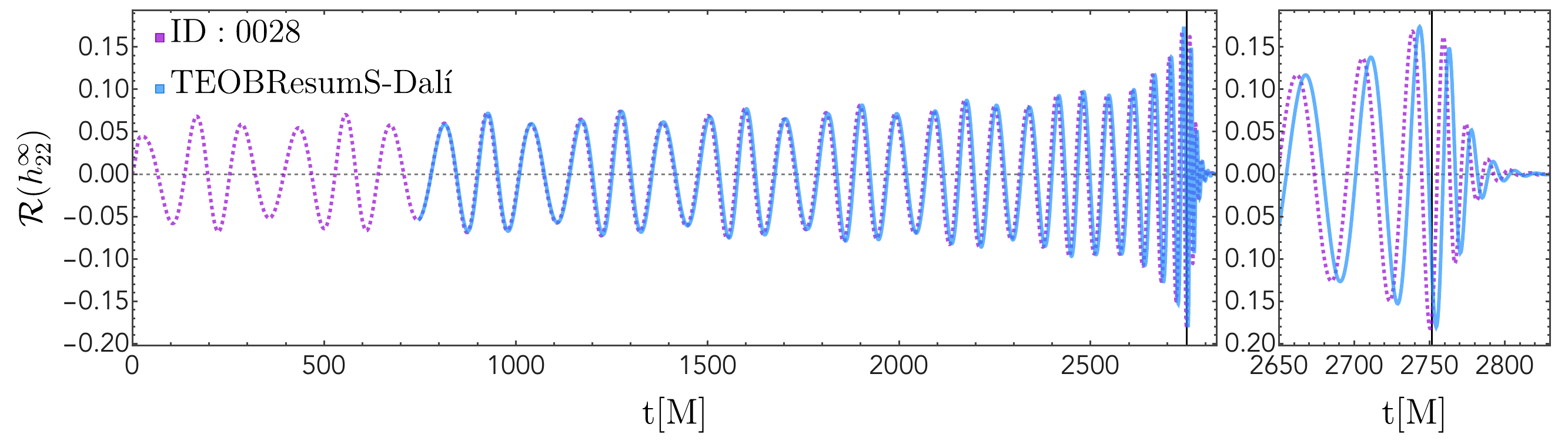} \\
    \includegraphics[width=\columnwidth]{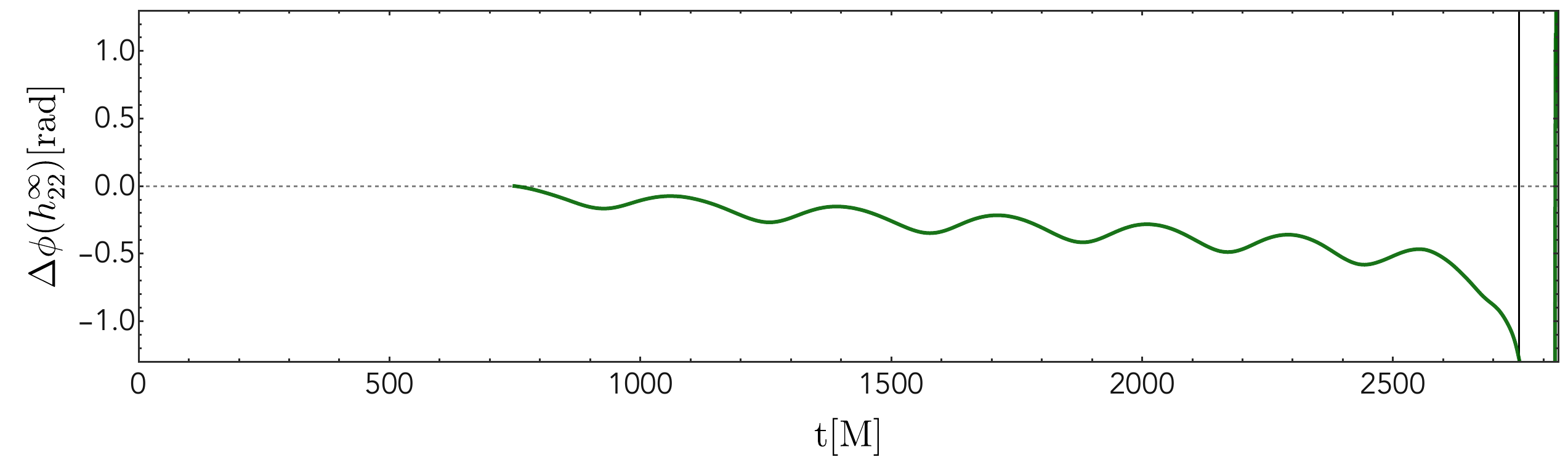}
\end{minipage}
}
\subfloat[\label{subfig:d}]{%
\begin{minipage}{\columnwidth}
    \includegraphics[width=\columnwidth]{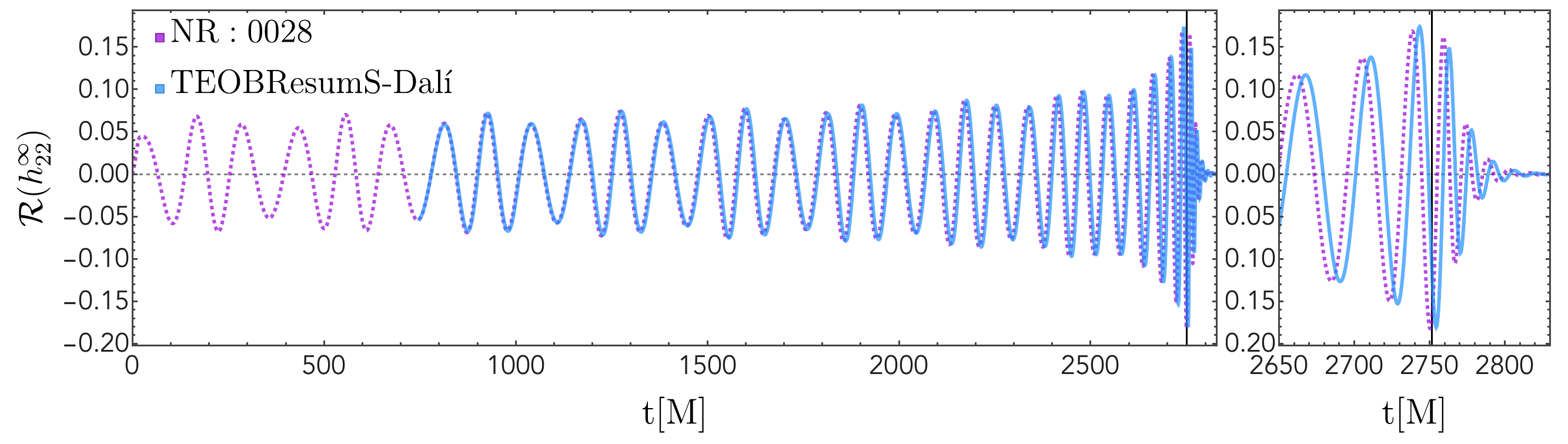} \\ 
    \includegraphics[width=\columnwidth]{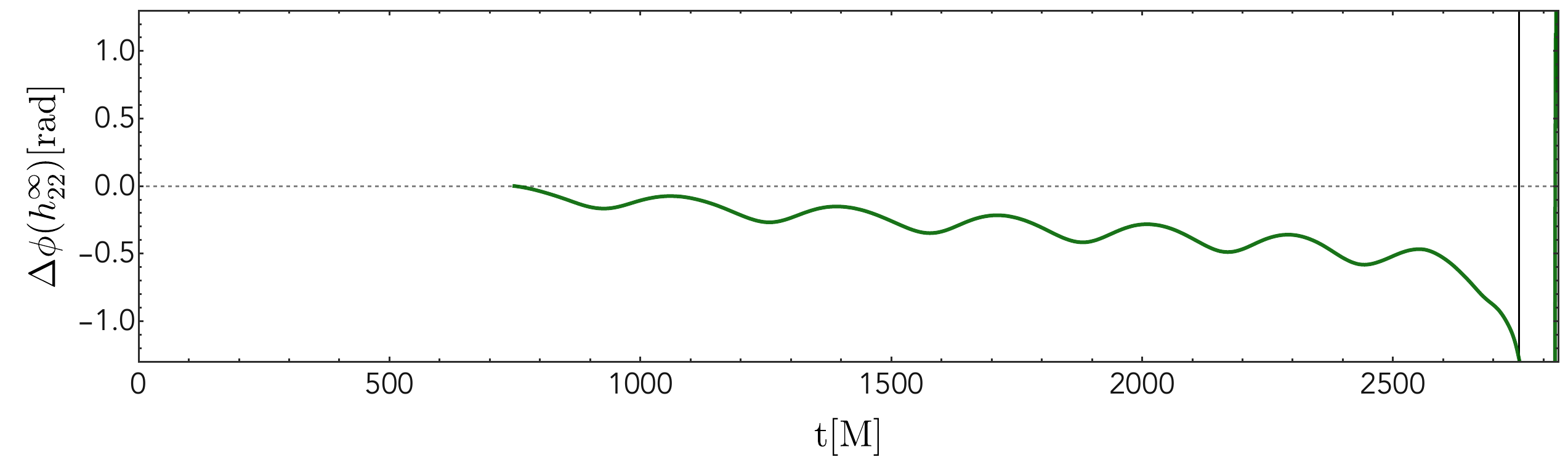}
\end{minipage}
}
\caption{\label{fig:peaks}
Here we explore the impact of including the last peak using simulation \texttt{ID:0028}. The two upper panels (a and b) show the waveforms and phases differences obtained when retaining the \textit{first} peak. The bottom two panels (c and d) are when we exclude the \textit{first} peak from the estimation. The left panels (a and c) are obtained by \textit{removing} the last peak, whereas the right panels (b and d) are obtained by \textit{including} the last peak. We generally observe larger oscillatory phase differences when we include the last peak, i.e., the right panels.
}
\end{figure*}

We verify the reliability of the described procedure by testing a range of different assumptions, including: i) accounting for the last peak, ii) employing $\Psi_4$ instead of the strain $h$, iii) determining the initial conditions using the second apastron, and iv) considering finite extraction radii.  
The obtained initial conditions are summarised in Tab.~\ref{tab:otherpeaks_psi4}. 
Figure~\ref{fig:peaks} illustrates the impact of the inclusion (exclusion) of the last peak, as well as the inclusion (exclusion) of the first peak for \texttt{ID:0028}. 
The two upper panels (\ref{subfig:a} and \ref{subfig:b}) show the waveforms and phases obtained by retaining the first peak, the two bottom panels (\ref{subfig:c} and \ref{subfig:d}) show the waveforms and phases obtained by removing the first peak. 
As expected from the study in Sec.~\ref{sec:sim_length}, in both cases oscillatory phase differences are more pronounced and a larger phase shift is needed (see Tab.~\ref{tab:otherpeaks_psi4}) when including the last peak (right panels) than when removing it (left panels). 
On the other hand, it can sometimes be advantageous to start the EOB dynamics at the second peak as the first peak can be affected by numerical noise, making the eccentricity estimation more prone to errors.
We demonstrate that our procedure is capable of handling such noise artifacts by applying it starting from the second apastron instead. As expected, we obtain slightly different initial conditions but find a very similar maximal dephasing at merger. As before, the largest phase error is seen when also including the last peak. 

% TABLE III
\begin{table*}[]
\centering
\renewcommand{\arraystretch}{1.3}
\begin{tabular}{l|>{\centering}p{2cm}|>{\centering}p{2cm}|>{\centering}p{2cm}|>{\centering}p{2cm}|c|c}
    \hline
    \hline
 \multicolumn{3}{c|}{Numerical Relativity}
  &\multicolumn{2}{c|}{\TEOBDALI{}}
  &\multicolumn{1}{c|}{$\rm \Delta{\phi}$ [rad]}
  &\multicolumn{1}{c}{$\Delta t\, [M]$}\\
\hline
ID & $M\bar{f}_{\rm NR}$ & $e_{\rm NR}$  & $M\bar{f}_{\rm DALI}$& $e_{\rm DALI}$& & \\
\hline
%BBH$\_$q1$\_$e0.05$\_$N80
\texttt{0001} & $0.0076203$ & $0.0510486$ &$0.00762334$ & $0.0510313$ & $0.439$ & $1.64$\\
\texttt{0001} (incl. last peak) & $0.00765019$ & $0.048717$ &$0.00762572$ & $0.0487785$ & $0.461$ & $-14.86$ \\
\hline 
%BBH$\_$q1$\_$e1$\_$N80$\_$v2 
\texttt{0002} & $0.00853042$ & $0.0980399$ & $0.00852527$ & $0.0980725$ & $0.346$ & $-1.78$\\ 
\texttt{0002} ($r_0 =$ 100 M) & $0.00851419$ & $0.0970785$ & $0.0085116$ & $0.0970969$ & $0.636$ & $-0.87$ \\
\texttt{0002} ($r_0 =$ 200 M) & $0.00851574$ & $0.0974921$ & $0.00851574$ & $0.0974921$ & $0.511$ & $1.56$\\ 
\texttt{0002} ($r_0 =$ 250 M) & $0.00851735$ & $0.0975702$ & $0.00852125$ & $0.0975467$ & $0.403$ & $1.31$\\ 
\hline
%BBH$\_$q2$\_$e0.05$\_$a0.5$\_$a0.5 
\texttt{0016} & $0.00736959$ & $0.0457956$ & $0.0073548$& $0.0458236$ & $-0.345$ & $-16.40$\\
%BBH$\_$q2$\_$e0.05$\_$a0.5$\_$a0.5
\texttt{0016} ($\Psi_4$) & $0.00738382$ & $0.046704$ & $0.00734072$& $0.0468821$ & $-0.483$ & $-38.83$\\
\hline
%q6$\_$a1$\_$0p0$\_$a2$\_$0p0$\_$e$\_$0p1$\_$D12p83 
\texttt{0028} & $0.00781405$ & $0.105506$ & $0.00777383$ & $0.105979$ & $1.326$ & $-44.02$\\
\texttt{0028} (incl. last peak) & $0.00827036$ & $0.0990475$ & $0.00820059$ & $0.0997272$ & $1.285$ & $-70.19$\\
\texttt{0028} ($2^{\rm nd}$ apastron) &$0.00822478$ & $0.101856$ & $0.00818341$ & $0.102409$ & $1.285$ & $-36.69$\\
\texttt{0028} ($2^{\rm nd}$ apastron, incl. last peak) & $0.0078723$ & $0.102723$ & $0.00779677$ & $0.103425$ & $1.504$ & $-96.02$\\
\hline
\texttt{SXS:BBH:1360} (earlier time) & $0.00598258$ & $0.228105$  & $0.00597789$ & $0.228307$ & $0.645$ & $-8.28$\\
\hline
\hline
\end{tabular}
\caption{\label{tab:otherpeaks_psi4} Summary of initial conditions calculated under different assumptions: Varying number of peaks, using $\Psi_4$ instead of $h$ and different finite extraction radii. The last row gives the initial conditions inferred when extrapolating to earlier times.}
\end{table*}

% FIGURE 10

As shown previously in Fig.~\ref{fig:egwextrradii} for \texttt{ID:0002}, waveforms extracted at a finite radius rather than extrapolated to infinity yield a slightly different eccentricity evolution, resulting in a small change in initial conditions. We find that the smallest extraction radii lead to the largest dephasing at merger, while the extrapolated waveforms give the best agreement with \TEOBDALI{}.

Finally, we also test the procedure using $\Psi_4$ instead of the strain. 
This can be useful when the waveform extraction procedure is particularly challenging or affected by numerical noise. 
This approach allows indeed to compute the eccentricity evolution directly from $\Psi_4$ by simply using the ratio Eq.~\eqref{eq:ratio}. 
Employing the latter we calculate the ratio $\mathcal{R}( \bar{\omega}_{\rm NR}/2 )$ where $\bar{\omega}_{\rm NR} = 2 \pi \bar{f}_{\rm NR}$ for \texttt{ID:0016} at the frequency of first apastron to be $1.385$. 
The initial eccentricity is then calculated as  
\begin{equation}
    e_{\rm NR}=\frac{e_{\rm gw}(\Psi_4)}{\mathcal{R}}. 
\end{equation} 
The phase differences obtained using $\Psi_4$ are very similar to the ones obtained from $h$, see Tab.~\ref{tab:otherpeaks_psi4}. This suggests that the procedure is equally applicable to $\Psi_4$ and $h$.

Overall, we find that our procedure is robust and gives self-consistent results for the various assumptions made.

% FIGURE 11
\begin{figure}[t]
\center
\includegraphics[width=\columnwidth]{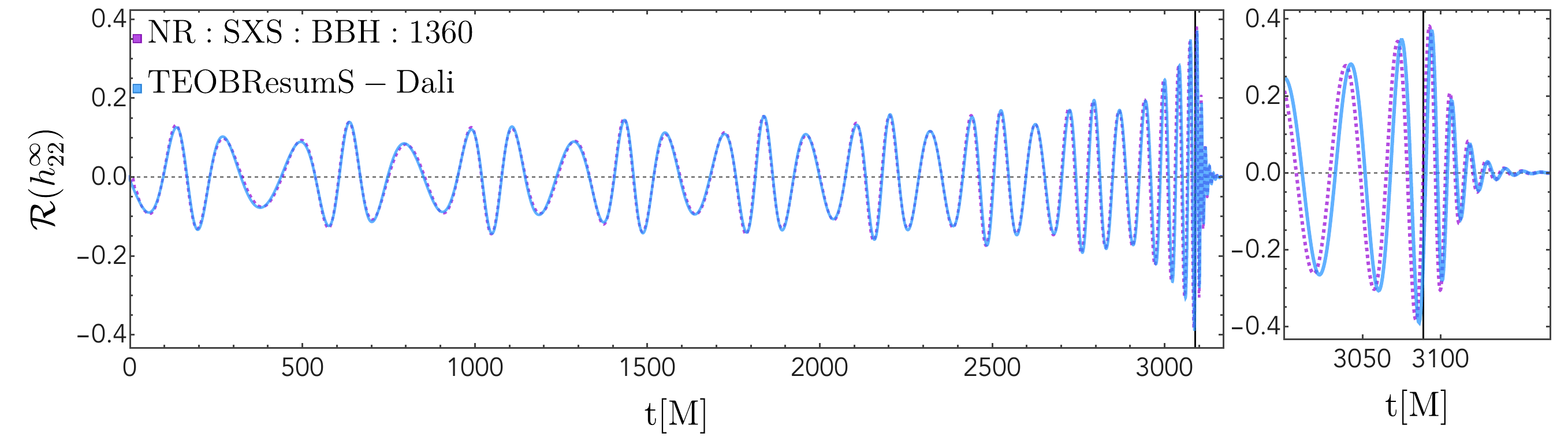}\\
\includegraphics[width=\columnwidth]{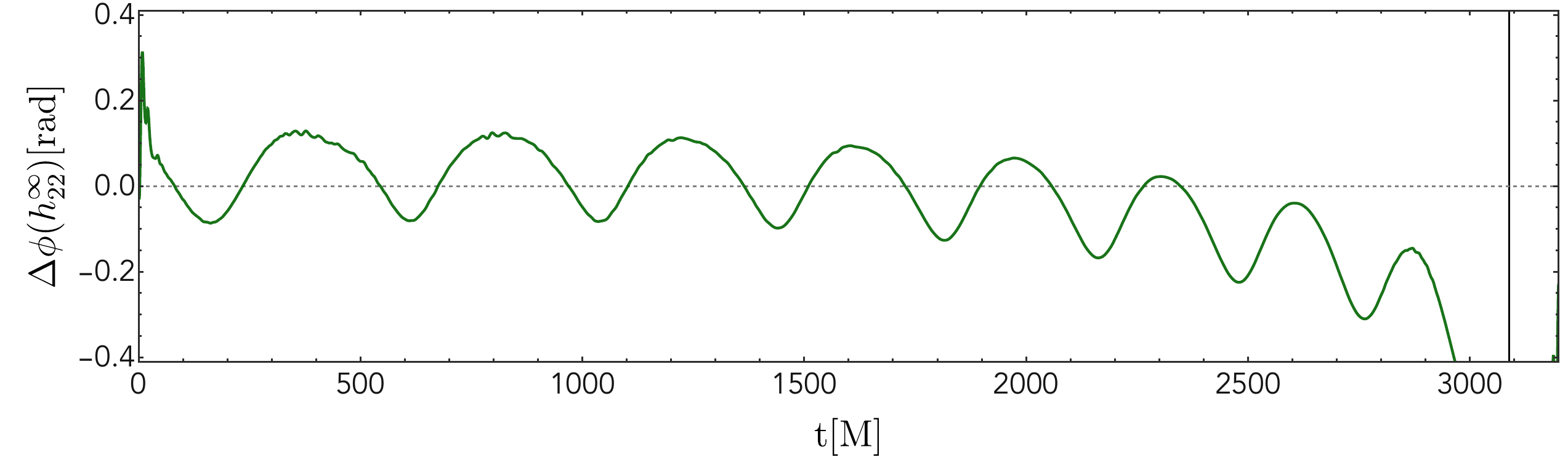}
\caption{
Comparison of \texttt{SXS:BBH:1360} (purple) and \TEOBDALI{} (blue) using the end-to-end pipeline to calculate the initial conditions at earlier times, as described in Sec.~\ref{sec:earlytimes}, together with the phase difference (lower panel).
}
\label{fig:earlytimes}
\end{figure}

%%%%%%%%%%%%%%%%%%%%%%%%%%%%%%%%%%%%%%%%%%%%%%
\subsection{Extrapolation to Earlier Times}
\label{sec:earlytimes}
%%%%%%%%%%%%%%%%%%%%%%%%%%%%%%%%%%%%%%%%%%%%%%
A limitation of this procedure is that we define a reference time based on the first apastron in the data. 
When using NR simulations, for example, this results in a loss of $\sim 300-400M$ at the start of the simulation. 
We could start the evolution from an earlier time using, e.g., a backwards PN evolution or, alternatively, we could try to extrapolate the initial conditions.
In this section we test the extrapolation of the initial conditions, demonstrating that we can robustly generate a waveform from earlier times. 

Using \texttt{SXS:BBH:1360}, we want our waveform to start at $t_{\rm NR} = 0$. However, this requires determining initial conditions for an apastron that occurs before $t_{\rm NR} = 0$. 
As we do not know the exact location of the apastron at earlier times, we empirically estimate it using the periastron advance as a guide, e.g., we find $t^a_{\rm NR} \sim - 63M$ for \texttt{SXS:BBH:1360}. The procedure laid out in Sec.~\ref{sec:ICs} is then slightly modified:

\begin{enumerate}
\item Read off the NR initial conditions at $ t^a_{\rm NR}+\Delta t$, taking $\Delta t$ to be the time difference already found and listed in Tab.~\ref{tab:comparison}. This constitutes our original initial conditions, as listed in columns two and three of Tab.~\ref{tab:otherpeaks_psi4}.
\item We then iterate the procedure to find new initial conditions based on the original values identified in 1) to help mitigate against the fact that the apastron was only estimated empirically, therefore allowing for a more precise measurement of the initial conditions. 
\end{enumerate}
The initial conditions, phase difference and the final $\Delta t$ are listed in the last row of Tab.~\ref{tab:otherpeaks_psi4}. 
The total time difference between the NR and the EOB evolution is given by the sum of the time differences listed in Tab.~\ref{tab:comparison} and Tab.~\ref{tab:otherpeaks_psi4}. The resulting waveforms and dephasing are shown in Fig.~\ref{fig:earlytimes}. We see that the phases differences are larger compared to Fig.~\ref{fig:waveformpanel} but the dephasing is still quite small, with a maximum of $0.645$ rad at merger. 
Whilst this procedure is promising, the robustness of extrapolating to earlier times needs to be carefully checked across the parameter space.

%%%%%%%%%%%%%%%%%%%%%%%%%%%
\subsection{ Mismatches}
\label{sec:match}
%%%%%%%%%%%%%%%%%%%%%%%%%%%
In order to quantify the (dis)agreement between the NR simulations and the \TEOBDALI{} waveforms, we use the match $\mathcal{M}$, defined in the usual way as
\begin{equation}
\label{eq:match}
\mathcal{M}(h_{\rm NR}, h_{\rm TEOB}) = \underset{t_c, \phi_c}{\rm max}  \frac{\langle h_{\rm NR} | h_{\rm TEOB} \rangle}{\sqrt{\langle h_{\rm NR} |h_{\rm NR} \rangle \langle h_{\rm TEOB} |h_{\rm TEOB} \rangle}},
\end{equation}
where $t_c$ and $\phi_c$ denote the time and phase of coalescence respectively, and $\langle h_{\rm NR} |h_{\rm TEOB} \rangle$ denotes a noise-weighted inner product
\begin{equation}
\label{eq:inner_prod}
\langle h_{\rm NR} |h_{\rm TEOB} \rangle = 4 \Re \int_{f_{\rm min}}^{f_{\rm max}} \frac{ \tilde{h}_{\rm NR}(f) \tilde{h}^*_{\rm TEOB} (f)}{S_n (|f|)} df,
\end{equation}
where $S_n (|f|)$ is the one-sided power spectral density (PSD) of the strain noise, $\tilde{h}$ the Fourier transform of $h$ and ${}^*$ denotes the complex conjugate.
In practice, we will use the mismatch, which is defined as
\begin{equation}
\overline{\mathcal{M}} (h_{\rm NR}, h_{\rm TEOB})  =  1-\mathcal{M}(h_{\rm NR}, h_{\rm TEOB}) 
\end{equation} where $h_{\rm NR} \equiv h^\infty_{22}$ from the numerical data.  

\begin{figure}[!tp]
\includegraphics[width=\columnwidth]{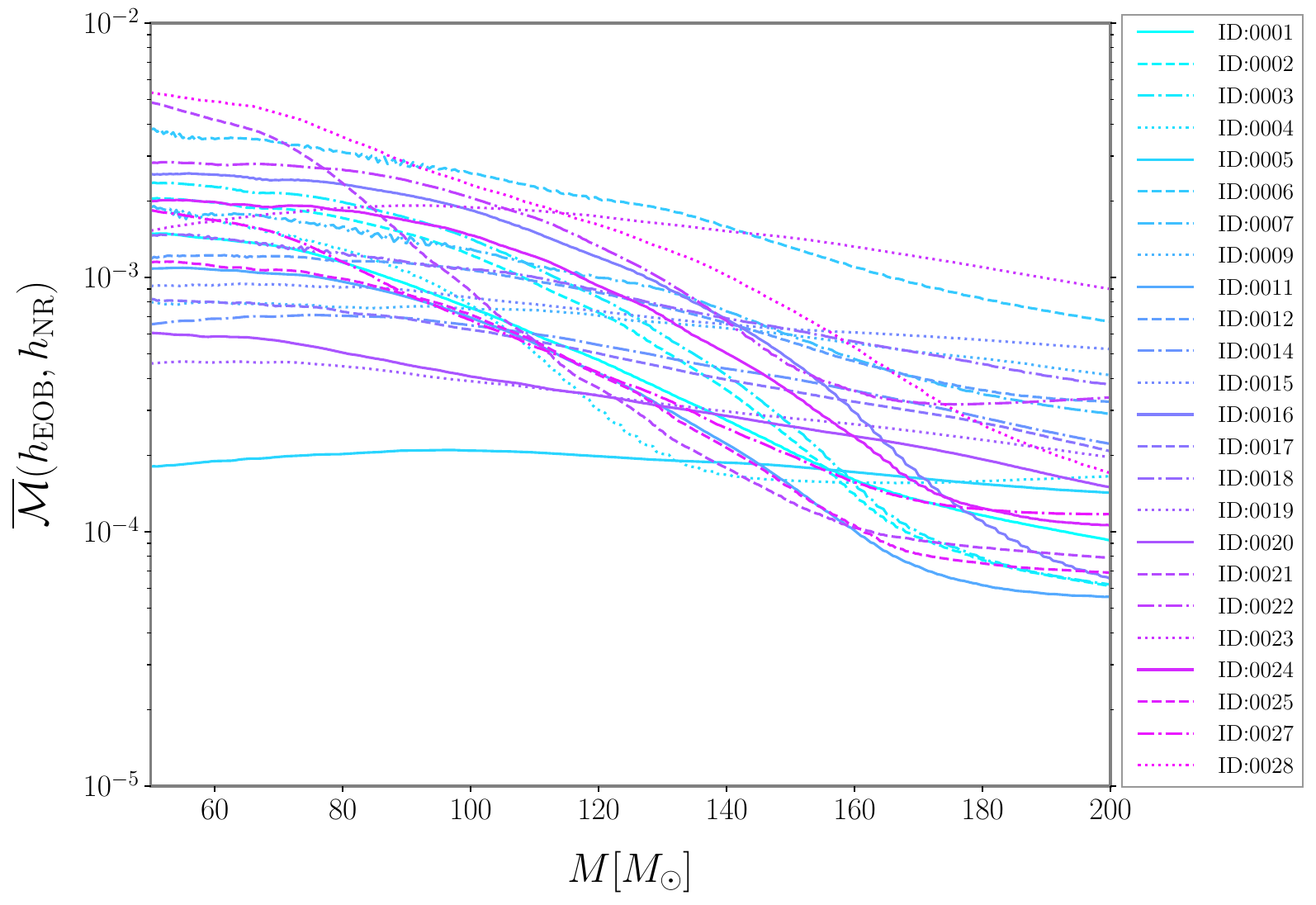}
\caption{\label{fig:mismatches} Mismatches between our suite of numerical simulations and \TEOBDALI{} across a range of total masses. The mismatches vary between a few times $10^{-3}$ to just below $10^{-4}$. For binaries with a total mass of $65 M_\odot$ the mean mismatch across our simulations is $0.0015$.}
\end{figure}

The mismatches are computed assuming $f_{\rm min} = 20 $ Hz and a total mass $M$ in the range $[50, 200] M_{\odot}$. 
We use a PSD that corresponds to the approximate sensitivity of the Advanced LIGO detector in the fourth observing run (O4) \cite{KAGRA:2013rdx,psddcc}. 
The resulting mismatches are shown in Fig.~\ref{fig:mismatches}. We find good agreement between the NR and EOB waveforms across the parameter space spanned by our simulations, with a typical mismatch on the order of $\sim \mathcal{O}(10^{-3})$. 
We do not find any strong correlations with $q$, $e_0$ and $\chi_{\rm eff}$ but note that the worst mismatches between our NR data and \TEOBDALI{} waveforms are found for unequal masses and large eccentricities. 
Our result is in broad agreement with \cite{Nagar:2021gss} but it is crucial to note that, as opposed to other works, we have not performed any optimisations over eccentricity parameters as this is not necessary due to the identified mapping between initial conditions. 
This allows us to treat the orbital eccentricity as a meaningful physical parameter that characterises the binary rather than as a nuisance parameter that is numerically optimised over.

% FIGURE 12

%~~~~~~~~~~~~~~~ Conclusions ~~~~~~~~~~~~~~~
\section{Conclusions}
\label{sec:discussion}
The primary goal of this work was to establish and validate a robust framework for defining a map between the eccentricity evolution as measured directly from a GW signal and eccentric initial conditions to allow for a direct comparison between different eccentric waveform models. 
Our framework builds on \cite{Bonino:2022hkj} in terms of the eccentricity estimator $e_{\rm gw}$ recently introduced in \cite{Ramos-Buades:2022lgf}, which has the correct Newtonian limit and is defined purely in terms of observables from the GW signal. 
The pipeline can be applied to any generic waveform data, including NR simulations and semi-analytic waveform models. 
In support of the analysis, we introduced a suite of \nnrsim{} aligned-spin, eccentric NR simulations for binary black hole mergers, designed to help us understand the dynamics of eccentric binaries.
We validated our pipeline against this catalog of NR simulations, as well as publicly available SXS simulations. 

A crucial ingredient is the measurement of the eccentricity evolution. 
To assess its robustness, we systematically tested a number of possible limitations. In particular, we investigated how the number of peaks in the GW signal, the data quality, extrapolation to future null infinity, and the use of $\Psi_4$ versus the strain $h$ impacted our ability to infer the eccentricity evolution from the NR simulations. 
As a result, we found that the minimum length of an NR simulation required to accurately measure the eccentricity evolution is $\sim 15$ GW cycles ($\sim 7$ orbits). 
Similarly, we found that the procedure breaks down as the binary approaches merger and is unreliable $\sim 4$ GW cycles ($\sim 2$ orbits) before merger, see also the discussion in \cite{Shaikh:2023ypz}. 
In testing the pipeline, we also found that dropping the last peak from the fitting procedure results in a more reliable determination of the eccentricity, as discussed in Sec.~\ref{sec:methodology}, see also Fig.~\ref{fig:peaks}. 
Furthermore, our procedure can be defined using various definitions of (monotonic) frequency, including the mean frequency and the average frequency, also in agreement with \cite{Shaikh:2023ypz}. 

With a reliable procedure for determining $e_{\rm gw}$ in place, we presented a simple framework for mapping eccentric initial conditions between NR simulations and EOB models in Sec.~\ref{sec:ICs}. 
When performing comparisons between NR simulations against \TEOBDALI{}, we found that we needed to account for a time shift between the NR and EOB waveforms before we could construct a map between the initial conditions. 
Nevertheless, after accounting for this time shift, we found excellent agreement between the NR and EOB initial conditions, resulting in waveforms with typical mismatches on the order of $\sim 10^{-3}$, as shown in Fig.~\ref{fig:mismatches}.
We also verified that the determination of the initial conditions is robust under different options for the input data such as waveform extrapolation, finite extraction radii or the use of $\Psi_4$, finding that the least dephasing is achieved when determining the eccentricity evolution from the extrapolated strain and excluding the last apastron peak in the fitting procedure. 
Utilising the periastron advance as an empirical estimator, we were also able to extrapolate our procedure to earlier times, allowing us to retain more of the NR data. 

A key advantage of our framework is that we no longer need to perform any numerical optimisation over eccentric parameters but it instead appeals to geometrically meaningful initial conditions. 
A key practical application of this approach is towards calibrating an EOB model, in which the eccentricity evolution is matched to that of a corresponding NR simulation in terms of the same $\lbrace M \bar{f}, e_{\rm gw} \rbrace$ pair. We will leave this for future work.

%~~~~~~~~~~~~~~~ Acknowledgments ~~~~~~~~~~~~~~~
\section*{Acknowledgements}

The authors thank Antoni Ramos-Buades, Md Arif Shaikh, Vijay Varma, Alessandro Nagar and Rossella Gamba for useful discussions.
We also thank Gregorio Carullo for useful comments. 
A.B. is supported by STFC, the School of Physics and Astronomy at the University of Birmingham and the Birmingham Institute for Gravitational Wave Astronomy. G.P. is very grateful for support from a Royal Society University Research Fellowship URF{\textbackslash}R1{\textbackslash}221500 and RF{\textbackslash}ERE{\textbackslash}221015. GP gratefully acknowledges support from an NVIDIA Academic Hardware Grant. G.P. and P.S. acknowledge support from STFC grant ST/V005677/1. P.S. also acknowledges support from a Royal Society Research Grant RG{\textbackslash}R1{\textbackslash}241327.
% Computing
Numerical simulations and computations were performed using the University of Birmingham's BlueBEAR HPC service, which provides a High Performance Computing service to the University's research community, the Sulis Tier 2 HPC platform hosted by the Scientific Computing Research Technology Platform at the University of Warwick funded by EPSRC Grant EP/T022108/1 and the HPC Midlands+ consortium, and on the Bondi HPC cluster at the Birmingham Institute for Gravitational Wave Astronomy.
This work also used the DiRAC@Durham facility managed by the Institute for Computational Cosmology on behalf of the STFC DiRAC HPC Facility (www.dirac.ac.uk). The equipment was funded by BEIS capital funding via STFC capital grants ST/P002293/1, ST/R002371/1 and ST/S002502/1, Durham University and STFC operations grant ST/R000832/1. DiRAC is part of the National e-Infrastructure. 
% Software
Processing of the NR data was performed with Wolfram's \texttt{Mathematica}~\cite{Mathematica} and \texttt{SimulationTools}, a package written by Ian Hinder and Barry Wardell, with contributions from Kyriaki Dionysopoulou and Aaryn Tonita. \texttt{Matplotlib}~\cite{Hunter:2007} was used for some of the visualisations. The version of \TEOBDALI{} used in the comparisons corresponds to git commit bac2c08~\cite{TEOBgit}.
% LIGO DCC
This manuscript has the the LIGO document number P2400141.

% %~~~~~~~~~~~~~~~ Appendix A
% ~~~~~~~~~~~~~~~
\appendix
 
\section{Filtering of $\Psi_4$ and choice of $\omega_0$}
\begin{figure}
\center
\includegraphics[width=0.48\columnwidth]{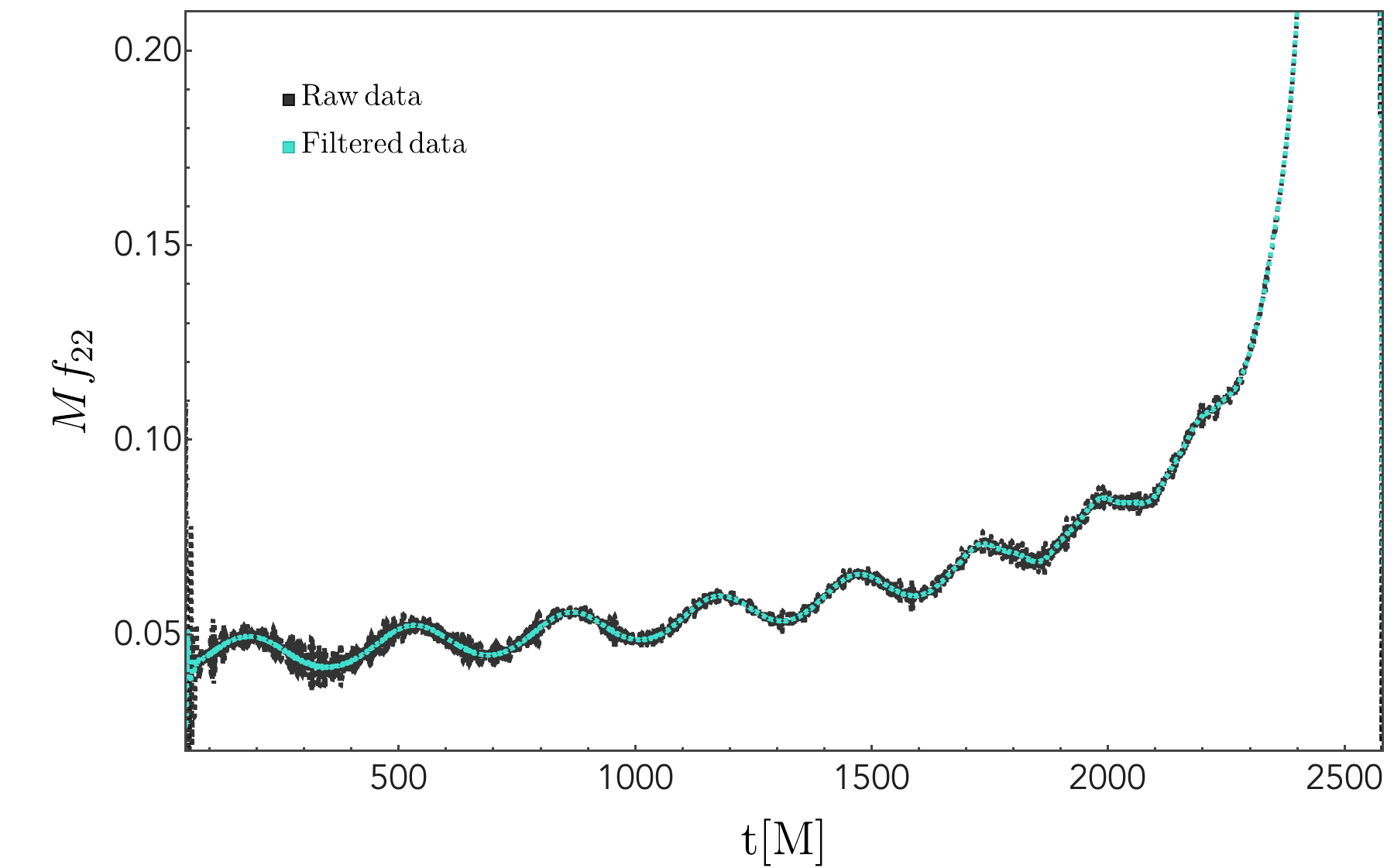}
\includegraphics[width=0.48\columnwidth]{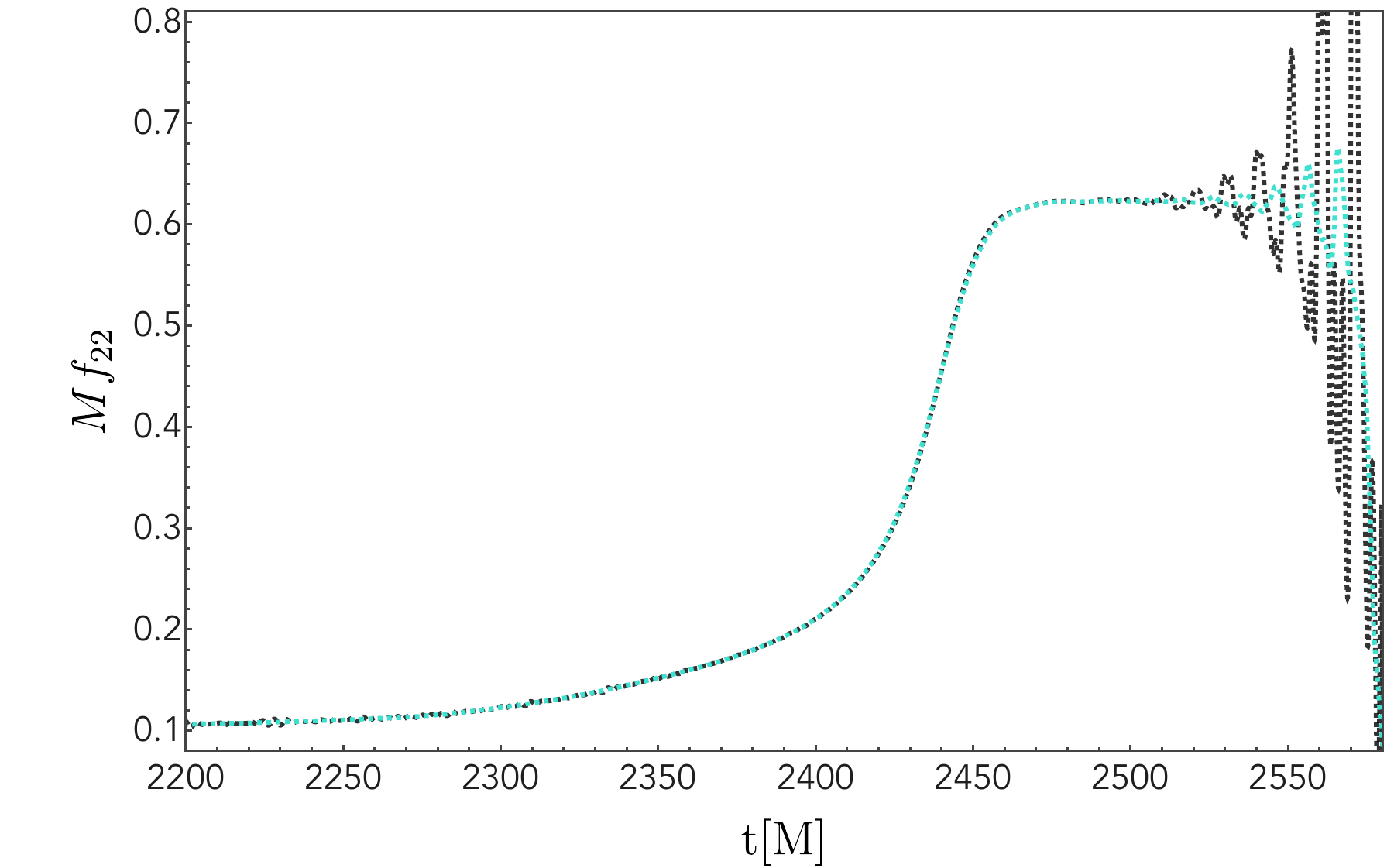}
\caption{\label{fig:freqfilter} 
Noise filtering of $\Psi_4$. When filtering the data, we need to be careful to not introduce unphysical features or behaviour.
}
\end{figure}

\begin{figure}
\center
\includegraphics[width=0.9\columnwidth]{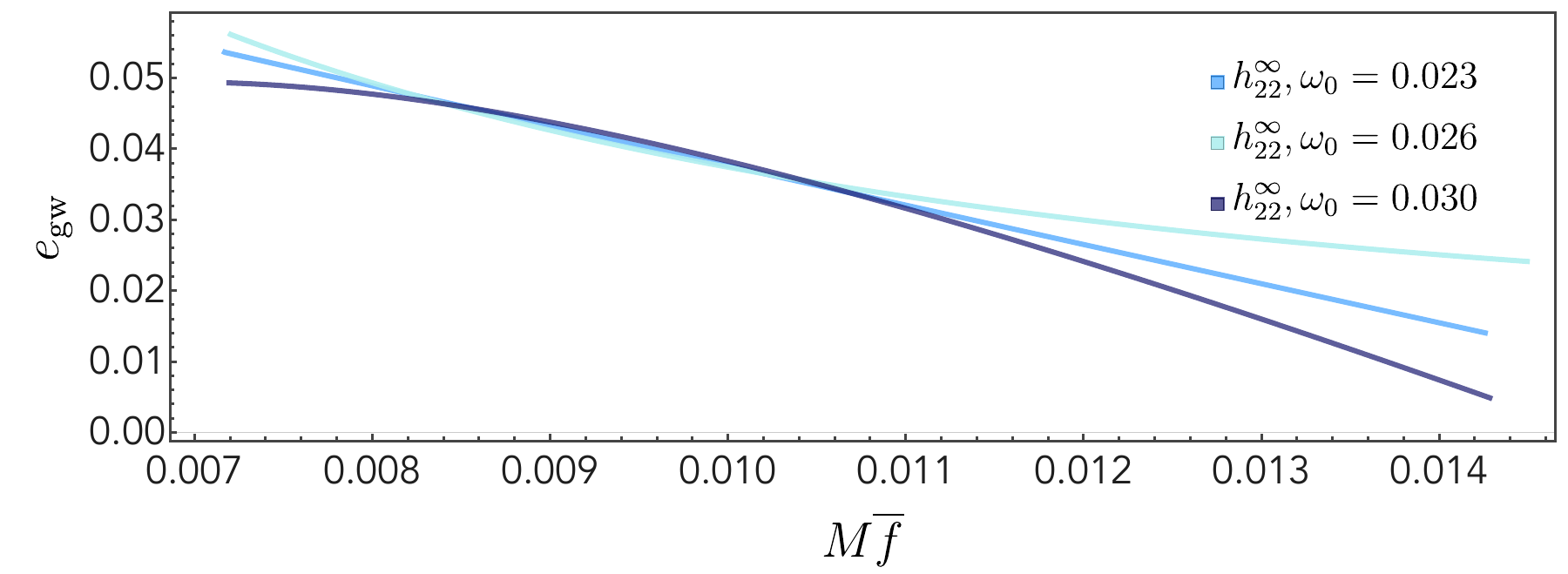}
\caption{\label{fig:FFIappendix} 
Eccentricity curves for \texttt{ID:0001} obtained by varying the FFI cutoff $\omega_0$. We note how the eccentricity evolution is dependent on the choice of the parameter.
}
\end{figure}

%Numerical errors}
\label{sec:appA}
In this section we illustrate other possible errors that may occur when we apply the procedure described in Sec.~\ref{sec:eccestimation}. 
In Fig.~\ref{fig:freqfilter}, we illustrate our filtering of $\Psi_4$, noting that due care needs to be taken not to introduce unphysical features into the data. 
Figure~\ref{fig:FFIappendix} shows how the eccentricity evolution depends on the choice of cutoff frequency $\omega_0$ for \texttt{ID:0001}. The evolution is strongly dependent on the choice of $\omega_0$ and this can introduce large numerical errors. As discussed in   
Sec.~\ref{sec:methodology}, taking $\omega_0 = \omega^{p}_{22}/2$ was found to be a robust choice for the cutoff frequency entering the FFI. 

\begin{figure}
\center
\includegraphics[width=0.48\columnwidth]{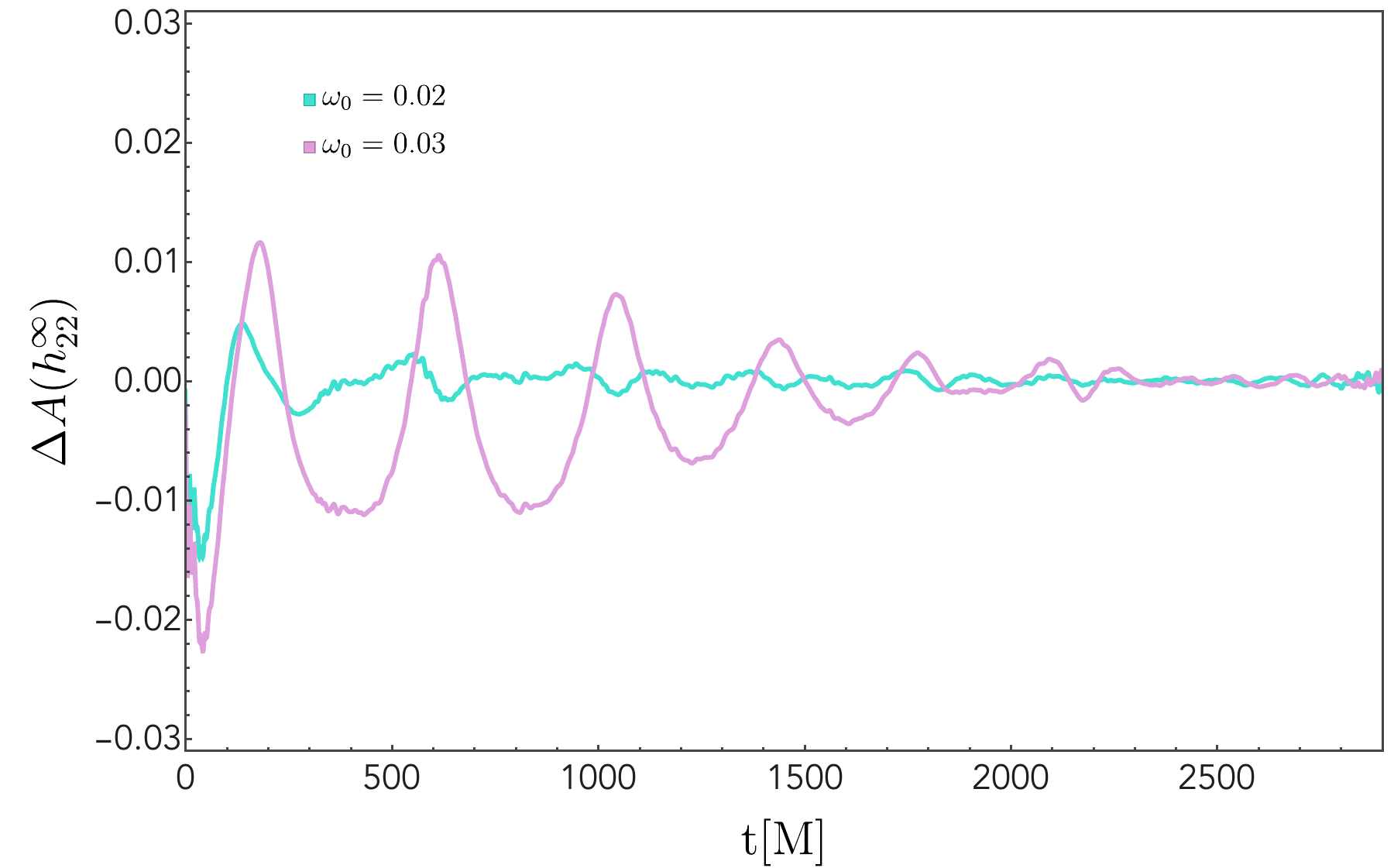}
\includegraphics[width=0.48\columnwidth]{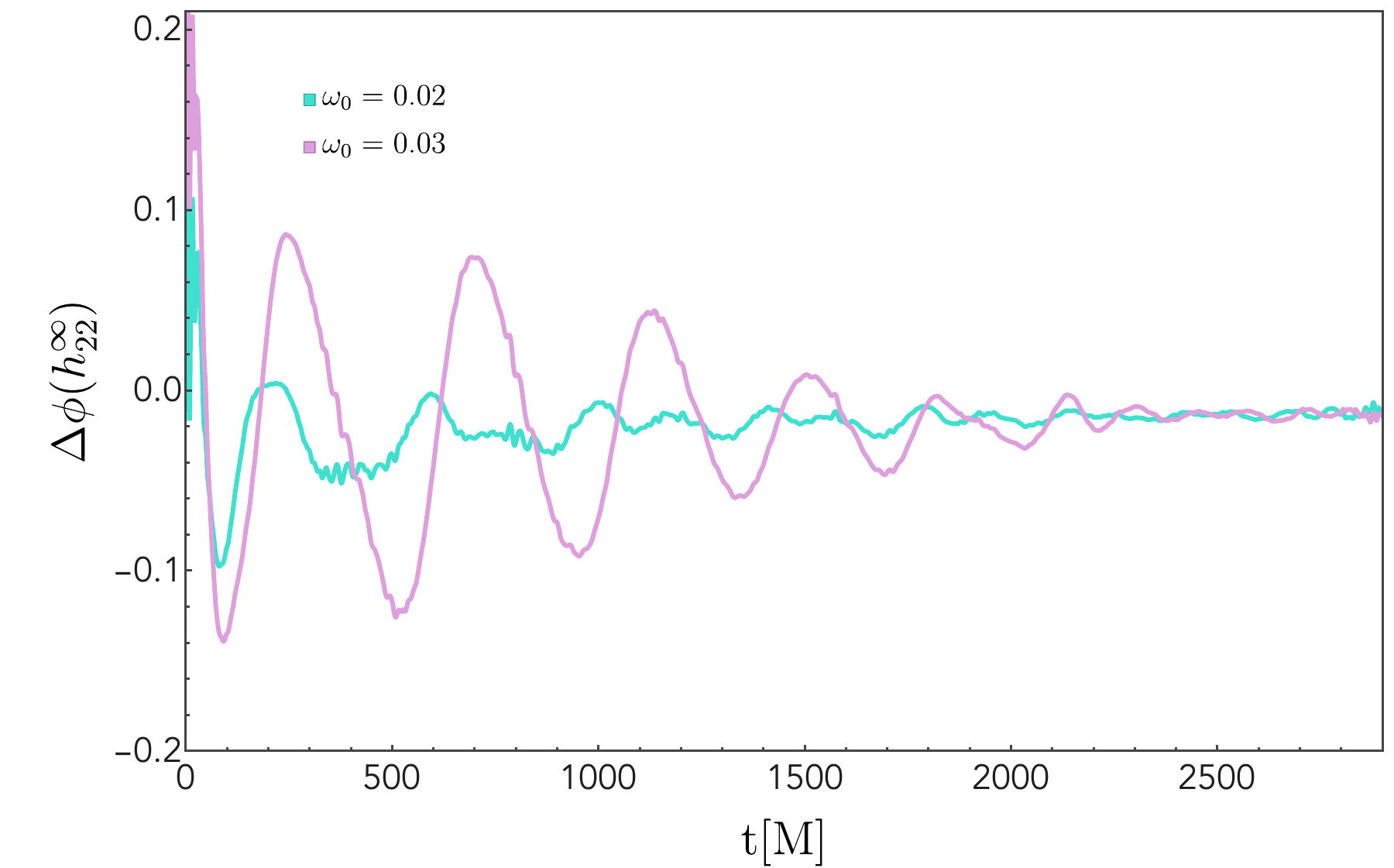}
\caption{\label{fig:SXScheck} 
Test of consistency between the strain extracted from the multipoles of the Newman-Penrose scalar, $\Psi_{4,\ell m}$, and the one calculated from the metric perturbations via the Regge-Wheeler-Zerilli equations (as discussed in Sec.~\ref{sec:methodology}), using \texttt{SXS:BBH:1360}. The left (right) panel shows the differences $\Delta A(h^{\infty}_{22}) = A^{\rm RWZ}_{22}- A^{\rm FFI}_{22}$ ($\Delta \phi(h^{\infty}_{22}) = \phi^{\rm RWZ}_{22}- \phi^{\rm FFI}_{22}$) between the amplitudes (phases) calculated using the RWZ formalism and the ones extracted from $\Psi_{4,\ell m}$ for different values of $\omega_0$. Different values of $\omega_0$ have a small impact on the $h^\infty_{22}$ inferred using the FFI, leading to a maximum difference after junk radiation of $\sim -0.02$ for the amplitudes and $\sim -0.15$ rad for the phases. We note that the waveform data were also extrapolated.}
\end{figure}

We also checked the agreement between the strain extracted from the multipole moments $\Psi_{4,\ell m}$ and the one directly
calculated from the metric perturbations via the Regge-Wheeler-Zerilli equations (as discussed in Sec.\ref{sec:methodology}) finding that small variations in $\omega_0$ only have a small impact on $h$ constructed using FFI as shown in Fig.~\ref{fig:SXScheck}.

\onecolumngrid

\newpage
% %~~~~~~~~~~~~~~~ Appendix B
% ~~~~~~~~~~~~~~~
\onecolumngrid
\section{Derivation of $e_{\Psi_4}$ in the quasi-Keplerian parametrization} 
\label{sec:appB}
In this appendix we derive the eccentricity estimator from the Newman-Penrose Weyl scalar $\Psi_4$ in  Eq.~\eqref{eq:egwpsi41pn} following the procedure to derive the eccentricity estimator from the waveform  in the quasi-Keplerian parametrization described in Appendix C of \cite{Ramos-Buades:2022lgf}. 
We start from the expression for the $(2,2)$-mode of the waveform $h_{22}^{\rm QK}$ given by Eq.(C26-C28) of \cite{Ramos-Buades:2022lgf} and compute $\Psi_4 \equiv - \ddot{h}
$. The time derivatives of $x$, $e_t$, the mean anomaly $l$ and the phase of the $(2,2)$-mode of the waveform $\phi \equiv \phi_{22}$ at 1PN are found in \cite{Hinder:2008kv, Konigsdorffer:2006zt} and the Kepler equation for the eccentric anomaly $u$ at Newtonian order is given by \cite{Ramos-Buades:2022lgf}
\begin{align}
\dot{u}  = \frac{ \dot{l} + \dot{e}_t \sin u}{1- e_t \cos u },
\end{align}
\newline 
with higher-order corrections entering at 2PN.

The time derivatives of $h_{22}^{\rm QK}(x, e_t, u, l, \phi)$ are expressed as
\begin{align}
\dot{h}_{22}^{\text{1PN}} &\equiv \frac{d h_{22}^{\rm QK}}{dt}= \frac{\partial h^{\rm QK}_{22}}{\partial x} \dot{x} + \frac{\partial h^{\rm QK}_{22}}{\partial e_t} \dot{e}_t + \frac{\partial h^{\rm QK}_{22}}{\partial u} \dot{u}+\frac{\partial h^{\rm QK}_{22}}{\partial l} \dot{l}  +\frac{\partial h^{\rm QK}_{22}}{\partial \phi} \dot{\phi},
\\
\ddot{h}_{22}^{\text{1PN}} &\equiv \frac{d \dot{h}_{22}^{\rm 1PN}}{dt}= \frac{\partial \dot{h}^{\rm 1PN}_{22}}{\partial x} \dot{x} + \frac{\partial \dot{h}^{\rm 1PN}_{22}}{\partial e_t} \dot{e}_t + \frac{\partial \dot{h}^{\rm 1PN}_{22}}{\partial u} \dot{u}+\frac{\partial \dot{h}^{\rm 1PN}_{22}}{\partial l} \dot{l}  +\frac{\partial \dot{h}^{\rm 1PN}_{22}}{\partial \phi} \dot{\phi}.
\end{align}
The 1PN expressions for  $\Psi_{4,22}^{\rm QK}$ can be written as 
\begin{align}
\Psi_{4, 22}^{\rm QK} & = 4 \nu x^4 \sqrt{\frac{\pi}{5}}\left[\hat{\Psi}_{4, 22}^{\text{0PN}}+\epsilon \hat{\Psi}_{4,22}^{\text{1PN}} \right] e^{-2 i \phi}, 
\label{eq:psi4QK}
\end{align}

\begin{align}
\hat{\Psi}_{4, 22}^{\rm 0PN}=\frac{-8 + e_t \cos (u)+ e_t^2 (6+\cos (2 u))+i 10  e_t \sqrt{1-e_t^2} \sin (u)}{\left(e_t \cos (u)-1\right){}^5},
\end{align}

\begin{align}
 \hat{\Psi}_{4, 22}^{\rm 1PN} & = \frac{x}{168 \left(1-e_t^2\right){}^{3/2} \left(e_t \cos (u)-1\right){}^7} \Bigg\{ 
  \nu  \left( e_t^2-1 \right) \Bigg[\sqrt{1-e_t^2} \left(e_t^4 (-338 \cos (2 u)+17 \cos (4 u)+149)-e_t^3 (485 \cos (u) \right.  \nonumber \\
  & \left.\left. +87 \cos (3 u))+4 e_t^2 (424 \cos (2 u)+431)  -4436 e_t \cos
   (u)+1760\right) \right.  \nonumber \\
   &  -14 i e_t \left(e_t \left[e_t \left(81-191 e_t^2\right) \sin (u)+5 e_t \left(5 e_t^2+1\right) \sin (3 u)+8 \left(16 e_t^2-31\right) \sin (2 u)\right] +260 \sin
   (u)\right)\Bigg] 
   \nonumber \\
   & +14 i e_t \left(e_t^2-1\right) \left[e_t \left(e_t \left(845-423 e_t^2\right) \sin (u)+45 e_t \left(e_t^2+5\right) \sin (3 u)+4 \left(17 e_t^2-287\right) \sin (2
   u)\right) +928 \sin (u)\right] \nonumber \\
   & +e_t \sqrt{1-e_t^2} \left[e_t^3 \left(-57 e_t^2 (\cos (4 u)+45)+1107 e_t \cos (3 u)+6 \left(515 e_t^2-2227\right) \cos (2 u)-279 \cos (4
   u)-5999\right) \right. \nonumber \\
   & \left. +e_t \left(-99 e_t \cos (3 u)+10944 \cos (2 u)+10180\right)+\left(6737 e_t^4+3851 e_t^2-16972\right) \cos (u)\right]+3424 \sqrt{1-e_t^2} \Bigg\}.
\end{align}

The phase of Eq.~\eqref{eq:psi4QK} can be written as 
\begin{align}
 \phi^{\rm 1PN}_{22}= \hat{\phi}^{\rm 0PN}(\Psi_{4, 22}^{\rm QK}) + \epsilon \hat{\phi}^{\rm 1PN}(\Psi_{4, 22}^{\rm QK}), 
 \label{eq:phasepsi4}
\end{align}

\begin{align}
\hat{\phi}^{\rm 0PN}(\Psi_{4, 22}^{\rm QK})=\text{Arctan}\left[\frac{20 e_t \left(e_t^2-1\right) \sin (u) \cos (2 \phi )+2 \sqrt{1-e_t^2} \sin (2 \phi ) \left(6 e_t^2+e_t (e_t \cos (2 u)+\cos
   (u))-8\right)}{20 e_t \left(e_t^2-1\right) \sin (u) \sin (2 \phi )-2 \sqrt{1-e_t^2} \cos (2 \phi ) \left(6 e_t^2+e_t (e_t \cos (2 u)+\cos
   (u))-8\right)}\right],
\end{align}

\begin{align}
\hat{\phi}^{\rm 1PN}(\Psi_{4, 22}^{\rm QK})&= \frac{A}{B},   
\end{align}
where
\begin{align}
A &= x e_t \left[2 \sin (u) \left((18339-4428 \nu) e_t^6+(5745 \nu -30328) e_t^4+(93 \nu -3965) e_t^2-2880 \nu +17424\right) \right. \nonumber \\
& \left. +e_t \left[2 \sin (2 u) \left(e_t^2 \left((2895 \nu
   -7804) e_t^2-3594 \nu +19202\right)+1854 \nu -12553\right) \right. \right. \nonumber \\
   & \left. \left. +e_t \left(15 e_t^2 \sin (5 u) \left((3 \nu -1) e_t^2+4 \nu -6\right)+e_t \sin (4 u) \left((753 \nu -3163) e_t^2-1068
   \nu +3478\right) \right. \right. \right. \nonumber  \\
   & \left. \left. \left. +\sin (3 u) \left(e_t^2 \left(3 (423 \nu -2759) e_t^2-5586 \nu +22582\right)+4002 \nu -13990\right)\right)\right]\right], \\
   B &= 42 \sqrt{1-e_t^2} \left(e_t \cos
   (u)-1\right){}^2 \left[e_t \left[\left(26 e_t^2-32\right) \cos (u)+e_t \left(e_t \left(e_t \cos (4 u)+2 \cos (3 u)\right) \right.\right. \right. \nonumber \\
   & \left.\left. \left.  +\left(124 e_t^2-131\right) \cos (2 u)\right)\right] -27
   e_t^4-91 e_t^2+128\right].
\end{align}
The frequency of $\Psi_{4, 22}^{\rm QK}$ can be expressed as the derivative of the Eq.~\eqref{eq:phasepsi4} as 

\begin{align}
\omega_{22}^{\text{1PN}} \equiv \frac{d\phi^{\rm 1PN}_{22}}{dt}= \frac{\partial \phi^{\rm 1PN}_{22}}{\partial x} \dot{x} + \frac{\partial \phi^{\rm 1PN}_{22}}{\partial e_t} \dot{e}_t + \frac{\partial \phi^{\rm 1PN}_{22}}{\partial u} \dot{u}+\frac{\partial \phi^{\rm 1PN}_{22}}{\partial l} \dot{l}  +\frac{\partial \phi^{\rm 1PN}_{22}}{\partial \phi} \dot{\phi}.
\end{align}
At the turning points, periastron ($u=0$) and apastron ($u=\pi$), the $(2,2)$-waveform frequency at 1PN can be written as
\begin{align}
 \omega_{22,p}^{\rm 1PN} &= 
 \frac{8 (e_t+1) (2+3 e_t)}{\sqrt{1-e_t^2} \left(7 e_t^2+e_t-8\right)} x^{3/2} \\ \nonumber &\qquad + \epsilon \, e_t \, x^{5/2} \frac{\left(9 (134 \nu -327) e_t^3+(2028 \nu -6094) e_t^2+(2238 \nu -823) e_t + 24 (52 \nu +125)\right)}{21 \sqrt{1-e_t^2} \left(7 e_t^2+e_t-8\right)^2}, 
 \label{eq:omg1pnper}
\end{align}
\newline 
and
\begin{align}
 \omega_{22,a}^{\rm 1PN} &= \frac{8 (2-3 e_t) \sqrt{1-e_t}}{(1+e_t)^{3/2} (7 e_t-8)} x^{3/2} \\ \nonumber &\qquad + \epsilon \, e_t \, x^{5/2} \frac{\sqrt{1-e_t^2} \left(-9 (134 \nu -327) e_t^3+(2028 \nu -6094) e_t^2+(823-2238 \nu ) e_t+24 (52 \nu +125)\right)}{21 \left(8-7 e_t\right){}^2
   \left(e_t-1\right) \left(e_t+1\right){}^3}. 
\label{eq:omg1pnap}
\end{align}

Substituting Eq.~\eqref{eq:omg1pnper} and Eq.~\eqref{eq:omg1pnap} into Eq.~\eqref{eq:e22}, and expanding up to 1PN, we find 

\begin{align}
 e_{\omega_{22}}(\Psi_4)& =\frac{21 e_t^4-\left(\sqrt{441 e_t^4-772 e_t^2+256}+15\right) e_t^2+\sqrt{441 e_t^4-772 e_t^2+256}-16}{32 e_t^3-42 e_t} \nonumber\\
 & -\frac{x \left(e_t \left(30 (54 \nu +269)
   e_t^4+(13194 \nu +1615) e_t^2-192 (52 \nu +125)\right)\right)}{84 \left(\sqrt{441 e_t^4-772 e_t^2+256} \left(21 e_t^4+\left(\sqrt{441 e_t^4-772 e_t^2+256}-15\right)
   e_t^2-\sqrt{441 e_t^4-772 e_t^2+256}-16\right)\right)}. 
\end{align}

Substituting in Eq.~\eqref{eq:egw} and Eq.~\eqref{eq:egw_psi} and expanding for small eccentricities we obtain Eq.~\eqref{eq:egwpsi41pn}

\begin{align}
e_{\Psi_4}= \frac{7}{4} e_t - \epsilon \, x \, e_t \bigg( \frac{52 \nu + 125}{168} \bigg).
\end{align}

\clearpage
\twocolumngrid

%~~~~~~~~~~~~~~~ Bibliography ~~~~~~~~~~~~~~~
\bibliography{ref.bib, local.bib}

\end{document}